\documentclass[10pt,a4paper,twocolumn,english,prb,aps,
showpacs,floatfix,groupedaddress,superscriptaddress,longbibliography]{revtex4-2}

\usepackage{graphicx}
\usepackage{epsfig}
\usepackage[english]{babel}
\usepackage{amsmath}
\usepackage{amssymb}
\usepackage{amsfonts}
\usepackage{physics}
\usepackage{siunitx}
\usepackage{fixmath}
\usepackage{color,array}
\usepackage{xcolor}
\usepackage{dsfont}
\usepackage{enumitem}
\usepackage{placeins}
\usepackage{nicefrac}
\usepackage[normalem]{ulem}
\usepackage{hyperref}

\def\dul#1{\underline{\underline{#1}}}
\newcommand{\emf}{^{\text{\tiny{(mf)}}}} 
\newcommand{\pemf}{^{\text{\tiny{(mf)}}}_{\phantom{\vec{V}}}} 
\newcommand{\esc}{^{\text{\tiny{(sts)}}}} 
\newcommand{\pesc}{^{\text{\tiny{(sts)}}}_{\phantom{\vec{V}}}} 

\newcommand{\be}{\begin{equation}}
\newcommand{\ee}{\end{equation}}
\newcommand{\ba}{\begin{align}}
\newcommand{\bs}{\begin{subequations}}
\newcommand{\es}{\end{subequations}}
\newcommand{\mc}{\mathcal}
\newcommand{\mb}{\boldsymbol}
\newcommand{\dagg}{\dagger}

\newcommand{\lnorm}{\left\vert\left\vert}
\newcommand{\rnorm}{\right\vert\right\vert}

\usepackage{mathtools}
\usepackage{bbold}

\usepackage{cleveref}

\begin{document}
\title{Dynamic mean-field theory for dense spin systems at infinite temperature}

\author{Timo Gr\"{a}\ss{}er}
\email{timo.graesser@tu-dortmund.de}
\affiliation{Condensed Matter Theory, TU Dortmund University,
Otto-Hahn Stra\ss{}e 4, 44221 Dortmund, Germany}

\author{Philip Bleicker}
\email{philip.bleicker@tu-dortmund.de}
\affiliation{Condensed Matter Theory, TU Dortmund University,
Otto-Hahn Stra\ss{}e 4, 44221 Dortmund, Germany}

\author{Dag-Bj\"{o}rn Hering}
\email{dag.hering@tu-dortmund.de}
\affiliation{Condensed Matter Theory, TU Dortmund University,
Otto-Hahn Stra\ss{}e 4, 44221 Dortmund, Germany}

\author{Mohsen Yarmohammadi}
\email{mohsen.yarmohammadi@tu-dortmund.de}
\affiliation{Condensed Matter Theory, TU Dortmund University,
Otto-Hahn Stra\ss{}e 4, 44221 Dortmund, Germany}


\author{G\"otz S.~Uhrig}
\email{goetz.uhrig@tu-dortmund.de}
\affiliation{Condensed Matter Theory, TU Dortmund University,
Otto-Hahn Stra\ss{}e 4, 44221 Dortmund, Germany}

\date{\rm\today}

\begin{abstract}
A dynamic mean-field theory for spin ensembles (spinDMFT) at infinite temperatures
on arbitrary lattices is established. The approach is introduced for an isotropic 
Heisenberg model with $S = \tfrac12$ and external field. 
For large coordination numbers, it is shown that the effect of the environment of
 each spin is captured by a classical time-dependent random mean-field which is normally distributed. Expectation values are calculated by averaging over {these
mean-fields}, i.e., by a path integral over the normal distributions.
A self-consistency condition is derived by linking the {moments defining
the normal distributions} to spin autocorrelations. In this framework, we explicitly show 
how the rotating wave approximation becomes a valid description for increasing
magnetic field. We also demonstrate that the approach can easily be extended.
Exemplarily, we employ it to reach a quantitative understanding of a dense
ensemble of spins with dipolar interaction which are distributed randomly on 
a plane including static Gaussian noise as well.
\end{abstract}

\maketitle

\section{Introduction}

Nuclear magnetic resonance (NMR) has been an extremely important field {for a long time}.
On the one hand, it constitutes a powerful analytical technique in physical chemistry
\cite{haebe76,ernst87,slich96,levit05} which helps to understand the structure of 
molecules on all levels from their primary structure
to their tertiary structure. One the other hand, it is a technique which has
enabled fundamental steps in quantum computing by taking spins $S={\tfrac12}$ as quantum bits
\cite{niels00}. The latter development illustrates that 
the dynamics of the spin degree of freedom has gained enormous attention in particular
in recent years. Closely related is the rapid development of the field of quantum 
sensing based on NV-centers in diamond 
\cite{dolde11,grotz11,lange11b,schaf11,stein13,sushk14,rossk14,grino14}
which behave {similarly} to an elementary spin \cite{jelez06}. 

A key phenomenon in this field is decoherence, i.e., the loss of coherence of a small quantum system
in contact with a larger environment, often called bath. A generic
approach to small systems in weak contact with a large bath is the theory of open
quantum systems \cite{breue06}. This is a powerful approach if the {energy
scales} of system and bath are very different. If the bath
correlations decay much faster than the system's dynamics quantum master equations
reliably capture the physics, for example
in radiative decay processes. If, however, the separation of energy scales is
not given and the back-action of the system onto the bath cannot be neglected
a quantitative description is notoriously difficult.

In the context of the
coherent control of spins, the small quantum system generically is a single spin.
The decoherence can result from a fluctuating environment, for instance from stray magnetic fields or from phonons
which may be fast. But very often it results from surrounding spins  of electronic or nuclear origin.
This is the case often relevant in NMR and in sensing by NV-centers. Then
the back-action of the considered spin onto its neighboring spins is important and
cannot be neglected. 

{While we cannot provide a comprehensive review over all techniques applicable to spin systems, we present a brief overview of the most commonly used techniques. 
This allows us to highlight differences to the approach we are proposing in this article.}

For {very small bath sizes of only} a few spins the resulting problem
can be tackled by exact diagonalization (ED). {The Chebyshev polynomial expansion technique (CET) allows for substantially larger but still comparably small finite bath sizes}
\cite{talez84,weiss06a}. If, however,
the degrees of freedom of the bath are too numerous then brute force
numerical approaches cannot be applied due to the exponential growth 
of the Hilbert space with the number of bath spins. For certain 
geometries such as chains and stars density-matrix renormalization \cite{schol05,stane13}
provides numerical alternatives. But the maximum times {which} can be reached
are limited. {
For approximately star-like topologies, like the one of the central spin model,
cluster expansions \cite{witze05,witze06} and related methods \cite{lindo18}, linked-cluster expansions \cite{saiki06,rigol14} and cluster-correlation expansions \cite{yang08a,yang09,zhang20} are prominent approaches. But, these approaches become cumbersome for lattices with many different bonds. In addition,
they represent expansions in time so that the reachable maximum time
is limited by the complexity of the tractable maximum clusters.
The eminent problem with (occasionally constrained) Monte Carlo sampling methods \cite{sandv91,janke96,assel10} is that statistical errors can become very substantial, see in particular Ref.~\cite{farib13a} concerning the effects for up to $N=48$ nuclear spins.
By means of semi-classical or quantum mechanical master equations for the density matrix of the whole system \cite{kawas66,prose11,sanche19,duboi21}, 
an access to the overall dynamics can be obtained which is typically easy to realize, but potentially suffers from the drawbacks of mean-field approaches 
if these are not based on small expansion parameters.}

Hence, alternative techniques are of significant interest. One useful observation is that the
dynamics of spins and in particular the effect of a larger
number of spins can  be captured fairly well by their classical equations of
motion \cite{erlin04,chen07,stane14b,scher18,lindo20}. This can be understood
as an application of the ideas of the truncated Wigner approximation
\cite{polko10,zhu19} whose foundations date back to the idea of
Wigner that a part of the quantumness is captured by averaging 
over distributions of initial conditions \cite{wigne32}. 
But it is conceptually very difficult
to extend this approach systematically to take more and more quantum effects
into account. Apart from the smallness of Planck's constant $\hbar$ no small
parameter is apparent.

In the present article we deal with dense spin systems where each spin interacts
with a large number of other spins. In the limit where this number of
interaction partners becomes infinite we derive a dynamic mean-field theory
for the spin dynamics (spinDMFT) at infinite temperature, i.e., for completely
 disordered spins. As in all mean-field theories,
spinDMFT comprises an effective single-site problem \emph{and} a self-consistency
condition. Similar to the case of the established fermionic dynamic mean-field theories
\cite{georg96} the {time dependence} of the mean-field is a crucial ingredient. 
It bears similarities to the mean-field approach for the Sherrington-Kirkpatrick quantum model
in which spin glass behavior has been established 
\cite{bray80,sachd93,gremp98,georg00a}.
A dynamic mean-field approach has also been used for ordered magnetic phases
\cite{otsu13} for which, however, the couplings between the spins have to be 
scaled differently.

After this introduction, we derive the spinDMFT in Sect.~\ref{sec:approach}
in consecutive steps for an isotropic Heisenberg model and 
discuss details of the numerical implementation. Subsequently,
we compare the results of spinDMFT for several systems with results
obtained by CET and iterated equations of motion \cite{kalth17,bleic18} in Sect.~\ref{sec:gauging}.
Sect.~\ref{sec:applicationdipolar}
is devoted to the application of spinDMFT to two-dimensional
spin ensembles in which the spins couple via dipolar interactions.
In particular, we can continuously show how the well-known {rotating wave approximation (RWA)} becomes more and more 
reliable for increasing external magnetic field. In Sect.~\ref{sec:conclusion} we conclude the article and give
an outlook to future directions of research. The appendices provide technical details of 
the derivation and the
numerical implementation of spinDMFT including an analysis of the 
achievable accuracy in the numerical simulations.

\section{Approach}
\label{sec:approach}

\subsection{Model and Definitions}
\label{subsec:model}

For concreteness, we consider an isotropic Heisenberg model for an ensemble
of spins with $S=\tfrac12$ at infinite temperature. The spins are subjected 
to a static and homogeneous magnetic field
aligned in the $z$ direction so that the Hamiltonian reads as
\begin{align}
	\mb{H} &= \sum_{i<j} J_{ij} \, 
	\vec{\mb{S}}_{i} \cdot \vec{\mb{S}}_{j} +
	\gamma_{\text{s}} B \sum_{i} 	\mb{S}^{z}_{i}.
	\label{eqn:isoHam}
\end{align}
Here and henceforth we use bold symbols for quantum mechanical operators
and three-dimensional vectors are indicated by the arrow above the symbol.
The properties of the underlying spin lattice are encoded in the 
couplings $J_{ij} = J_{ji}$.
It is useful to introduce the operators of the local environments of each spin
\be
\label{eq:Vdef}
	\vec{\mb{V}}_{i} := \sum_{j, j\neq i} J_{ij} \vec{\mb{S}}_{j}.
\ee
Using them the Hamiltonian can be expressed as
\ba
	\mb{H} &= \frac12 \sum_{i} \vec{\mb{S}}_{i}
	\cdot \vec{\mb{V}}_{i} + 	\gamma_{\text{s}} B \sum_{i} \mb{S}^{z}_{i}.
	\label{eqn:isoHamV}
\end{align}
The prefactor $\tfrac12$ occurs here to avoid double counting of the couplings.
For later purposes, we also introduce the {moments} of the coupling constants
\ba
\label{eq:momenta_def}
	\mc{J}_{m,i} &:= \Bigl(\sum_{j} {\vert}J_{ij}{\vert}^{m} \Bigr)^{1/m}
\end{align}
and the effective coordination numbers depending on the site $i$
\ba
\label{eq:coord_num_def}
	z_{i} &:= \frac{\mc{J}_{1,i}^2}{\mc{J}_{2,i}^2}, & z'_{i} &:= \frac{\mc{J}_{2,i}^4}{\mc{J}_{4,i}^4}.
\end{align}
Note that we do not restrict the model to periodic lattices, but
include arbitrary clusters. Both coordination numbers assess the number of spins that constitute the environment of site $i$. Considering only 
constant nearest-neighbor (NN) interactions both numbers $z_i$ and
$z'_i$ equal the number of nearest neighbors
$z_{\text{NN},i}~=~z_{i}~=~z_{i}'$ which is the usual coordination number.

A common property of mean-field approaches is that they become exact in the 
limit $z \to \infty$ \cite{georg96}. Therefore,
$1/z$ serves as a control parameter allowing us to systematically neglect terms in 
non-leading order in $1/z$.
Hence, we examine several quantities with respect to their scaling with the effective coordination numbers.
We will show that the spinDMFT becomes exact 
in the limit of infinite $z_{i}$ and $z'_{i}$.

As a consequence, the approach is not optimum for
 low-dimensional systems. However, since we consider \emph{effective} coordination numbers instead of the standard one, not only the dimensionality but also the 
overall behavior of the coupling constants matters.
Obviously, long-range couplings will increase the effective coordination 
numbers at given, fixed dimension.
 In Sect.~\ref{sec:applicationdipolar}, we demonstrate that in case of dipolar
couplings, i.e., for weakly decreasing couplings with {the} distance,
our approach is successful even in two dimensions.

We establish the dynamic mean-field theory for spins (spinDMFT) 
for the introduced model. This is done in four steps:
\begin{enumerate}[label=(\roman*)]
	\item \label{item:step1} We replace the local-environment operators 
	by classical time-dependent random local mean-fields.
	\item \label{item:step2} We argue that the dynamics of 
	the local mean-field at site $i$ does not depend on the dynamics of the 
	\emph{single} spin at site $i$.	
	\item \label{item:step3} We show that the mean-fields are normally distributed.
		\item \label{item:step4} The defining moments of the normal distributions 
		are linked to spin autocorrelations yielding a closed set of self-consistency equations.
\end{enumerate}

Since we consider infinite temperature, the density operator is given by $\mathds{1}/d$, 
where $d$ is the dimension of the Hilbert space. Thus, any quantum expectation values {are}
given by
\bs
\be
	\langle \mb{A} \rangle = \frac1{d} \text{Tr} \left( \mb{A} \right)
\ee
and, consequently, any correlation by
\be
	\langle \mb{A}(t) \mb{B}(0) \rangle = \frac1{d} \text{Tr} \left( e^{it{\mb{H}}}\mb{A} e^{-it{\mb{H}}} \mb{B}\right) ,
\ee
\es
where we {have set $\hbar=1$}.
In the next section, we undertake the first and the second step, \ref{item:step1}
and \ref{item:step2}.

\subsection{From the spin ensemble to an effective single site}
\label{subsec:lattimp}

\subsubsection{Step (i)}
We justify the substitution of the local-environment operators $\vec{\mb{V}}_{i}(t)$ 
by classical fields. For this to hold, it is crucial that the spin ensemble is \emph{dense}
so that quantum fluctuations of the environment are negligible relative to the classical
dynamics. The argument is adapted from Ref.\ \onlinecite{stane13} and runs as follows.
We consider the Frobenius norm of an operator defined by
\be
	\lnorm \mb{A} \rnorm^2 = \frac1{d} \text{Tr} \left( \mb{A}^{\dagg} \mb{A} \right)
\ee
and apply it to 
\be
	\lnorm \mb{V}^{\alpha}_{i} \rnorm^2 = \frac{\mc{J}_{2,i}^2}{4}
\ee
for each component of the local-environment operator. To assess the 
role of the coordination numbers, we assume that the $\mc{J}_{2,i}$
are of the same order of magnitude at every site. They set the
relevant energy scale which one should think of being constant
when scaling the coordination numbers, i.e., the individual
couplings scale roughly like $J_{ij} \propto 1/\sqrt{z}$.

For a classical
variable, any commutator would vanish. Hence, we study the commutators of
the local-environment operators and compare their norm to the one
of the $\vec{\mb{V}}_{i}$ themselves
\be
	\lnorm [\mb{V}^{\alpha}_{i},\mb{V}^{\beta}_{i}] \rnorm^2 = 
	\frac{\mc{J}_{4,i}^4}4 = \frac1{4} \frac{\mc{J}_{2,i}^4}{z'_{i}}
\ee
for $\alpha \neq \beta$; for $\alpha=\beta$ the commutator vanishes.
Clearly, for diverging effective coordination number $z'_i\to\infty$
the commutator vanishes relative to the norm of the operator. Hence 
its quantumness becomes negligible and the local-environment operators
can be replaced by classical mean-fields 
$\vec{\mb V}_i \to \vec{V}_i$. Note that this is a very
common phenomenon in quantum mechanics. Quantities which represent
the collective properties of a large number of constituents
behave classically. We stress, however, that this argument does
\emph{not} imply that the classical field is static. Hence,
we avoid this oversimplification and take the mean-fields
as classical, but time-dependent and dynamic. A potential correlation
between $\vec{V}_i$ and $\vec{\mb S}_i$ is not ruled out at this stage.

\subsubsection{Step (ii)}

Here, the aim is to show that the dynamics of the individual spin at site $i$ does not 
influence the dynamics of $\vec V_i$ in the limit of $z_i\to \infty$. The basic
idea is simple: a single spin contributes only negligibly to the large
sum defining $\vec V_i$. But it is not straightforward to cast this idea into
a formal argument. What we want to show is that the dynamics of
$\vec{\mb{S}}_i$ does not influence the dynamics of $\vec V_i$, i.e., that
no back-action needs to be taken into account. Indeed, we show
in Appendix A that the correlation between
the spin at site $i$ and at an adjacent site $j$ scales like $1/z$
for the special case of a Bethe lattice with NN
interaction, where $z=z_i=z'_i \; \forall i$ holds.
Hence the correlation between the spin at site $i$ and its local environment
$\vec V_i$ scales like $J\propto 1/\sqrt{z}$ and becomes negligible for
$z\to\infty$. The number of spins in $\vec V_i$ scales like
$z$ compensating the factor $1/z$ from the correlations.

We stress that this conclusion is subtle. It is valid if the dynamics of
$\vec V_i$ is induced by a process of order $z^0$ because 
the relative error then is indeed of order $z^{-1/2}$. 
But if there is no process inducing a dynamics of order $z^0$
this does not hold true. Indeed, the central spin model (CSM)
provides an instructive example. In this model, a central spin is
coupled to a large number of bath spins, but the bath spins
are not coupled among themselves
\be
	\mb{H}_{\text{CSM}} = \vec{\mb{S}}_{0} \cdot \sum_{i=1}^{n} C_{i} \vec{\mb{S}}_{i} = \vec{\mb{S}}_{0} \cdot \vec{\mb{P}},
\ee
wherein $C_{i}$ are arbitrary coupling constants and $\vec{\mb{P}}$ denotes the 
so-called Overhauser field.  This looks like a perfect scenario for replacing
the $\vec{\mb{P}}$ by a classical Overhauser field $\vec P$ with a given
dynamics imposed on the central spin. Yet, this approach fails
\cite{stane13,stane14b,stane14c} because the Overhauser field has \emph{no}
dynamics if the central spin is taken out and treated separately.
This reasoning shows us that it is  essential at this stage to deal
with a \emph{dense} spin system where each spin interacts with 
many others so that excluding one of them from the dynamics
hardly changes the dynamics of all others. This is the case if
the coordination number at \emph{each} site is large.
We point out that this is clearly not the case in the CSM where
excluding the central spin brings all dynamics to a halt and where the
coordination number $z$ for each bath spin is only 1.

On the basis of the above arguments, we replace each local-environment 
field $\vec{\mb{V}}_{i}$ in the Hamiltonian \eqref{eqn:isoHamV} by an
\emph{a priori} given dynamic
mean-field $\vec{V}_{i}(t)$ so that the spin dynamics is given by
\bs
\ba
	\mb{H}_{\text{mf}}(t) &= \sum_{i} \mb{H}_{\text{mf},i}(t)
	\\
	\mb{H}_{\text{mf},i}(t) &= 
	\vec{V}_{i}(t) \cdot \vec{\mb{S}}_{i} + \gamma_{\text{s}} B \mb{S}^{z}_{i}.
	\label{eqn:localmfHamiltonian}
\end{align}
\label{eqn:mfHamiltonian}
\es
These mean-field Hamiltonians, labeled by the subscript ``mf'', 
only contain linear spin-operator terms so that 
the spin dynamics at a given site is decoupled from the one at other sites
once {the mean-fields $\vec{V}_{i}(t)$ are given as function of time}. In order to 
emphasize that not only single values $\vec{V}_{i}(t)$ are meant, but
the whole time dependence we introduce the shorthand $\vec{\mc{V}}_{i}$
for it. These time series
are drawn from a so far unknown probability functional 
$p\left[\vec{\mc{V}}_{1},...,\vec{\mc{V}}_{N}\right]$ 
which we will determine in the next section.
In conclusion, the original $2^{N}$-dimensional ensemble is mapped to 
$N$ two-dimensional quantum impurity systems
each capturing a single spin subjected to 
a {time-dependent} mean-field and the external field.

\subsection{Distribution of the mean-fields $\vec{\mc{V}}_{i}$}
\label{subsec:distribution}

Here we carry {out step} \ref{item:step3} and step \ref{item:step4} from the list in Sect.\ 
\ref{subsec:model}. 

\subsubsection{Step (iii)}

The central argument is again that two spins at
site $i$ and $j$ are only weakly correlated. This is difficult to
show for an arbitrary cluster with arbitrarily linked spins. But for
the Bethe lattice with NN interaction we demonstrate
in  Appendix A that the correlation 
$\langle \mb{S}^\alpha_i(t) \mb{S}^\beta_j(0)\rangle$ scales like
$z^{-||i-j||}$ where $ ||i-j||$ is the number of NN links needed 
 to go from site $i$ to $j$. {Moreover, we show that} the correlation
$\langle V^\alpha_i(t) V^\beta_j(0)\rangle$ {is suppressed
at least like} $1/z$ for $i\ne j$. Thus, we conclude that the
time series of the local mean fields $\vec{\mc{V}}_i$ 
are independent at different sites
\be
	p\left[\vec{\mc{V}}_{1},\dots ,\vec{\mc{V}}_{N}\right] 
	= \prod_{i} p_{i}\left[\vec{\mc{V}}_{i}\right] \ .
	\label{eqn:factorization}
\ee
This allows us to compute any local expectation value
by
\bs
\label{eq:singletime}
\ba
\langle \mb A_i(t) \rangle &= 
\int \mc{D}\vec{\mc{V}}_{i} \, p_{i}\left[\vec{\mc{V}}_{i}\right] \, 
\langle 	\mb{U}_{i}^{\dagg}\left(t,t_0\right) \mb{A}\, 
\mb{U}_{i}\left(t,t_0\right) \rangle\esc_{\vec{\mc{V}}_i}
\\
&= \frac12 \text{Tr} \left( \mb{A} \right).
\label{eq_secondA}
\end{align}
\es
where $\langle\ldots\rangle\esc_{\vec{\mc{V}}_i}$ stands
for the expectation value for a given {\bf s}ingle {\bf t}ime {\bf s}eries 
$\vec{\mc{V}}_i$. This
contribution is weighted by the probability $p_{i}\left[\vec{\mc{V}}_{i}\right]$
to reach the total average. The unitary time evolution $\mb U_i(t,t_0)$ 
is the solution of the Schr\"odinger equation
\be
\frac{d}{dt}\mb U_i(t,t_0) = -i\mb{H}_{\text{mf},i}(t) \mb U_i(t,t_0)
\ee
for the initial condition $\mb U_i(t_0,t_0)=\mathds{1}$. 
For future use, it is worth mentioning that the unitary evolution
operator $\mb U_i(t,t_0) $ only depends on the mean-field time 
series between $t_0$ and $t$ and not on all times.
Hence, the computation of a time-dependent expectation 
value only requires to average over mean-field time series 
within the relevant time interval.

For the case in \eqref{eq:singletime}, it turns {out} that 
no averaging over time series is necessary at all.
 Since we assume that the system is completely disordered at $t_0$, i.e.,
the density matrix is proportional to the identity, the
unitary evolution cancels out and one arrives at the
second line \eqref{eq_secondA}. Averaging actually
does not matter here.

For correlations, the evolution in time does matter and hence
does the averaging over the time series. We consider
\begin{widetext}
\bs
\label{eq:correlation}
\ba
 \langle \mb{A}_{i}(t_1) \mb{B}_{i}(t_2) \rangle\pemf &:= 
\int \mc{D}\vec{\mc{V}}_{i} \, p_{i}\left[\vec{\mc{V}}_{i}\right]  \langle
	\mb{U}_{i}^{\dagg}\left(t_1,t_0\right) \mb{A}_i\, \mb{U}_{i}\left(t_1,t_0\right) 
	\mb{U}_{i}^{\dagg}\left(t_2,t_0\right)
	\mb{B}_i\, \mb{U}_{i}\left(t_2,t_0\right) \rangle\esc_{\vec{\mc{V}}_i}
	\\
	&= \int \mc{D}\vec{\mc{V}}_{i} \, p_{i}\left[\vec{\mc{V}}_{i}\right] \langle
	\mb{U}_{i}^{\dagg}\left(t_1,t_2\right) \mb{A}_i\, \mb{U}_{i}\left(t_1,t_2\right) 
		\mb{B}_i \rangle\esc_{\vec{\mc{V}}_i}
\label{eq:correlation_simple}		
\end{align}
\es
\end{widetext}
which clearly only depends on the time interval $[t_1,t_2]$. Note that
temporal homogeneity  is not given for a \emph{single} time series, but it
 holds on average so that 
$\langle \mb{A}_{i}(t_1) \mb{B}_{i}(t_2) \rangle\pemf
= \langle \mb{A}_{i}(t_1-t_2) \mb{B}_{i}(0) \rangle\pemf $ holds.

The next, important conclusion is that each local environment
$\vec{\mc{V}}_i$ is the sum of a large number of essentially independent
spins \eqref{eq:Vdef}. This number becomes infinite for diverging
coordination number so that the central limit theorem applies
and we conclude that the $\vec{\mc{V}}_i$ are normally distributed.
This means that we need only two moments, {the} first and second, to determine
the distribution. This brings us to the fourth and final step.

\subsubsection{Step (iv)}
\label{sss:selfconscond}

We establish self-consistency conditions which link the
first and second moments of the normal distribution to
 quantum expectation values and correlations.

For the first moment, it is straightforward to see from Eq.~\eqref{eq:singletime}
that it vanishes 
\be
\langle \mb{V}^\alpha_i(t) \rangle = \frac12 \text{Tr} \left(\mb{V}^\alpha_i \right) = 0 
\ee
for all sites $i$, all components $\alpha\in\{x,y,z\}$, and all times $t$
because we start from the disordered, $T=\infty$ case where the expectation
values of all spin operators vanish. Hence, we conclude that
the distribution $p_i(\vec{\mc{V}})$ is a normal distribution with vanishing 
first moments
\be
\overline{{V}^\alpha_{{i}}{(t)}} := \int \mc{D}\vec{\mc{V}} p_i\left[\vec{\mc{V}}\right]  
{V}^\alpha =0 .
\ee
This is the first self-consistency condition which is easy to fulfill.
The spinDMFT can also be extended to include non-vanishing first moments
and even time-dependent moments, but this is not the scope of the present
article.

For the second {moments}, we consider
\bs
\label{eqn:selfcons}
\begin{align}
	 \langle \mb{V}_{i}^{\alpha}(t_1)\mb{V}_{i}^{\beta}(t_2) \rangle &= 
	\sum_{j,k \neq i} J_{ji} J_{ki} \langle \mb{S}^{\alpha}_{j}(t_1) 
	\mb{S}_{k}^{\beta}(t_2) \rangle
	\\
	&= \sum_{k\neq i} J_{ki}^2 \langle \mb{S}^{\alpha}_{k}(t_1) \mb{S}_{k}^{\beta}(t_2) \rangle\pemf
	\\
&= 	\sum_{k\neq i} J_{ki}^2 \langle \mb{S}^{\alpha}_{k}(t_1-t_2) \mb{S}_{k}^{\beta}(0) 
\rangle\pemf ,
\end{align}
\es
where the second line results from the fact that the spin-spin correlations
between different sites vanish in the limit of infinite coordination number.
The third line, finally, results from \eqref{eq:correlation} on average.
Self-consistency requires that the second moment computed 
above equals the correlations of the local mean-fields, i.e.,
\bs
\label{eq:selfcon}
\ba
\overline{V_{i}^{\alpha}(t_1)V_{i}^{\beta}(t_2)} &\overset{!}{=}
\sum_{k\neq i} J_{ki}^2 \langle \mb{S}^{\alpha}_{k}(t_1-t_2) \mb{S}_{k}^{\beta}(0) \rangle\pemf
\\
&= \overline{V_{i}^{\alpha}(t_1-t_2)V_{i}^{\beta}(0)}.
\end{align}
\es
This closes the set of self-consistency conditions. If the second moments only
depend on the time difference $t_1-t_2$ the resulting two-time spin expectation values
only depend on $t_1-t_2$. Hence solutions homogeneous in time
exist. Whether they are the only conceivable solutions is
an additional question which we do not study in this paper and leave 
for future research.


{
At this stage, we observe the interesting feature that
the resulting mean-field theory is the same that one
would obtain for classical spins of the same average length.
This is so since the effective single-site problem in Eq.~\eqref{eqn:mfHamiltonian}
only contains the spin operator linearly. According to the
Ehrenfest theorem the quantum mechanical expectation values 
behave identical to classical variables. We conclude that 
the classical and quantum mechanical spin system converge to
the same spinDMFT for infinite coordination number.
We emphasize, however, that for spins larger than $1/2$ non-linear
local terms may arise, for instance from quadrupolar 
couplings \cite{glazo18}. Then there is a difference
between the quantum and the classical spinDMFT.
}


Henceforth, we use the term `autocorrelation' to denote the local spin-spin correlation
$\langle \mb{S}^{\alpha}_{i}(t_1) \mb{S}_{i}^{\beta}(t_2) \rangle$. Later, when numerical results are presented, 
we will also distinguish between diagonal autocorrelations ($\alpha=\beta$) and cross autocorrelations ($\alpha\neq \beta$).

The message of Eq.~\eqref{eq:correlation} is that one can 
compute the autocorrelations at each site if one knows the moments
$\overline{V_{i}^{\alpha}(t)V_{i}^{\beta}(0)}$ defining the normal distribution
of $\vec{\mc{V}}_i$. In return, Eq.~\eqref{eq:selfcon} tells us
that the knowledge of the autocorrelations of the spins linked to site $i$
 {yields} the second moments of  $\vec{\mc{V}}_i$.
It is to be expected that this closed set of self-consistency equations
can be solved iteratively and our numerical results 
confirm that this is true. Numerical aspects will be discussed in the
next section.

In the present paper, we do not intend to use spinDMFT for problems with
spatial dependence even though the general formalism derived so far
allows for such spatial dependencies. But the concomitant numerical 
task is quite demanding. Our goal here is first to introduce the
approach of spinDMFT and to illustrate its performance. To this end, we
opt to consider homogeneous spin ensembles where each site is equivalent
to every other site. Certainly, this is the case for periodic lattices
but it can also hold for dense random spin ensembles where each spin
is interacting \emph{on average} with the same number of spins and 
with the same interaction strength. Then, all 
autocorrelations are the same and hence all second moments
of the local mean-fields. Then the self-consistency condition
\eqref{eq:selfcon} simplifies considerably because the autocorrelations
on the right hand side can be taken out of the sum. The site-independent
second moments read as
\be
	\overline{V^{\alpha}(t) V^{\beta}(0)} = 
	\mc{J}_{2}^2 \langle \mb{S}^{\alpha}(t) \mb{S}^{\beta}(0) \rangle\pemf.
	\label{eqn:final_selfcons}
\ee
Since all sites are equivalent, no site indices need to be
denoted. Interestingly, the \emph{only} energy constant governing the
spin dynamics aside from the external magnetic field
is the root-mean-square $\mc{J}_2$ of the couplings.

\subsection{Numerical implementation}
\label{subsec:numproc}

\begin{figure}[ht]
	\centering
	\includegraphics[width=0.48\textwidth]{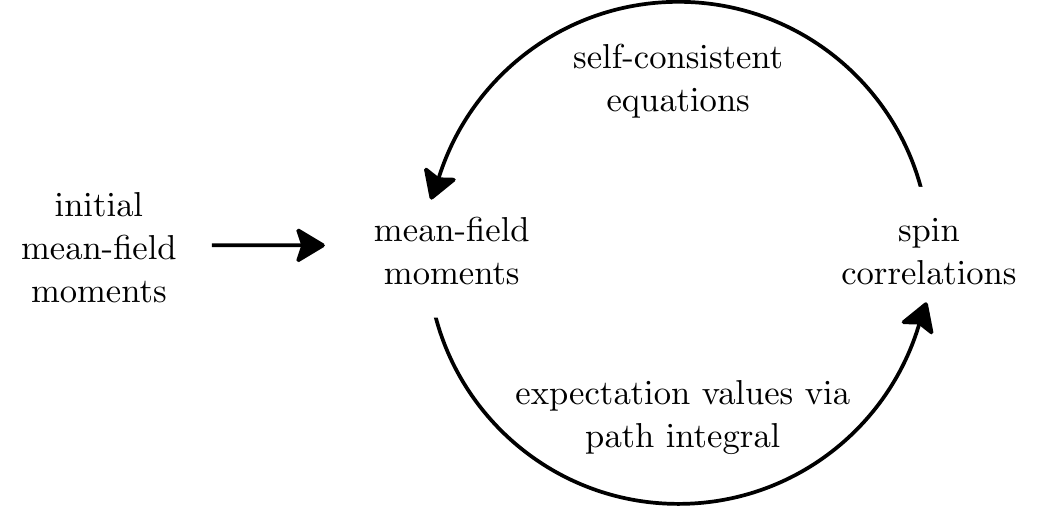}
	\caption{Scheme of the iteration procedure.}
	\label{fig:iteration_illustration}
\end{figure}

Our aim is to implement a numerical procedure by which the mean-field moments 
determined by the self-consistent equations can be evaluated.
The basic idea is to start with some arbitrary initial function and to 
converge iteratively to the solution. In each iteration step, one
computes the autocorrelations for a single spin via the path integral 
\eqref{eq:correlation_simple} and subsequently the mean-field moments 
via the self-consistent equations \eqref{eqn:final_selfcons}. This scheme is 
illustrated in Fig.~\ref{fig:iteration_illustration}.

The computation of the path integral constitutes the numerical challenge.
First, we need to discretize the time so that the number of 
second moments becomes finite. Henceforth, we set $t_0 = 0$ and choose an 
equidistant discretization $[t_0=0,... t_{L}]$, i.e., $t_l = l\delta t$. 
The numerical error resulting from the discretization is discussed and analyzed in
appendix \ref{app:discerr}. We obtain a $[3(L+1)\times 3(L+1)]$-dimensional covariance matrix $\dul{M}$ 
of which the matrix elements are 
\be
	M^{\alpha,\beta}_{t_1,t_2} = \overline{V^{\alpha}(t_1) V^{\beta}(t_2)}.
	\label{eqn:covmat}
\ee
No site label $i$ occurs because we treat a homogeneous spin ensemble
so that the covariance matrix is the same at each site. But a possible generalization
to a spatial dependence is obvious. If $\dul{M}$ is known the 
corresponding normal distribution reads as
\be
	p\left[\vec{\mc{V}}\right] = \frac1{\sqrt{\text{det} 2\pi \dul{M}}} 
\exp\left(-\frac12 \vec{\mc{V}}^{\top} \dul{M}^{-1} \vec{\mc{V}}\right).
	\label{eqn:Gaussdis}
\ee
To be specific, the vector-matrix-vector product in the argument
of the exponential function stands for
\be
	\vec{\mc{V}}^{\top} \dul{M}^{-1} \vec{\mc{V}} = \sum_{\alpha,\beta} \sum_{t_1, t_2} 
	V^{\alpha}(t_1) \left(\dul{M}^{-1}\right)^{\alpha,\beta}_{t_1,t_2} V^{\beta}(t_2).
\ee

Second, we use a Monte-Carlo method to carry out the path integral: 
we draw a time series $\vec{\mc{V}}$ from the distribution function,
compute the expectation value for each time series and 
finally calculate the arithmetic mean of the results. 
This is done sufficiently often to achieve a small enough
statistical error which is studied in detail in App.\ \ref{app:staterr}.
The strategy to determine the mean-field moments is set up as follows:
\begin{enumerate}[label=\arabic*)]
	\item Choose arbitrary functions for the initial second moments 
	of the mean-fields in the studied time interval.
	\item \label{item:constructCov} Construct the $[3(L+1) \times 3(L+1)]$-dimensional 
	covariance matrix as in Eq.\ \eqref{eqn:covmat}.
	\item \label{item:DrawMF} Draw a large number of time series
	for the mean-field according to the distribution \eqref{eqn:Gaussdis}.
	\item \label{item:EstimateSpin} Compute the time evolution operator 
	$\mb{U}\left(t_{l},0\right)$ at all times for the drawn time series. This
	allows one to calculate the individual spin autocorrelations 
	$\langle \mb{S}^{\alpha}(t_l) \mb{S}^{\beta}(t_{l'}) \rangle\esc_{\vec{\mc{V}}}$
	for each time series and all pairs of $t_l, t_{l'}$.
	If one assumes homogeneity in time, only the time
	difference matters and one can set $t_{l'}=0$. Numerical issues
	arising for this assumption are clarified in App.\ \ref{app:timetrans}.
	\item \label{item:SampleAverage} Determine the autocorrelations 
	by averaging over the individual autocorrelations computed in the previous step.
	\item \label{item:Convergence} Evaluate the iterated mean-field moments 
	from the self-consistency conditions \eqref{eqn:final_selfcons}
	and return to step \ref{item:constructCov} or stop if convergence of the
	second order mean-field moments is achieved within a given tolerance.
\end{enumerate}

For step \ref{item:constructCov}, it is convenient to set up the covariance matrix in blocks depending on the spin components $\alpha, \beta\in\{ x,y, z\}$:
\be
	\dul{M} =
	\begin{pmatrix}
		\dul{M}^{xx} & \dul{M}^{xy} &\dul{M}^{xz} \\
		\dul{M}^{yx} & \dul{M}^{yy} &\dul{M}^{yz} \\
		\dul{M}^{zx} & \dul{M}^{zy} &\dul{M}^{zz} \\
	\end{pmatrix}.
	\label{eqn:Mblocks}
\ee
Spin symmetries of the system can easily be exploited to  reduce the numerical effort. 
For instance, for zero magnetic field any
block with $\alpha\neq\beta$ vanishes, so that the covariance matrix becomes block-diagonal.
Furthermore, we stress that $\dul{M}$ is symmetric. This is actually required for
a covariance matrix. Here, it results from the physics at infinite temperature: 
the quantum expectation values and hence the mean-field moments are symmetric 
\be
	\langle \mb{V}^{\alpha}(t_1) \mb{V}^{\beta}(t_2) \rangle = 
	\langle \mb{V}^{\beta}(t_2) \mb{V}^{\alpha}(t_1) \rangle
\ee
due to the cyclic invariance of the trace.
Another crucial property of covariance matrices is their positive semidefiniteness. 
In App.\ \ref{app:definitecov} we show that
$\dul{M}$ is automatically positive definite because it results from 
the quantum expectation values of Hermitian operators. In App.~\ref{app:timetrans}
we explain how including time-translation invariance in the algorithm 
 reduces the numerical effort further.

Another algorithmic issue is the sampling procedure in step \ref{item:DrawMF}. Since the covariance matrix is generally non-diagonal, the
mean-fields at different times cannot be drawn independently of each other. 
Hence, it is indicated to first change basis such that $\dul{M}$ is diagonal 
in the new basis. In this basis, for each vector component an
independent random variable can be drawn
from a one-dimensional normal distribution. Subsequently, we transform back
into the original basis obtaining the desired autocorrelation in time. 
We recommend the following strategy:
\begin{enumerate}[label=\alph*)]
	\item\label{item:samplea} Diagonalize the symmetric, non-negative
	covariance matrix by the orthogonal transformation $\dul{O}$
	\begin{align}
		\dul{D} &= \dul{O}^{\top} \dul{M}\, \dul{O}.
	\end{align}
	\item \label{item:sampleb} Sample a $3(L+1)$-dimensional vector $\vec{\mc{R}}$ of uncorrelated Gaussian random numbers in the diagonal basis.
	Each component has a zero average and a variance given by the 
	corresponding eigenvalue of $\dul{M}$, i.e., 
	the corresponding diagonal element of $\dul{D}$.
	\item \label{item:samplec} 
	Transform the random vector $\vec{\mc{R}}$ to the original basis 
	\be
		\vec{\mc{V}} = \dul{O} \vec{\mc{R}}.
	\ee
\end{enumerate}
The diagonalization needs to be performed only once in each iteration step
 since all drawn time series belong to the same covariance matrix.
In contrast, steps \ref{item:sampleb} and \ref{item:samplec} have to be performed
for each drawn time series.

To compute the time evolution operator, or propagators,
in step \ref{item:EstimateSpin} numerically
we split it into a product of propagators over the short time interval
between consecutive $t_l$, i.e., over $\delta t = t_{l+1}-t_l$ 
\be
	\mb{U}\left(t_{l},0\right) = \mb{U}\left(t_{l},t_{l-1}\right) 
	\ldots \mb{U}\left(t_2,t_{1}\right)	\mb{U}\left(t_{1},t_{0}\right).
\ee
These propagators  can be computed efficiently by commutator-free exponential time propagation (CFET)\cite{alver11a}. Since we do not have
information about $H\emf$ at times between two consecutive $t_l$
any integral can only be approximated by trapezoidal rule. Therefore, the error of each
propagator is at best of order $\delta t^3$ so that CFETs of orders larger than two appear pointless. From our numerical experience, we recommend a
second-order and an optimized fourth-order CFET \cite{alver11a}
\bs
\begin{align}
	\mb{U}^{(2)}_{\text{CF}}\bigl(t_{k},t_{k-1}&\bigr)	= \mathrm{e}^{\mb{A}_1}
	\\ 
	\begin{split}
    \mb{U}^{(4\text{Opt})}_{\text{CF}}\bigl(t_{k},t_{k-1}&\bigr) = 
		\mathrm{e}^{\frac{11}{40} \mb{A}_1 + \frac{20}{87}
    \mb{A}_2 + \frac7{50} \mb{A}_3} \cdot \\
		& \mathrm{e}^{\frac{9}{20} \mb{A}_1 - \frac7{25} \mb{A}_3} \, 
		\mathrm{e}^{\frac{11}{40} \mb{A}_1 - \frac{20}{87} \mb{A}_2 + \frac7{50} \mb{A}_3},
	\end{split}
	\intertext{where}
	\mb{A}_{j} = -\texttt{i}(2j-1) &\frac{\delta t}{2} 
	\Bigl( \mb{H}(t_{k}) - (-1)^{j} \mb{H}(t_{k-1}) \Bigr).
\end{align}
\es
{In step} \ref{item:Convergence}, one requires an exit condition 
to decide when a sufficiently converged result has been found. 
A possible choice is to compute the deviation between the results
of current and previous iterations and {compare} it to {a} chosen tolerance threshold. 
If the deviation falls below the tolerance threshold, the iteration is stopped.
We discuss the definition of the deviation and the choice of the tolerance 
in App.\ \ref{app:abortcon}. 
In general, when we graphically show numerical results of spinDMFT, we choose the numerical parameters
such that the resulting errors are not larger than the thickness of the lines, if not explicitly discussed otherwise.
As mentioned before, Appendix \ref{app:errors} provides a closer insight {into} the error sources.
In the following sections, we examine the validity of spinDMFT by comparing
its results to the ones of established numerical techniques.


\section{Comparison of $\text{spin}$DMFT to other approaches}
\label{sec:gauging}

Before applying the advocated spinDMFT to various physical systems it is 
advisable to compare results of spinDMFT with results of different well-established methods.
Since the main idea of spinDMFT is based on a large
number of interaction partners we expect the agreement to become the better
the larger the coordination number of the spin ensemble is.

\subsection{Methods for comparison}

Below, we use two methods to obtain results for comparison. The first method is the  Chebyshev expansion technique \cite{talez84,weiss06a}\,(CET), the second method is the iterated equations of motion\,(iEoM) approach \cite{kalth17,bleic18}. 
The CET is numerically exact up to a systematically controlled 
error threshold. The iEoM approach is an approximate approach controlled by
the number $m$ of iterations performed.

To obtain the {time dependence} $\mb{O}(t)$ of an observable using CET 
we expand the unitary time evolution operator $\mb{U}=e^{-i\mb{H}t}$ 
in terms of Chebyshev polynomials defined recursively by
\begin{subequations}
\label{eq:chebyshev_polynomials_recursion_relation}
\begin{alignat}{3}
	T_0(x) &= 1, \qquad T_1(x)=x \\
	T_{n+1}(x) &= 2x T_n(x)-T_{n-1}(x).
\end{alignat}
\end{subequations}
All polynomials $T_n$ are defined on the closed interval $I=\left[-1;1\right]$. To ensure that the energy spectrum of a given Hamiltonian $\mb{H}$ lies in $I$ 
we rescale the Hamiltonian according to
$\mb{H}\to \mb{H}' = (\mb{H}-b)/a$. Then, the Chebyshev polynomials
 can be used as an orthogonal functional basis. In order to perform the rescaling
 an estimate of the extremal eigenvalues \cite{lancz50,arnol51,kuczy92}
of $\mb{H}$ is needed to obtain $a=\left(E_\mathrm{max}-E_\mathrm{min}\right)/2$
and $b=\left(E_\mathrm{max}+E_\mathrm{min}\right)/2$.
Rough estimates in the form of upper (lower) bounds for 
$E_\mathrm{max}$ ($E_\mathrm{min}$) are sufficient because the rescaling 
only has to ensure that the rescaled eigenvalues lie within $I$. 
Subsequently, the expanded time evolution operator reads as
\begin{subequations}
\label{eq:cet_time_dependent_series_final}
\begin{alignat}{3}
	\mb{U} &= \sum_{n=0}^\infty \alpha_n(t)T_n(\mb{H}')
	\\
	\alpha_n(t) &= (2-\delta_{n,0}) i^n e^{-i b t} J_n(at)
	\label{eq:cet_time_dependent_series_final_coefficients}
\end{alignat}
\end{subequations}
where the time-dependent coefficients contain the
Bessel functions of first kind $J_n(at)$.
Given an initial state $\ket{\psi_0}$ its time evolution reads as
\begin{equation}
	\label{eq:cet_time_evolution_state}
	\ket{\psi(t)}=\mb{U}\ket{\psi_0}=\sum_{n=0}^\infty \alpha_n(t)
	\underbrace{T_n(\mb{H}')\ket{\psi_0}}_{=:\,\ket{\phi_n}}.
\end{equation}
Here, the basis states of the expansion are $\ket{\phi_0}\!:=\!\ket{\psi_0}$,
$\ket{\phi_1}:=\mb{H}'\ket{\psi_0}$, and 
$\ket{\phi_{n+1}}:=2\mb{H}'\ket{\phi_n}-\ket{\phi_{n-1}}$.

In the {numerical} implementation, the infinite series \eqref{eq:cet_time_evolution_state} 
must be terminated at some finite value $N_\mathrm{c}<\infty$. 
The time dependence of the  prefactors
is essentially determined by the time dependence of the Bessel functions 
$J_n(t)$ \cite{olver19}. The higher the order $n$ the longer it takes the Bessel function $J_n(t)$ 
to contribute noticeably to the series. Given the  cut-off $N_\mathrm{c}$ of the series 
 the truncation error of the CET series is estimated by
\begin{equation}
	\epsilon \lessapprox \left(\frac{a t \cdot e }{2 N_\mathrm{c}}\right)^{N_\mathrm{c}}.
	\label{eqn:CETerrortolerance}
\end{equation}
Note that the truncation error is not only related to the cut-off $N_\mathrm{c}$, but
also depends on the maximum time up to which results are calculated as well as on the
parameter $a$ which equals half the width of the energy spectrum. 
The important property of the CET is that  $N_\text{c}$, required to
keep the error low, increases only linearly with
the time~$t$ up to which one intends to compute the evolution.


The second method we employ for comparison is the iEoM approach
\cite{kalth17,bleic18} which approximates the time dependence of an operator
in the Heisenberg picture.
Starting with an arbitrary operator $\mb{A}_1$ 
one expands
\be
\mb{A}(t) = \sum_i h_i(t) \mb{A}_i 
\label{eqn:iEoM:expansion}
\ee
where all time dependence is incorporated in the complex prefactors $h_i(t)$.
The constant operators $\mb{A}_i$ form an operator basis $\{\mb{A}_i\}$.
The expansion \eqref{eqn:iEoM:expansion} is unique if the $\mb{A}_i$
are linearly independent.
For a Hamiltonian constant in time the
Heisenberg equation of motion reads as
\be
 \frac{\mathrm{d}}{\mathrm{d}t} \mb{A}(t) =
\texttt{i} \left[ \mb{H}(t), \mb{A}(t) \right]
\eqqcolon
\texttt{i}  \mathcal{L}\left( \mb{A}(t) \right)
\label{eqn:iEoM:heisenberg_eq}
\ee
with the Liouville superoperator $\mathcal{L}$.
Expanding the result of $\mathcal{L}(\mb{A}_i)$ in terms of the chosen basis 
$\{\mb{A}_i\}$ by means of
\be
\mathcal{L}(\mb{A}_i) = \sum_j L_{ij} \mb{A}_j 
\label{eqn:iEoM:L_expansion}
\ee
leads to the Liouvillian matrix $\dul{L}$, also called dynamic matrix.
For a compact notation the time-dependent prefactors $h_i(t)$
are combined to a vector $\vec{h}(t)$ of which the dynamics is obtained by
inserting both expansions \eqref{eqn:iEoM:expansion} and 
 \eqref{eqn:iEoM:L_expansion} in  \eqref{eqn:iEoM:heisenberg_eq}
yielding
\be
    \frac{\mathrm{d}}{\mathrm{d}t}\vec{h}(t) = \texttt{i} \dul{L} \vec{h}(t). 
		\label{eqn:iEoM:h_dgl}
\ee

The Liouvillian matrix is most easily computed for an orthonormal operator basis 
$\{A_i\}$ (ONOB) so that each matrix element is given by
\be
    L_{ij} = \left(\mb{A}_i \vert \mathcal{L}(\mb{A}_j)  \right).
\ee
As previously argued \cite{kalth17,bleic18}, it is crucial to
achieve Hermiticity of $\dul{L}$ to avoid exponentially diverging 
solutions which are unphysical. 
The Hermiticity of $\dul{L}$ is equivalent to the self-adjointness 
of $\mathcal{L}$ which depends on the used operator scalar product.
A convenient choice is the Frobenius scalar product
\be
    (\mb{A}|\mb{B}) \coloneqq \frac1{d} \Tr(\mb{A}^{\dagg}\mb{B}) .
		\label{eqn:iEoM:frobenius_scalar_product}
\ee
Due to the invariance of the trace under cyclic permutations
$\mathcal{L}$ is indeed self-adjoint and 
thus $\dul{L}$ is assured to be Hermitian \cite{kalth17,bleic18}.

\begin{table}[ht]
    \centering
    \begin{tabular}{| c | c | c | c |}
                    \hline
         $\mathbb{1}_i$ & $\sigma^z_i$ & $\sqrt{2}\sigma^{+}_i$ & $\sqrt{2}\sigma^{-}_i$\\
         \hline
        \end{tabular}
    \caption{Local operator basis for a two-dimensional local Hilbert space. All operators are orthonormal with respect to the operator scalar product
		\eqref{eqn:iEoM:frobenius_scalar_product}.}
    \label{tab:iEoM:localoperators}
\end{table}

The ONOB is found iteratively by applying the Liouville superoperator
$m$ times which is called the loop order. Starting from a spin operator at
a given site, the application of $\mc{L}$ 
creates more and more increasingly complicated expressions
which are sums of operator monomials, i.e., sums of products of
local operators. The number of such monomials is finite
for all $m$, but grows exponentially for increasing $m$.
For a spin $S={\tfrac12}$ the site
local operators are those given in Tab.\ \ref{tab:iEoM:localoperators}.
After $m$ iterations monomials involving up to  $m+1$ sites occur.
This  means that in these monomials Pauli matrices at up to $m+1$ sites
can occur. The identity operator is trivial and does not need to
be tracked. If $\mb{A}=\mb{A}_1$ is the initial spin operator the
initial vector $\vec h$ has the components
 \be
    h_i(0) =
    \begin{cases}
        1 \hfill & \text{ if $i=1$} \\
        0 \hfill & \text{ otherwise.}
    \end{cases}
    \label{eqn:iEoM:h_initial_condition}
\ee



\subsection{Observables and symmetries}
\label{subsec:numres}

Since we consider infinite temperature, any expectation value of a 
single-time observable is actually time-independent, see Eq.\ \eqref{eq:singletime}.
Therefore, the primarily interesting observables are the 
spin autocorrelations which we denote by 
\be
	g^{\alpha \beta}(t) := \langle \mb{S}^{\alpha}(t) \mb{S}^{\beta}(0)\rangle.
\ee
There are  nine different autocorrelations of the above type due to
the choices for $\alpha, \beta\in\{x,y,z\}$. 
However, the symmetries of the Hamiltonian imply a number
of relations between them so that only a small number needs to be
considered. We briefly discuss the symmetries of the system in the following.

The original Hamiltonian \eqref{eqn:isoHam} is invariant under any spin rotation 
 around the $z$ axis, in particular  about the angle $\pi/2$ implying
\begin{align}
	\mb{S}_{i}^{x} &\to \mb{S}_{i}^{y} & \mb{S}_{i}^{y} &\to -\mb{S}_{i}^{x}.
\end{align}
As a consequence any correlation between the transversal and 
longitudinal spin components disappear
\begin{align}
	g^{\alpha z}(t) &= g^{z \alpha}(t) = 0, & &\forall \alpha \neq z,
\end{align}
while the transversal cross autocorrelations $g^{xy}$ and $g^{yx}$ 
fulfill
\be
g^{xy}(t) = -g^{yx}(t) .
\ee
By means of cyclic permutations in the trace and homogeneity {in} time
we additionally derive
\begin{subequations}
\begin{align}
	g^{xy}(t) &= -g^{xy}(-t)  \\
	g^{yx}(t) &= -g^{yx}(-t) .
\end{align}
\end{subequations}
The transversal diagonal autocorrelations are equal
\be
g^{xx}(t) = g^{yy}(t) .
\ee

In case of zero magnetic field, the system is also invariant under
time reversal because the Hamiltonian is bilinear in spin operators
which implies 
\begin{align}
	g^{xy}(t) &= g^{yx}(t) = 0,
\end{align}
so that all cross autocorrelations vanish in this case. 
Furthermore, the diagonal autocorrelations $g^{\alpha\alpha}$ are equal due to 
complete isotropy of the model.

\begin{figure}[ht]
	\centering
	\includegraphics[width=\columnwidth]{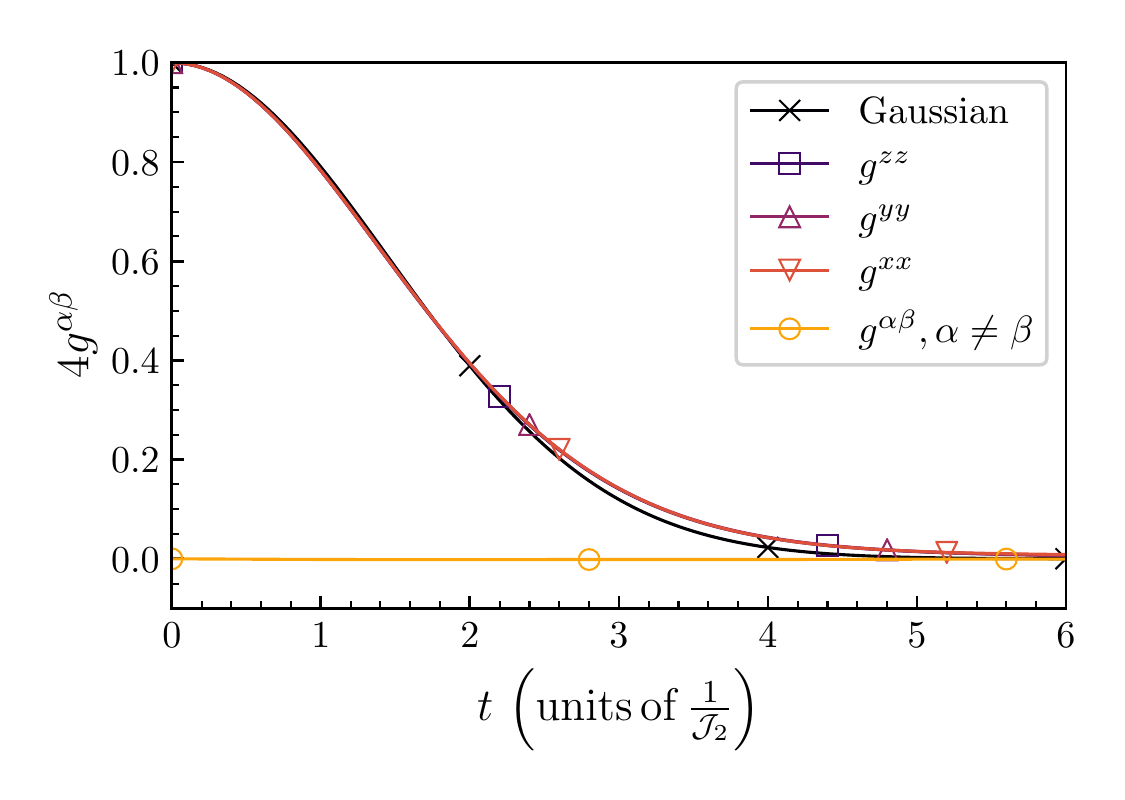}
	\caption{spinDMFT results for the isotropic Heisenberg model with zero magnetic field. 
	The Gaussian fit for short times is best with a
	standard deviation $\sigma~=~1.46/\mc{J}_{2}$.}
	\label{fig:iso_plot_A}
\end{figure}

\begin{figure}[ht]
	\centering
	\includegraphics[width=\columnwidth]{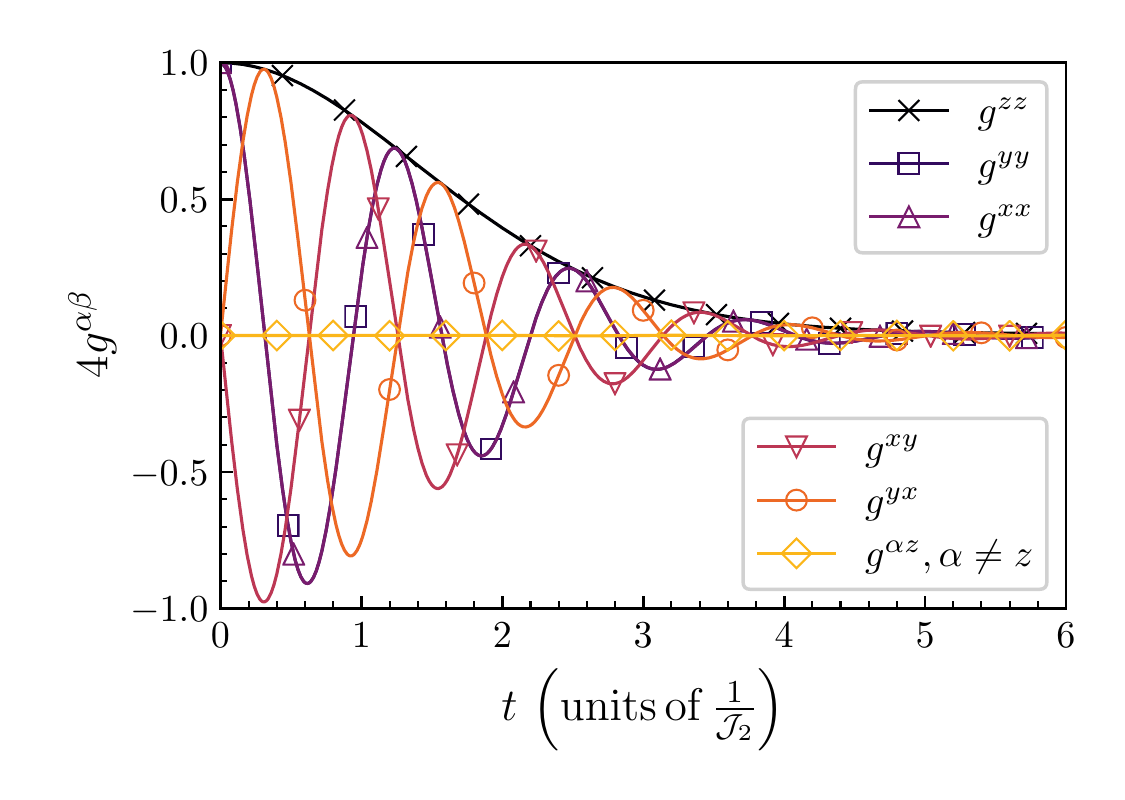}
	\caption{spinDMFT results for the isotropic Heisenberg model with finite magnetic field 
	$\gamma_\text{s} B=5.0\,\mathcal{J}_2$.}
	\label{fig:iso_plot_B}
\end{figure}

A first validation of spinDMFT consists of the successful check that the derived symmetry relations hold in the framework of spinDMFT.
The results of the self-consistency problem \eqref{eqn:final_selfcons} for zero and finite magnetic field are depicted in Figs.~\ref{fig:iso_plot_A} and \ref{fig:iso_plot_B}.
The spinDMFT fulfills the symmetry relations for both cases. 
Moreover, the Larmor precession with frequency 
$\omega_{\text{L}}~=~\gamma_{\text{s}}~B$ is clearly visible
in the transversal components for finite magnetic field.

\begin{figure}[ht]
	\centering
	\includegraphics[width=\columnwidth]{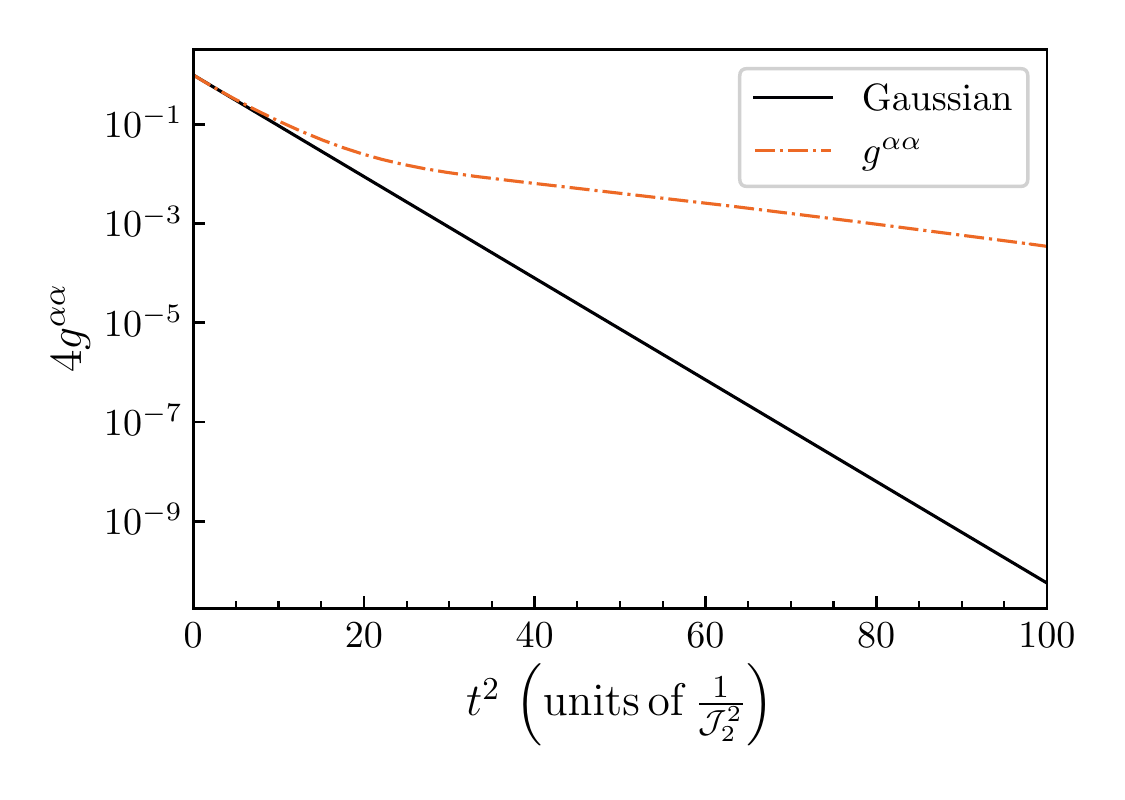}
	\caption{Diagonal autocorrelations on a log-scale for zero magnetic field 
	compared to the Gaussian short-time fit as functions of $t^2$.}
	\label{fig:iso_logplot}
\end{figure}

For short times and zero magnetic field, a Gaussian fit describes 
 $g^{\alpha\alpha}$ very well in the linear plot in Fig.~\ref{fig:iso_plot_A}.
Some deviation is discernible from intermediate times onwards. To
analyze this deviation in more detail the functions are plotted
 in Fig.~\ref{fig:iso_logplot} on a logarithmic scale vs.\ $t^2$. 
Interestingly, the diagonal autocorrelations appear to show
Gaussian behavior at short and at long times, but with different
standard deviations.  For longer times, the decay is slowed down.

\subsection{Comparison to results of other approaches}

We compare results from the spinDMFT to results from exact diagonalization (ED), 
iterated equations of motion (iEoM), and Chebyshev expansion technique (CET).
ED is a very well-known technique and the latter two approaches have been
explained above. A conceptual difficulty lies in the fact that these alternative 
techniques work best for small and low-dimensional systems while spinDMFT
is rather justified in large, high-dimensional systems. But comparing results from
spinDMFT to these alternatives is the best option at hand. Note that such 
comparisons are particularly challenging for spinDMFT.

\begin{figure}[ht]
	\centering
	\includegraphics[width=\columnwidth]{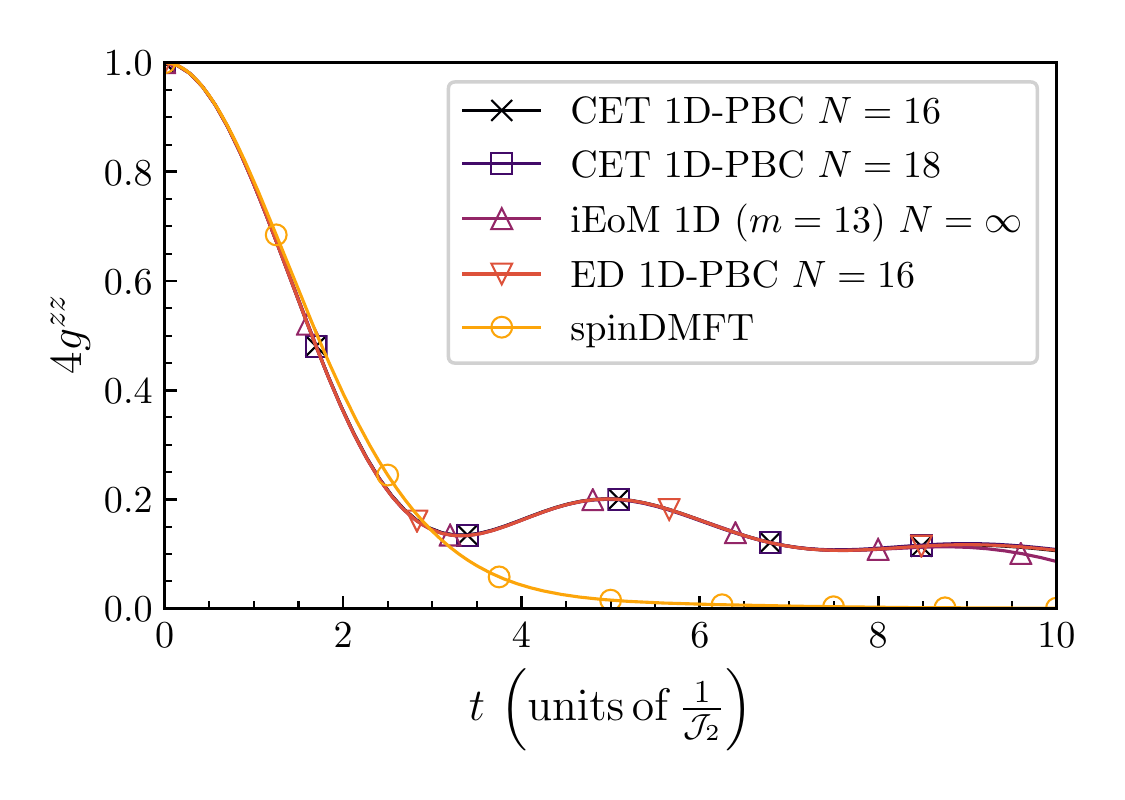}
	\caption{Results for the isotropic diagonal autocorrelation
	in the 1D Heisenberg chain as calculated by ED, CET, and iEoM compared to the
	data of spinDMFT. {We emphasize that the CET operates on finite cluster systems with
    periodic boundary conditions (PBC) only. The number of sites considered here is denoted by $N$.}
	The relative error tolerance \eqref{eqn:CETerrortolerance} of each CET time evolution is $\epsilon = \num{1e-3}$.}
	\label{fig:1D}
\end{figure}

Considering the self-consistency problem \eqref{eqn:final_selfcons}, we stress
that all lattice properties are embodied in a single coupling
constant, namely the root-mean-square $\mathcal{J}_2$. No site index appears
because we deal with homogeneous systems. Time is naturally measured in units of 
$1/\mathcal{J}_2$. First, we consider one-dimensional (1D) spin chains with $S={\tfrac12}$.
For finite pieces of chains with $N$ sites, periodic boundary conditions (PBC) are taken.
The Hamiltonian in the isotropic case reads
\be
	\mb{H}_{1\text{D}} =  J \sum_{i=1}^{N} \vec{\mb{S}}_{i} \cdot \vec{\mb{S}}_{i+1},
\ee
which entails
\be
	\mathcal{J}_2 = \sqrt{2} J.
\ee
Figure \ref{fig:1D} compares the results from the above mentioned methods
to the data obtained from spinDMFT. 
The ED data are taken from Ref.\ \onlinecite{fabri97a}. 
The results from ED and CET coincide very nicely in the
considered time interval. Moreover, no finite-size effects are discernible
in this interval. For not too long times, the iEoM result also matches
very well. It has the advantage to consider the infinite system by construction.
These results almost coincide with the spinDMFT data until roughly $t\approx 3/\mc{J}_2$.
The subsequent deviations can be attributed to the small coordination number
of the 1D chain with $z=2$ which, obviously, is a challenge for 
a mean-field approach.

The spinDMFT shows quick and rather complete decoherence while the 
genuine 1D results show weak coherent revivals at $t\approx 5/\mc{J}_2$
and $t\approx 9/\mc{J}_2$. This is not surprising because the integrable
1D system is strongly constrained in its dynamics due to its 
extensive number of conserved quantities \cite{bethe31}.

\begin{figure}[ht]
	\centering
	\includegraphics[width=0.5\textwidth]{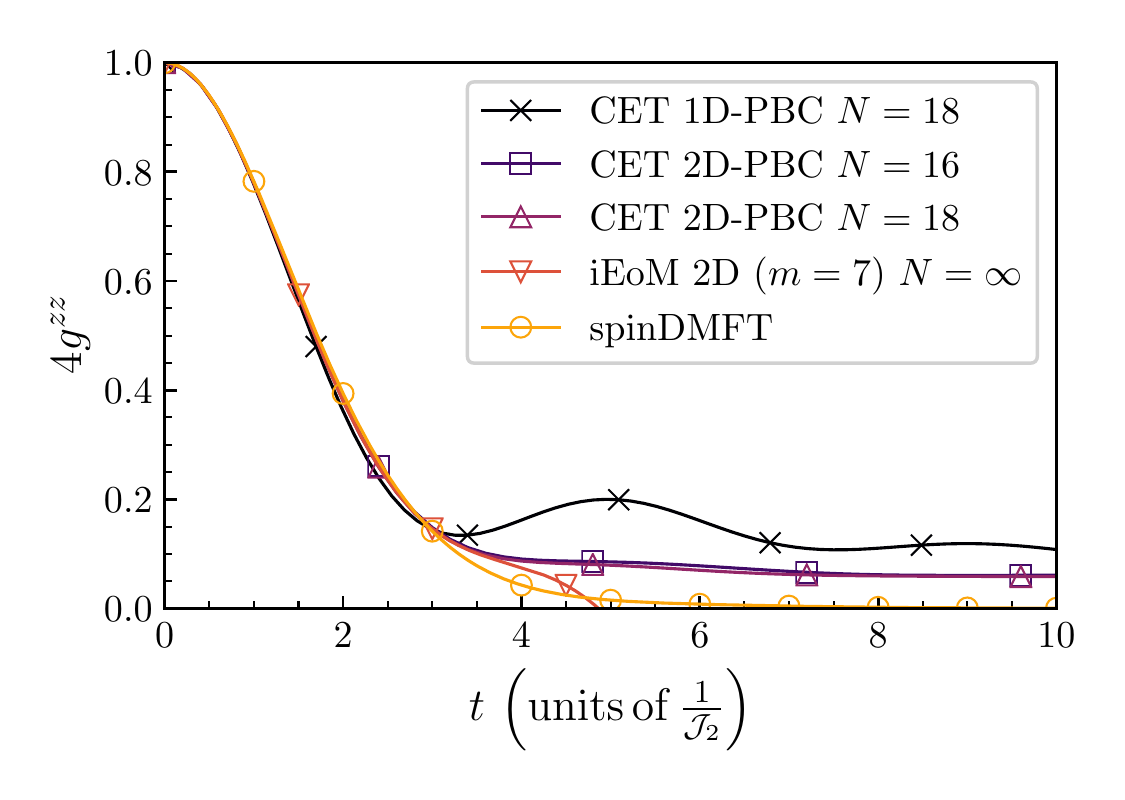}
	\caption{Results for the isotropic diagonal autocorrelation
	in the 2D Heisenberg square lattice as calculated by CET and iEoM compared to the
	data of spinDMFT. PBC stands for periodic boundary conditions {and $N$ is the number of sites}. The relative error tolerance \eqref{eqn:CETerrortolerance}
	of each CET time evolution is $\epsilon = \num{1e-3}$.}
	\label{fig:2D}
\end{figure}

Figure \ref{fig:2D} compares the results of CET and iEoM 
in 2D, i.e., for the isotropic Heisenberg model on the square lattice
\be
	\mb{H}_{2\text{D}} = 
	J \sum_{\langle i,j\rangle} \vec{\mb{S}}_{i} \cdot \vec{\mb{S}}_{j}
\ee
with NN coupling $J$ to the data obtained from spinDMFT. In this case, 
$\mc{J}_2= 2J$ holds. The agreement between CET and iEoM is good
up to $t\approx 3.5/\mc{J}_2$; then, the effects of finite loop order $m$
kick in. In 2D, it is unfortunately not possible to reach larger values of $m$.
Up to this range, the spinDMFT is in nice agreement with the other approaches
as well. What is even more interesting is to see the 
evolution from 1D to 2D. To this end, we include the CET result in 1D.
Clearly, passing from 1D to 2D improves the agreement between the 
genuine numerical results and spinDMFT. This is exactly what one
had to expect in view of the derivation of spinDMFT as mean-field theory
which becomes exact for infinite coordination number. Hence, 
this observation constitutes a good confirmation of the validity
of spinDMFT.

\begin{figure}[ht]
	\centering
	\includegraphics[width=0.5\textwidth]{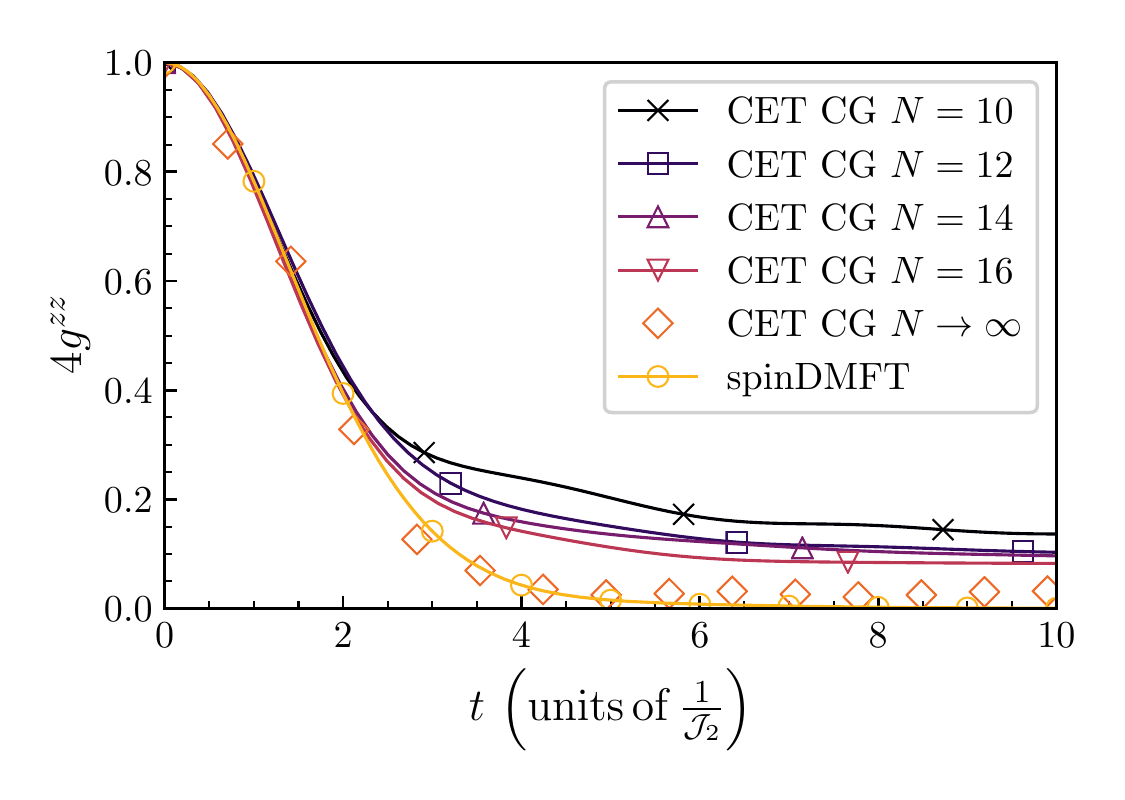}
	\caption{CET results of the isotropic diagonal autocorrelation
	in the random Heisenberg model on various complete graphs 
	compared to the data of spinDMFT. {$N$ is the number of sites.} In addition, 
	the extrapolation to $N=\infty$ is shown.
	The relative error tolerance \eqref{eqn:CETerrortolerance}
	of each CET time evolution is $\epsilon = \num{1e-3}$.}
	\label{fig:CG}
\end{figure}

In order to corroborate the foundation of spinDMFT further we consider
the Heisenberg model on complete graphs, i.e., graphs where each site is 
connected to all other sites
\be
	\mb{H}_{\text{CG}} = 
	 \sum_{i < j} J_{ij} \vec{\mb{S}}_{i} \cdot \vec{\mb{S}}_{j}.
\ee
Such graphs or clusters are called infinite-range clusters in the physical literature. 
The Heisenberg model on such graphs is highly symmetric if the coupling
is the same for all links. This leads to rather special autocorrelations.
In order to avoid features from non-generic high symmetries we
consider a random model where the couplings are drawn from a Gaussian
distribution. Then, they are normalized, i.e., multiplied by 
a suitable constant $\lambda_j$, such that 
\be
\label{eq:normalized}
\mc{J}_2^2 = \sum_i J_{ij}^2
\ee
holds for all $j$. The results for the autocorrelations are 
averaged over 100 sets of random couplings.
Figure \ref{fig:CG} compares the CET results to the data from spinDMFT
for various values of $N$. The symbols display the data extrapolated to 
$N=\infty$ by a linear fit in 
$1/N^{3/2}$ of the data for the three largest values of $N$.
This particular power law fit is chosen in view of the scaling of
each $J_{ij} \propto 1/\sqrt{N}$ stemming from the normalization
\eqref{eq:normalized}. This {suggests} to use power law fits $\propto 1/N^{p/2}$
with some integer $p$. We found that $p=3$ is most robust.

The obvious trend is that the data for finite $N$ appear to converge
to the spinDMFT. This observation again justifies the systematic
derivation of the advocated dynamic mean-field theory.

Finally, we mention that the Ising model on complete graphs
\be
	\mb{H}_{\text{CG,Ising}} = 
	J \sum_{i< j} {\mb{S}}_{i}^z {\mb{S}}^z_{j},
\ee
in the limit $N\to\infty$ has the autocorrelations \cite{dekey91}
\bs
\begin{align}
		4g^{xx} &= \mathrm{e}^{-\frac18 \mathcal{J}_2^2 t^2} 
		\\
		4g^{zz} &= 1 .
	\end{align}
	\es
The energy scale is $\mc{J}_2=\sqrt{N-1}J$.
These results are  reproduced by spinDMFT; we refrain
from displaying them because the graphs coincide perfectly. 
Due to the spin anisotropy the self-consistency conditions are
changed to
	\bs
	\begin{align}
		\overline{V^{\alpha}(t_1)V^{\alpha}(t_2)} &= 
		\mc{J}_2 \langle \mb{S}^{\alpha}(t_1) \mb{S}^{\alpha}(t_2) \rangle & \alpha&=x,y
		\\
		\overline{V^{z}(t_1)V^{z}(t_2)} &= \mc{J}_2 \langle \mb{S}^{z}(t_1) \mb{S}^{z}(t_2) \rangle & &\\
		\overline{V^{\alpha}(t_1)V^{\beta}(t_2)} &= 0 & \alpha&\neq\beta .
	\end{align}
	\es

On the basis of the above results, we conclude that spinDMFT is a systematically controlled dynamic mean-field approach to disordered spin systems at infinite
temperature which becomes exact for infinite coordination number.
It is designed to provide quantitative information of the  local spin dynamics.
 It is a valid approximation
for finite coordination numbers which nicely captures essential physics such 
as rapid decoherence, spin anisotropies, and Larmor precession.
Due to the required moderate computational resources it is an
attractive tool to understand spin dynamics in various setups.
We illustrate this last point by applying spinDMFT to a spin ensemble
with dipolar interactions.

\section{Application to a dipolar surface spin ensemble}
\label{sec:applicationdipolar}

In this section, we want to show that spinDMFT can be adapted to 
models which describe experimental setups or are very close to 
experimental questions. We illustrate that spinDMFT can be applied to 
complex physical situations because of its flexibility and, furthermore,
that the resulting numerical task is feasible and does not require
excessive compute resources.

The model which we will address is motivated
by intensive studies of localized defect spins of electronic origin
with $S={\tfrac12}$ and a $g$-factor of $g\approx 2$ 
on diamond surfaces which are observed by NV centers \cite{grotz11, grino14,sushk14}.
These spins were seen and examined in recent studies 
\cite{rossk14,myers14,romac15}.
The precise origin of the defect spins is still a matter of debate
although recent progress indicates that they are formed by trapped
electrons very close to the surface \cite{stace19}. They appear to
be inhomogeneously distributed over the surface 
\cite{sangt19,grotz11,grino14}. Driving these spins reduces decoherence
in shallow NV centers \cite{bluvs19}. The surface spins interact with one
another and they are subject to additional, slow noise. The origin of the
latter is not yet clarified:  nuclear proton spins are candidates \cite{sushk14,staud15}
which agrees with the importance of the precise chemical and morphological
conditions at the surface \cite{sangt19}. A second candidate is phonons which 
also play a role \cite{romac15}. In addition, NV centers can also measure
the dynamics of $^{13}$C nuclear spin baths which are distributed three-dimensionally
in the bulk of diamonds \cite{larao13}.

Here, we do not aim at describing one of the above exciting experiments
in detail, but to address a generic model comprising the essential features.
To this end, we consider a random ensemble of localized electronic 
spins $S={\tfrac12}$ on a planar surface interacting by dipolar couplings. 
Aside from these interactions, the spins are subjected to a global magnetic field 
as well as to local static magnetic field noise. The latter can be viewed as 
being generated
{by slowly} fluctuating nuclear spins stemming, for instance, from protons.

\begin{figure}[ht]
	\centering
	\includegraphics[width=0.48\textwidth]{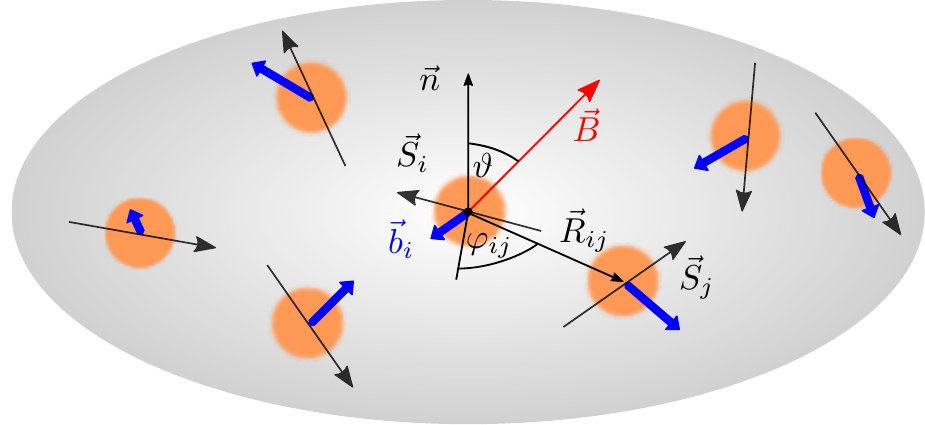}
	\caption{Sketch of the considered dipolar spin ensemble on a planar surface. The magnetic field $B$ (red) defines the $z$-direction; 
	the isotropic random local magnetic fields $\vec b_{i}$	(blue) vary
	in strength and direction from site to site.}
	\label{fig:dipole_system}
\end{figure}

\subsection{Model and Definitions}
\label{ss:model}

We consider an ensemble of spins $S=\tfrac12$ distributed randomly
over a planar surface which interact via dipole-dipole interaction.
Each spin is subjected to an external  magnetic field $\vec B$ 
whose direction defines the $z$ direction which forms an angle
$\vartheta$ with the normal $\vec n$ of the plane. Additionally,
the spins see local random magnetic noise, i.e., local 
magnetic fields $\vec{b}_{i}$ with zero average. While the system
is inhomogeneous we assume that the random distribution is such
that each spin sees the same environment on average, i.e., the
system behaves on average like a homogeneous system. This means
that the average quantities do not depend on the site.

The Hamiltonian is given by
\begin{align}
\nonumber
	\mb{H} &= \sum_{i<j} J\left(R_{ij}\right) 
	\Big[ (\vec{\mb{S}}_{i} \cdot \vec{\mb{S}}_{j}) 
	- \frac3{R_{ij}^2} (\vec{R}_{ij} \cdot \vec{\mb{S}}_{i})	
	(\vec{R}_{ij} \cdot \vec{\mb{S}}_{j}) \Big] 
	\\
	& \qquad + \gamma_{\text{s}} B \sum_{i} \mb{S}^{z}_{i} + 
	\gamma_{\text{s}} \sum_{i} \vec{b}_{i} \cdot \vec{\mb{S}}_{i},
	\label{eqn:dipHam}%
\end{align}
where
\be
\label{eq:dipolar}
	J(R) = \frac{\mu_0 \gamma_{\text{s}}^2}{4 \pi R^3}.
\ee
is the dipolar coupling. We apply spinDMFT which relies on 
average dynamic mean-fields. Since we want to treat
the random local magnetic fields $\vec{b}_{i}$ in addition we have to
distinguish three types of averages: (i) the one from spinDMFT
which we denote by an overline as before, (ii) the one {solely} due
to the average over the local magnetic fields which we denote
by an overline with index `n' for `noise', and (iii) the complete
average comprising (i) and (ii) which we denote
by an overline with index `c' for `complete'.

For simplicity, we assume that the local magnetic fields
are distributed isotropically according to a normal distribution
defined by the moments
\begin{align}
	\overline{b_{i}^{\alpha}}^{\text{n}} &= 
	\mu_{\text{N}} = 0, & \overline{\left(b_{i}^{\alpha}\right)^2}^{\text{n}} 
	&= \sigma_{\text{N}}^2, & \forall i,\alpha.
\end{align}
and
\begin{align}
	\overline{b_{i}^{\alpha}b_{j}^{\beta}}^{\text{n}} &= \delta_{ij} \delta_{\alpha\beta} \sigma_{\text{N}}^2.
\end{align}
The latter implies that the local fields are independent from one another.
The distribution reads as
\begin{align}
	p_{\text{n}}\left(\vec{b}_{i}\right) = \prod_{\alpha} \frac1{\sqrt{2\pi \sigma_{\text{N}}^2}} \, \text{exp} \left(-\frac{\left(b^{\alpha}_{i}\right)^2}{2\sigma_{\text{N}}^2}  \right).
	\label{eqn:noisedist}
\end{align}

As mentioned above, we allow for  an angle $\vartheta$ {between} the surface normal 
$\vec{n}$ and the external magnetic field $\vec B$. We introduce the in-plane polar coordinates $R_{ij}, \varphi_{ij}$ to express the distance vectors between site $i$ and $j$ by 
\begin{align}
	\vec{R}_{ij} &= R_{ij}
	\begin{pmatrix}
		\cos\left( \varphi_{ij} \right) \\
		\sin\left( \varphi_{ij} \right) \cos( \vartheta ) \\
		\sin\left( \varphi_{ij} \right) \sin( \vartheta ) \\
	\end{pmatrix}.
\end{align}
The complete system is sketched in Fig.~\ref{fig:dipole_system}
including the various introduced quantities.

\subsection{spinDMFT}
\label{ss:generalfield}

As motivated in the previous section where we introduced spinDMFT
we define local operators describing the  environments of the spins
\be
	\vec{\mb{V}}_{i} = \sum_{j\neq i} J\left(R_{ij}\right) \, 
	\dul{D}\left(\varphi_{ij}, \vartheta\right) \vec{\mb{S}}_{j},
	\label{eqn:locenv}
\ee
where the anisotropies are incorporated in the matrix
\begin{widetext}
\be
	\dul{D}\left(\varphi_{ij}, \vartheta\right) =
	\begin{pmatrix}
		1-3\cos^2(\varphi_{ij}) & -3\cos(\varphi_{ij}) \sin(\varphi_{ij}) \cos(\vartheta) 
		& -3\cos(\varphi_{ij}) \sin(\varphi_{ij}) \sin(\vartheta) 
		\\
		-3\cos(\varphi_{ij}) \sin(\varphi_{ij}) \cos(\vartheta) & 1-3\sin^2(\varphi_{ij})\cos^2(\vartheta) 
		& -3\sin^2(\varphi_{ij}) \cos(\vartheta) \sin(\vartheta) 
		\\
		-3\cos(\varphi_{ij}) \sin(\varphi_{ij}) \sin(\vartheta) 
		& -3\sin^2(\varphi_{ij}) \cos(\vartheta) \sin(\vartheta) & 1-3\sin^2(\varphi_{ij})\sin^2(\vartheta) 
		\end{pmatrix}.
\ee
\end{widetext}
With their help, the Hamiltonian can be rewritten
\be
	\mb{H} = \frac12 \sum_{i} \vec{\mb{S}}_{i} \cdot \vec{\mb{V}}_{i} 
	+ \gamma_{\text{s}} B \sum_{i} \mb{S}^{z}_{i} 
	+ \gamma_{\text{s}} \sum_{i} \vec{b}_{i} \cdot \vec{\mb{S}}_{i},
	\label{eqn:dipHamV}
\ee
where a factor $\frac12$ is introduced to avoid double counting.

From the derivation of spinDMFT in Sect.\ \ref{sec:approach} 
we know that large coordination numbers provide the justification
for this dynamic mean-field theory. Thus, we consider the effective
coordination numbers
$z$ and $z'$ defined in \eqref{eq:coord_num_def} for various lattices
and dipolar coupling \eqref{eq:dipolar}, see Tab.\ \ref{tab:coordination}.

\begin{table}[ht]
	\begin{tabular}{| p{2.5cm} |c |c| c|}
		\toprule
		{lattice} & {$z$} & {$z'$} & {$z_{\text{NN}}$}\\
		\colrule
		triangular	& 19.1 & 6.8 & 6 \\
		square			& 17.5 & 5.3 & 4 \\
		hexagonal		& 13.1 & 3.6 & 3 \\
		rectangular (A) & 12.0 & 2.8 & 2 \\
		rectangular (B) & 7.8 & 2.2 & 2 \\
		\botrule
	\end{tabular}
		\caption{Effective coordination numbers for various two-dimensional lattices
		and dipolar coupling $\propto R^{-3}$. 
		The effective coordination numbers were computed
	 taking $4 \cdot 10^6$ unit cells into account. 
	 The conventional coordination number $z_{\text{NN}}$ equals the number of nearest neighbors at each lattice site. For comparison,
	we include rectangular lattices with the ratios $2:3$ (A) and $1:2$ (B)
	of their lattice constants in $x$ and $y$ direction, respectively.}
		\label{tab:coordination}
\end{table}

As expected, the long range of dipolar interactions increases
{the effective coordination number to considerably larger values compared to the} 
NN coupling, see for instance the triangular lattice.
This effect is larger for $z$ than for $z'$
because the sums for higher moments converge faster than those
for the second or first moment. We point out that the first
moment ``just'' converges like $\int \frac{RdR}{R^3} \propto \frac{1}{R}$.

For randomly distributed spins the issue is more complicated 
and depends on how the spins are located on the surface. In case of
random positions without any restrictions, the effective coordination numbers 
are fairly small because there is a high probability for each spin to have a single neighbor close to it which dominates $z$ and $z'$. We found for
completely random distributions $z\approx 1-{10}$. 
Then, this close partner governs the  dynamics and the
 application of spinDMFT is not well justified.
However, considering restrictions for the location of the spins, in particular
a minimum distance between the spins, the effective coordination numbers
increase substantially to about the values of the triangular lattice
which is the closest-packed lattice in two dimensions. 
We found $z\approx 5-15$.
We emphasize that such restrictions are very plausible:
a minimum distance can result from the surface structure which 
does not allow the spin-carrying adatoms to be located very
close to one another. In addition, a local repulsion between them
would also ensure a minimum distance between the spins.
We conclude that the lattice is dense enough so that spinDMFT is justified.

Next, we  replace the local-environment fields by dynamic mean-fields
\be
	\vec{\mb{V}}_{i} \to \vec{V}_{i}(t)
\ee
considering the local mean-field model 
\be
	\mb{H}_{\text{mf},i}(t) = \vec{V}_{i}(t) \cdot \vec{\mb{S}} + 
	\gamma_{\text{s}} B \mb{S}^{z} + \gamma_{\text{s}} \vec{b}_{i} 
	\cdot \vec{\mb{S}}.
	\label{eqn:localmfHamiltonianD}
\ee
The self-consistency conditions need to be complemented
by the effect of the random noise fields $\vec b_i$. Averaging
has to be done over the random time series for $\vec V_i$ \emph{and}
the random local magnetic fields. We perform this in a single step 
and thus pass to combined fields
\be
	{\vec{W}}_{i}(t) = \vec{V}_{i}(t) + \gamma_{\text{s}} \vec{b}_{i}
\ee
and perform a single average over the combined distribution
\be
	{p}_{i}[{\vec{\mc{W}}}_{i}] = 
	\frac1{\sqrt{\text{det} 2\pi \dul{M}_{i}}} \mathrm{e}^{-\frac12 {\vec{\mc{W}}}_{i}^{\top} \dul{M}_{i}^{-1} 
	{\vec{\mc{W}}}_{i}},
\ee
where the modified covariance matrix is given by
\bs
\begin{align}
	M_{i}^{\alpha\beta}(t_1,t_2) &= 
	\overline{{W}_{i}^{\alpha}(t_1) {W}_{i}^{\beta}(t_2)}^{\text{c}} \\
	&= \overline{V_{i}^{\alpha}(t_1) V_{i}^{\beta}(t_2)} + \gamma_{\text{s}}^2 \overline{b^{\alpha}_{i} b^{\beta}_{i}}^{\text{n}}.
\end{align}
\es
Then, the noise leads only to an offset in the 
second moments which is constant in time.

The self-consistency condition of the first moment is still
trivial
\be
	\overline{{W}^{\alpha}_{i}(t)}^{\text{c}} = 
	\langle \mb{V}^{\alpha}_{i}(t) \rangle + \gamma_{\text{s}} 
	\overline{b^{\alpha}_{i}}^{\text{n}} = 0, \qquad \forall \alpha, i.
\ee
For the second moments we consider the self-consistency  
\be
 \overline{V_{i}^{\alpha}(t_1) V_{i}^{\beta}(t_2)} = 
\langle \mb{V}_{i}^{\alpha}(t_1) \mb{V}_{i}^{\beta}(t_2) \rangle,
\ee
and thus the complete self-consistency reads as
\begin{align}
	&\overline{W_{i}^{\alpha}(t_1) W_{i}^{\beta}(t_2)}^{\text{c}}
	= \gamma_{\text{s}}^2 \overline{b^{\alpha}_{i} b^{\beta}_{i}}^{\text{n}} + \sum_{k \neq i} \sum_{\rho \gamma} J^2 \left( R_{ik}\right)
	\\ \nonumber
	 & \qquad\quad \times D_{\alpha \rho}\left(\varphi_{ik}, \vartheta\right) 
	D_{\beta \gamma}\left(\varphi_{ik}, \vartheta\right) \langle 
	\mb{S}_{k}^{\rho}(t_1) \mb{S}_{k}^{\gamma}(t_2) \rangle\pemf.
	\label{eqn:selfconsD}
\end{align}
This equation still constitutes a challenging numerical issue 
because it amounts up to a  self-consistency condition for each spin.
But, as stated at the beginning of Sect.\ \ref{ss:model}, 
we assume that the system is dense enough to be treated on average
as a homogeneous system. Essentially, this means that $\mc{J}_{2,i}$
takes the same value at {each site $i$}. Then, the site
indices can be omitted and we obtain much simpler self-consistency
conditions
\begin{align}
	\overline{W^{\alpha}(t_1) W^{\beta}(t_2)} &= \\
\mc{J}^2 \sum_{\rho \gamma} &\chi^{\alpha \beta}_{\rho \gamma}
	(\vartheta) \langle \mb{S}^{\rho}(t_1) \mb{S}^{\gamma}(t_2) \rangle\emf  
	+ \delta_{\alpha\beta} \gamma_{\text{s}}^2 \sigma_{\text{N}}^2. 
	\nonumber
\end{align}
The constants $\mc J$ and $\chi^{\alpha \beta}_{\rho \gamma}$
embody the energy scale and the spin anisotropies. The key
idea is to approximate the discrete sums in the self-consistency
conditions by integrals assuming a continuous distribution of spins
with density $n_0=1/r^2_\text{min}$. Of course, this is not exact, 
but it provides a well-justified quantitative relation between 
the dipolar interaction in Eq.\ \eqref{eq:dipolar} 
and the prefactors of the self-consistency condition \eqref{eqn:selfconsD}.
We replace the sum by the integration
\be
	\sum_{k\neq i} \approx n_{0} \int_{r_{\text{min}}}^{\infty} 
	\mathrm{d}R \, R \int_{0}^{2\pi} \mathrm{d} \varphi
\ee
yielding
\bs
\begin{align}
	\mc{J}^2 &= 2\pi n_0 \int_{r_{\text{min}}}^{\infty} 
	\mathrm{d}R \, R J^2(R) = \frac{\mu_0^2 \gamma_{\text{s}}^4}{32 \pi r_{\text{min}}^6} 
	\label{eqn:natenergy} 
	\\
	\chi^{\alpha \beta}_{\rho \gamma} (\vartheta) &= 
	\frac1{2\pi} \int_{0}^{2\pi} \mathrm{d} \varphi \, 
	D_{\alpha \rho}\left(\varphi, \vartheta\right) 
	D_{\beta \gamma}\left(\varphi, \vartheta\right).
	\label{eqn:chi}
\end{align}
\es

Since we are dealing with a system constant in time we
study self-consistent solutions which depend only on the
time difference $t_1-t_2$. Hence, from now on we
only consider correlation functions with $t_1 = t$ and $t_2 = 0$,
\bs
\begin{align}
	g^{\alpha \beta} (t) &:= \langle \mb{S}^{\alpha}(t) \mb{S}^{\beta}(0) \rangle\emf \\
	w^{\alpha \beta} (t) &:= \overline{{W}^{\alpha}(t) {W}^{\beta}(0)}.
\end{align}
\es

In the remainder of this section, we specialize the above general equations to 
the case of a perpendicular magnetic field, i.e.,  $\vartheta = 0$, for simplicity. 
From Eq.\ \eqref{eqn:chi} we obtain
\bs
\begin{align}
	\chi^{xx}_{xx} &= \chi^{yy}_{yy} = \tfrac{11}{8} \\
	\chi^{xy}_{xy} &= \chi^{yx}_{yx} = - \tfrac{7}{8}  \\
	\chi^{xx}_{yy} &= \chi^{yy}_{xx} = \chi^{xy}_{yx} = \chi^{yx}_{xy} = 
	\tfrac{9}{8} \\
	\chi^{xz}_{xz} &= \chi^{yz}_{yz} = \chi^{zx}_{zx} = \chi^{zy}_{zy} = 
	-\tfrac{1}{2} \\
	\chi^{zz}_{zz} &= 1,
\end{align}
\es
by straightforward analytic calculation while the 
other coefficients vanish.

For a brief symmetry discussion, we consider the original Hamiltonian 
\eqref{eqn:dipHam}. Since $\vartheta=0$, the transversal $x$ and $y$ spin components lie 
in the plane of the surface. Thus, the dipolar interaction term and the 
magnetic-field term are invariant under a rotation in spin \emph{and}
real space about the angle $\pi/2$ around the $z$ axis. 
This does not hold true for the noise term $\propto \vec b_i$.
But the noise distribution \eqref{eqn:noisedist} 
is isotropic so that on average this term also remains invariant and we 
have a rotational symmetry of the total system. In particular, this implies
\bs
\begin{align}
	g^{xx}(t) &= g^{yy}(t) \\
	g^{xy}(t) &= -g^{yx}(t) \\
	g^{xz}(t) &= g^{zx}(t) = g^{yz}(t) = g^{zy}(t) = 0.
\end{align}
\es
Summarizing, we obtain the self-consistency equations
\bs
\begin{align}
w^{xx}(t) &= w^{yy}(t) = \tfrac52 \mc{J}^2 g^{xx}(t) + 
\gamma_{\text{s}}^2 \sigma_{\text{N}}^2 
\label{eqn:weakfield_transversal} 
	\\
	w^{xy}(t) &= -w^{yx}(t) = -2 \mc{J}^2 g^{xy}(t) 
	\\
	w^{xz}(t) &= w^{zx}(t) = w^{yz}(t) = w^{zy}(t) = 0 
	\\
	w^{zz}(t) &= \mc{J}^2 g^{zz}(t) + \gamma_{\text{s}}^2 \sigma_{\text{N}}^2. 
	\label{eqn:weakfield_longitudinal}
\end{align}
\label{eqn:weakfield}%
\es
It is worth mentioning that the noise explicitly appears in these equations
because we included it in the mean-field $\vec{W}$. The magnetic field, 
on the other hand, enters the physical problem in the Hamiltonian 
\eqref{eqn:localmfHamiltonianD}. 
Another important observation is that the transversal and longitudinal equations 
\eqref{eqn:weakfield_transversal} and \eqref{eqn:weakfield_longitudinal} show different prefactors. Certainly, this leads to different behavior of the corresponding
autocorrelations. For zero magnetic field, where the cross autocorrelations
$g^{xy}=-g^{yx}$ vanish due to time reversal symmetry, 
we expect $g^{xx}=g^{yy}$ to decay slower than $g^{zz}$, 
since the transversal mean-field is stronger so that the $z$ components
are more strongly precessing.

Henceforth, we measure the time in units of $\frac1{\mathcal{J}}$
and it makes sense to
define a dimensionless noise strength and a dimensionless magnetic field
\bs
\begin{align}
\label{eq:Cdef}
	C := \frac{\gamma_{\text{s}}^2 \sigma_{\text{N}}^2}{\mc{J}^2} \\
	\widetilde{B} := \frac{\gamma_{\text{s}} B}{\mc{J}}.
\end{align}
\es

\begin{figure}[ht]
	\centering
	\includegraphics[width=1.0\columnwidth]{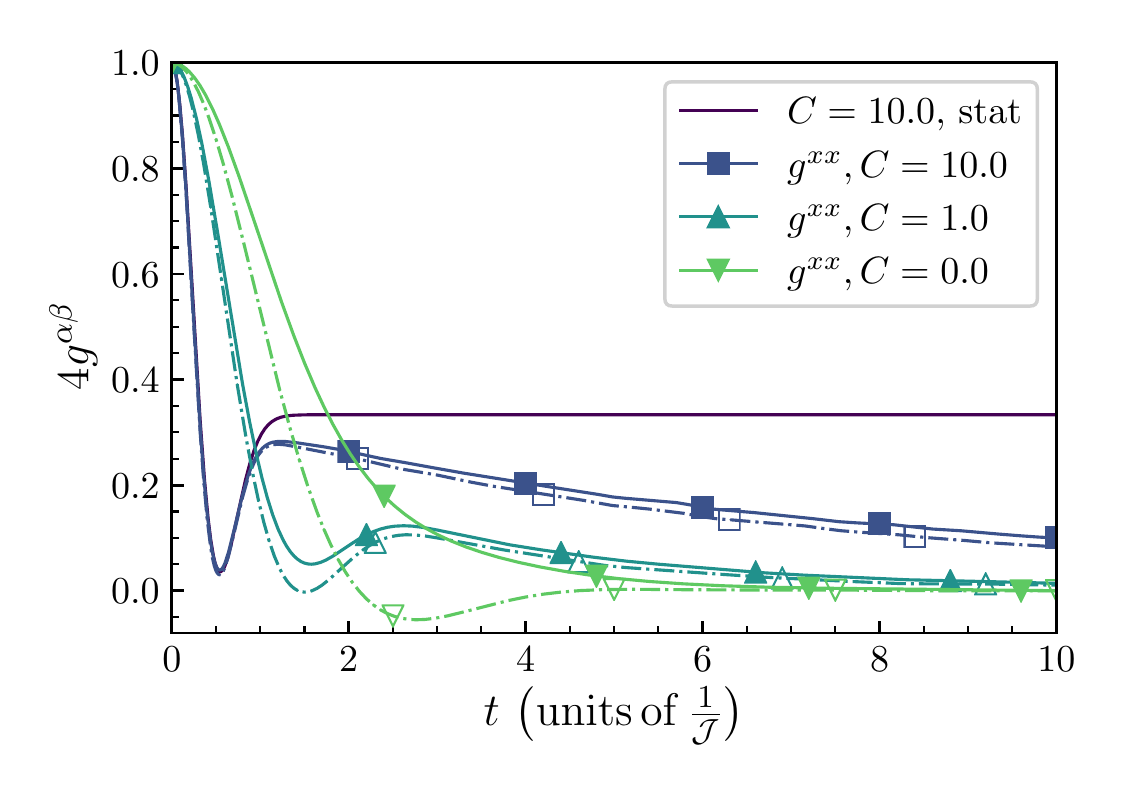}
	\caption{Numerical results of the self-consistency problem \eqref{eqn:weakfield} for zero magnetic field and various noise strengths. The
	transversal diagonal autocorrelations $g^{xx}$ (solid line, filled markers) as well as the longitudinal autocorrelations $g^{zz}$ (dashed-dotted line, {open} markers) are depicted. Moreover, 
	we plotted the analytical result \eqref{eqn:merku} for $C=10.0$ where only 
	the static noise (with subscript `stat') is considered without spin-spin coupling.} 
	\label{fig:dip_lab_zero_field}
\end{figure}

Fig.~\ref{fig:dip_lab_zero_field} shows the numerical results of the self-consistent equations \eqref{eqn:weakfield} for various noise strengths and 
zero magnetic field. Considering $C=0.0$, we observe a clear difference between the transversal and longitudinal signal. This anisotropy is expected due to the different prefactors 
in \eqref{eqn:weakfield_transversal} and \eqref{eqn:weakfield_longitudinal} as mentioned before. For $C=1.0$, 
the difference is still present, however, both curves show a similar trend: 
 a local minimum at the beginning followed by a rather slow decay.
As we increase $C$, the anisotropy further diminishes because the isotropic noise 
contributions  in \eqref{eqn:weakfield_transversal} and \eqref{eqn:weakfield_longitudinal}
dominate more and more over the dipolar terms. In case of a very large noise strength, the mean-field contributions can be neglected and 
the remaining dynamics can be solved analytically \cite{merku02,stane13}
\begin{align}
	4 g^{\alpha\alpha}_{\text{stat}}(t) &= \frac1{3} \left( 1 + 2 \left[ 1 - t^2 C \mathcal{J}^2 \right] \mathrm{e}^{- \frac12 t^2  C \mathcal{J}^2} \right).
	\label{eqn:merku}
\end{align}
Inspecting Fig.~\ref{fig:dip_lab_zero_field}, the data approach the curve 
\eqref{eqn:merku} for large values of $C$ and short times.
At larger times, i.e., beyond the local minimum, the dynamics due to the
dipolar couplings makes itself felt and the signal decays
below the analytical plateau. This is not surprising since the analytical consideration
 only includes static noise neglecting any mean-field dynamics. The corresponding
coupling may be weak compared to the noise, but it certainly affects the 
long-time behavior of the signals.

\begin{figure}[ht]
	\centering
	\includegraphics[width=1.0\columnwidth]{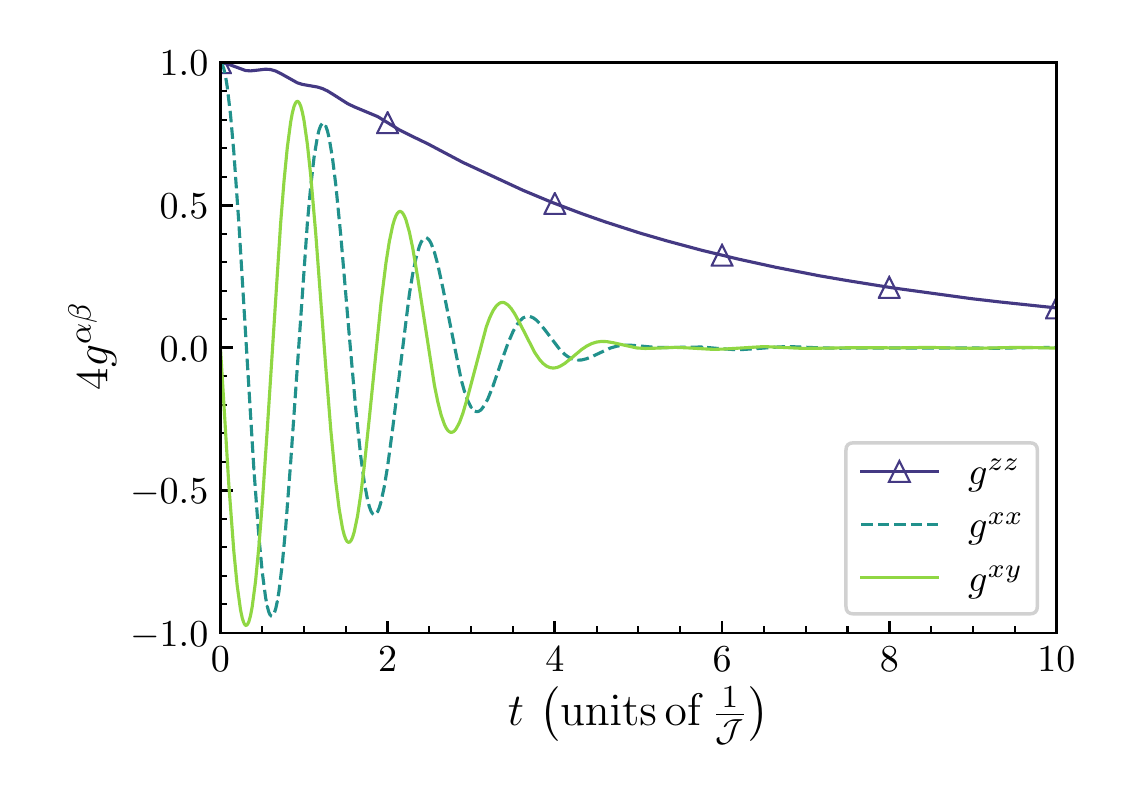}
	\caption{Numerical results of the self-consistency problem \eqref{eqn:weakfield} for 
	$C=0.0$, $\widetilde{B}=5.0$.}
	\label{fig:dip_lab_finite_field_A}
\end{figure}

\begin{figure}[ht]
	\centering
	\includegraphics[width=1.0\columnwidth]{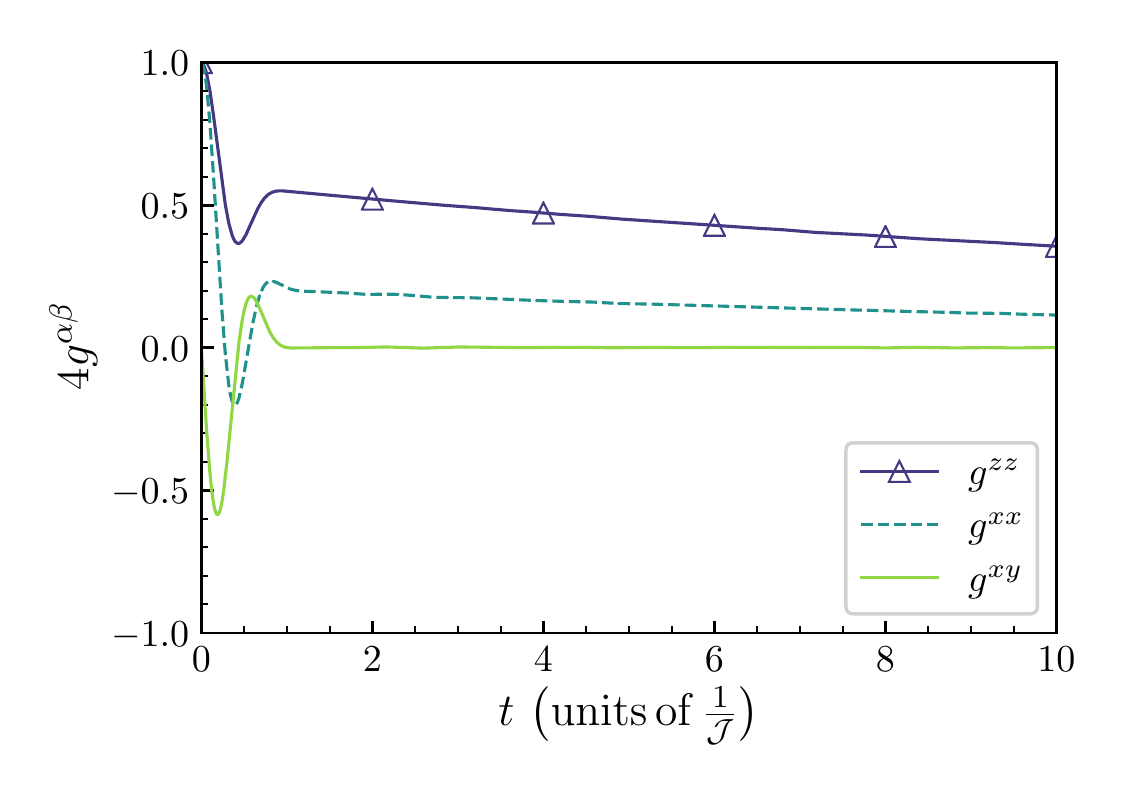}
	\caption{Numerical results of the self-consistency problem 
	\eqref{eqn:weakfield} for $C=10.0$ and $\widetilde{B}=5.0$.}
	\label{fig:dip_lab_finite_field_B}
\end{figure}

Figs.~\ref{fig:dip_lab_finite_field_A} and \ref{fig:dip_lab_finite_field_B} 
show  numerical results for finite magnetic field. 
What catches the eye is that the transversal autocorrelations in both figures 
show typical Larmor precessions with $\omega_{\text{L}} = \gamma_{\text{s}} B$.
For $C=0.0$, the precession persists until the transversal signals have decayed completely. 
For $C=10.0$, in contrast, the oscillations of $g^{xx}$ disappear very early 
although the signal is still finite. 
Subsequently, this correlation shows a slow long-time decay without 
discernible precession. We attribute this behavior to the presence of transversal noise
stabilizing $g^{xx}=g^{yy}$. This noise component is certainly weakened by the longitudinal magnetic field, but it appears to be still strong enough to keep the signal finite for 
quite a while.
Remarkably, the combination of noise and magnetic field causes a slow down of the longitudinal autocorrelation. This behavior is studied in detail in
the next section 
where we use the {RWA} to tackle the problem for considerably
larger magnetic fields. Since the transversal noise vanishes for such large fields, 
we expect the transversal long-time signal to vanish very quickly. 

\subsection{Strong-Field Regime and {RWA}}
\label{ss:rwa}

In the preceding section, we treated a general external magnetic field
of arbitrary strength, weak or strong, but perpendicular to the plane.
In this respect, the situation was specific. In the present section, we
choose the angle $\vartheta$ of the external field with the surface normal
in an arbitrary way, but consider a strong field so that the 
{RWA} is valid.

First, we switch from the laboratory frame to the frame rotating with the 
Larmor frequency of precession
\be
	\omega_{\text{L}} = \gamma_{\text{s}} B.
\ee
 This leads to a time-dependent effective Hamiltonian
with oscillating terms. They oscillate the faster the stronger the magnetic field is.
The {RWA} consists in averaging these fast oscillations
yielding an effective time-independent Hamiltonian again.
The spinDMFT is applied to this effective time-independent Hamiltonian.
This requires to solve
 a closed set of self-consistency equations capturing the spin dynamics in the
rotating frame.

In order to consider the system in the Larmor rotating frame 
one has to rotate any observables of the lab frame backwards
by the unitary evolution operator
\be
	\mb{U}_{\text{Z}}(t,t_0) := \mathrm{e}^{\texttt{i} \mb{H}_{\text{Z}} (t-t_0)}.
\ee
where the Zeeman term is the last-but-one term in Eq.\ \eqref{eqn:dipHam}:
\begin{align}
	\mb{H}_{\text{Z}} &:= \gamma_{\text{s}} B \sum_{i} \mb{S}^{z}_{i}.
\end{align}
The spin operators are given in the {rotating} frame by
\bs
\begin{align}
	\mb{S}^{\alpha}_{i,\text{rot}}(t) &= \mb{U}^{\dagg}_{\text{Z}}(t,t_0) \mb{S}^{\alpha}_{i,\text{lab}}(t) \mb{U}_{\text{Z}}(t,t_0) 
	\\
	&= \mb{U}^{\dagg}_{\text{Z}}(t,t_0) \mb{U}^{\dagg}(t,t_0) \mb{S}_{i}^{\alpha}(t_0) \mb{U}(t,t_0) \mb{U}_{\text{Z}}(t,t_0).
\end{align}
\es
The full time evolution in the rotating frame results from
the complete time evolution operator
\be
	\mb{U}^{\text{rot}}(t,t_0) := \mb{U}(t,t_0) \mb{U}_{\text{Z}}(t,t_0).
	\label{eqn:effTEO}
\ee
Its Sch\"odinger equation is derived by inserting \eqref{eqn:effTEO} in
the original Sch\"odinger equation 
\be
	\texttt{i} \partial_t \mb{U}^{\text{rot}}(t,t_0) 
	= \mb{H}^{\text{rot}}(t) \mb{U}^{\text{rot}}(t,t_0)
\ee
with the  Hamiltonian
\be
	\mb{H}^{\text{rot}}(t) = \mb{U}_{\text{Z}}(t,t_0) 
	\left( \mb{H} - \mb{H}_{\text{Z}}\right) \mb{U}^{\dagg}_{\text{Z}}(t,t_0).
\ee
Clearly, the Zeeman term $\mb{H}_{\text{Z}}$ in $\mb{H}$ is canceled in this way
which was the goal of this transformation.

The remaining terms, i.e.,
the dipole interaction and the noise are rotated by $\mb{U}_{\text{Z}}$. 
As a consequence, the Hamiltonian $\mb{H}^{\text{rot}}(t)$ is strongly 
time-dependent comprising fast oscillating terms such as 
$\cos (\omega_{\text{L}} t)$ or $\sin (\omega_{\text{L}} t)$
which are averaged in the sense of a Magnus expansion \cite{blane09} in first order.
The neglected terms are smaller by a factor $1/\omega_\text{L}$.
Put simply, the larger the  Larmor frequency $\omega_{\text{L}}$ is relative to the 
typical dipolar interaction frequency $\omega_{\text{DD}}=\mc J/\hbar$, the
better  the RWA is justified. Thus, the strong-field regime is realized for
\be
	B \gg B_{\text{DD}} = \gamma_{\text{s}} \omega_{\text{DD}}.
\ee
In this regime, one replaces all fast oscillating terms in the Hamiltonian 
by their average values, i.e.,
\bs
\begin{align}
	\cos (\omega_{\text{L}} t), \,\sin (\omega_{\text{L}} t) &\to 0 \\
	\cos (2\omega_{\text{L}} t), \,\sin (2\omega_{\text{L}} t) &\to 0.
\end{align}
\es
In our case, we obtain 
\begin{widetext}
\be
	\mb{H}^{\text{rot}} = 
	\frac12 J\left( R_{ij} \right)\sum_{i<j}  
	\left( 1 - 3\sin^2(\varphi_{ij})\sin^2(\vartheta)\right) 
	\left(2 \mb{S}_{i}^{z}\mb{S}_{j}^{z} - \mb{S}_{i}^{x}\mb{S}_{j}^{x} - 
	\mb{S}_{i}^{y}\mb{S}_{j}^{y} \right) 
	+ \gamma_{\text{s}} \sum_{i} b_{i}^{z} \mb{S}_{i}^{z},
	\label{eqn:rotframeHam}
\ee
\end{widetext}
which is again {time-independent} by construction. 
Note that the transversal components of the magnetic field noise are
 eliminated while the longitudinal component remains unchanged.
Therefore, one expects growing differences between transversal and
longitudinal autocorrelations upon increasing the noise strength $\sigma_\text{N}$.

Next, spinDMFT is applied to the Hamiltonian \eqref{eqn:rotframeHam}.
We only highlight expressions which differ from the previous
application. Expressed in local-field operators, the Hamiltonian reads as
\bs
\begin{align}
	\mb{H}^{\text{rot}} &= \frac12 \sum_{i} \vec{\mb{S}}_{i} \cdot \vec{\mb{V}}_{i} 
	+ \gamma_{\text{s}} \sum_{i} b_{i}^{z} \mb{S}_{i}^{z} 
	\\
	\vec{\mb{V}}_{i} &= \sum_{j, j\neq i} J\left(R_{ij}\right) \, \dul{D}^{\text{rot}}\left(\varphi_{ij}, \vartheta\right) \vec{\mb{S}}_{j},
\end{align}
\es
where $\dul{D}^{\text{rot}}$ is given by
\begin{align}
	\dul{D}^{\text{rot}}(\varphi_{ij},\vartheta) &= 
	\left( 3\sin^2(\varphi_{ij})\sin^2(\vartheta)-1\right)
	\begin{pmatrix}
		\tfrac12 &  0 & 0 \\
		 0 & \tfrac12 & 0 \\
		 0 &  0 & -1 \\
	\end{pmatrix}.
\end{align}
Replacing the local-field operators by the corresponding
mean-fields leads to the local mean-field Hamiltonian
\be
	\mb{H}^{\text{rot}}_{\text{mf},i} = 
	\vec{V}_{i}(t) \cdot \vec{\mb{S}} + \gamma_{\text{s}} b_{i}^{z} \mb{S}^{z}
	\label{eqn:RWA-mf-Hamiltonian}
\ee
and we again combine noise and mean-field in
\be
	{\vec{W}}_{i}(t) = \vec{V}_{i}(t) + \gamma_{\text{s}} \vec{e}_{z} b_{i}^{z},
\ee
since both follow from Gaussian distributions. 
While the first {moment} is still zero, the second moments obey
\begin{subequations}
\begin{align}
	w^{\alpha\beta}(t_1-t_2) :&= \overline{{W}^{\alpha}(t_1) {W}^{\beta}(t_2)} \\
	& = \delta_{\alpha\beta} \delta_{\alpha z} \gamma_{\text{s}}^2 \sigma_{\text{N}}^2 \\
	{+}\sum_{\rho \gamma} &\mc{J}^2  
	\left(\chi^{\text{rot}}\right)^{\alpha \beta}_{\rho \gamma} 
	(\vartheta) \langle \mb{S}^{\rho}(t_1) \mb{S}^{\gamma}(t_2) \rangle\emf. \nonumber
\end{align}
\label{eqn:selfconsRF}%
\end{subequations}
The coupling $\mc{J}$ is given by 
Eq.~\eqref{eqn:natenergy} and the 
coefficients $\chi^{\text{rot}}$ result from Eq.\ \eqref{eqn:chi}
with $\dul{D}$ replaced by $\dul{D}^{\text{rot}}$.

Defining the  function of the polar angle $\vartheta$
\bs
\label{eq:polar}
\begin{align}
	I(\vartheta) &:= \frac1{2\pi} \int_{0}^{2\pi} \mathrm{d} \varphi \, \left( 1 - 3\sin^2(\varphi)\sin^2(\vartheta)\right)^2  \\
	&= \frac{27}{8} \sin^4(\vartheta) - 3 \sin^2(\vartheta) + 1,
\end{align}
\es
we can express the prefactors concisely by
\bs
\begin{align}
	\left(\chi^{\text{rot}}\right)^{xx}_{xx} &= \left(\chi^{\text{rot}}\right)^{yy}_{yy} = \tfrac14 I(\vartheta) \\
	\left(\chi^{\text{rot}}\right)^{xy}_{xy} &= \left(\chi^{\text{rot}}\right)^{yx}_{yx} = \tfrac14 I(\vartheta) \\
	\left(\chi^{\text{rot}}\right)^{xz}_{xz} &= \left(\chi^{\text{rot}}\right)^{yz}_{yz} = -\tfrac12 I(\vartheta) \\
	\left(\chi^{\text{rot}}\right)^{zx}_{zx} &= \left(\chi^{\text{rot}}\right)^{zy}_{zy} = -\tfrac12 I(\vartheta) \\
	\left(\chi^{\text{rot}}\right)^{zz}_{zz} &= I(\vartheta).
\end{align}
\es
Any other coefficient vanishes because $\dul{D}^{\text{rot}}$ is diagonal.

We reconsider the symmetries of the underlying system
to be able to formulate the minimum set of 
 self-consistency conditions.
 The original rotating-frame Hamiltonian 
\eqref{eqn:rotframeHam} is invariant under spin rotation around the $z$
axis by construction: all transversal spin components only occur in pairs. 
The dipolar part is invariant under time reversal.
While this does not hold for the noise term for each individual 
$\vec b_i$, their distribution remains unchanged under time reversal.
Spin-rotation symmetry and time-reversal symmetry  allow us to conclude
\bs
\begin{align}
	g^{xx}(t) &= g^{yy}(t) & &\\
	g^{\alpha\beta}(t) &= 0, & \alpha &\neq \beta.
\end{align}
\es
This enables us to reduce the general self-consistency conditions
\eqref{eqn:selfconsRF} to
\bs
\begin{align}
	w^{xx}(t) &= w^{yy}(t) 
	= \tfrac14 \mc{J}^2 I(\vartheta) g^{xx}(t) & &\\
	w^{\alpha\beta}(t) &= 0, & \alpha &\neq \beta \\
	w^{zz}(t) &= \mc{J}^2 I(\vartheta) g^{zz}(t) 
	+ \gamma_{\text{s}}^2 \sigma_{\text{N}}^2. & &
\end{align}
	\label{eqn:strongfield}%
\es
Note that there is a natural anisotropy between the transversal and longitudinal equation again, but this time by a factor of four.
Furthermore, the noise only acts in the $z$-direction. 
Considering the results of the previous section, we expect even bigger 
differences between the autocorrelations here.
Henceforth, we no longer use the terms ``diagonal'' and ``cross'' because no cross autocorrelations appear in the RWA dipole model.
If not stated otherwise, we set $\vartheta$ to the so-called magic angle
\begin{align*}
	\vartheta_{\text{magic}} &:= \arcsin \sqrt{\frac23} \\
	I(\vartheta_{\text{magic}}) &= \frac12
\end{align*}
in our numerical calculations.

\begin{figure}[ht]
	\centering
	\includegraphics[width=1.0\columnwidth]{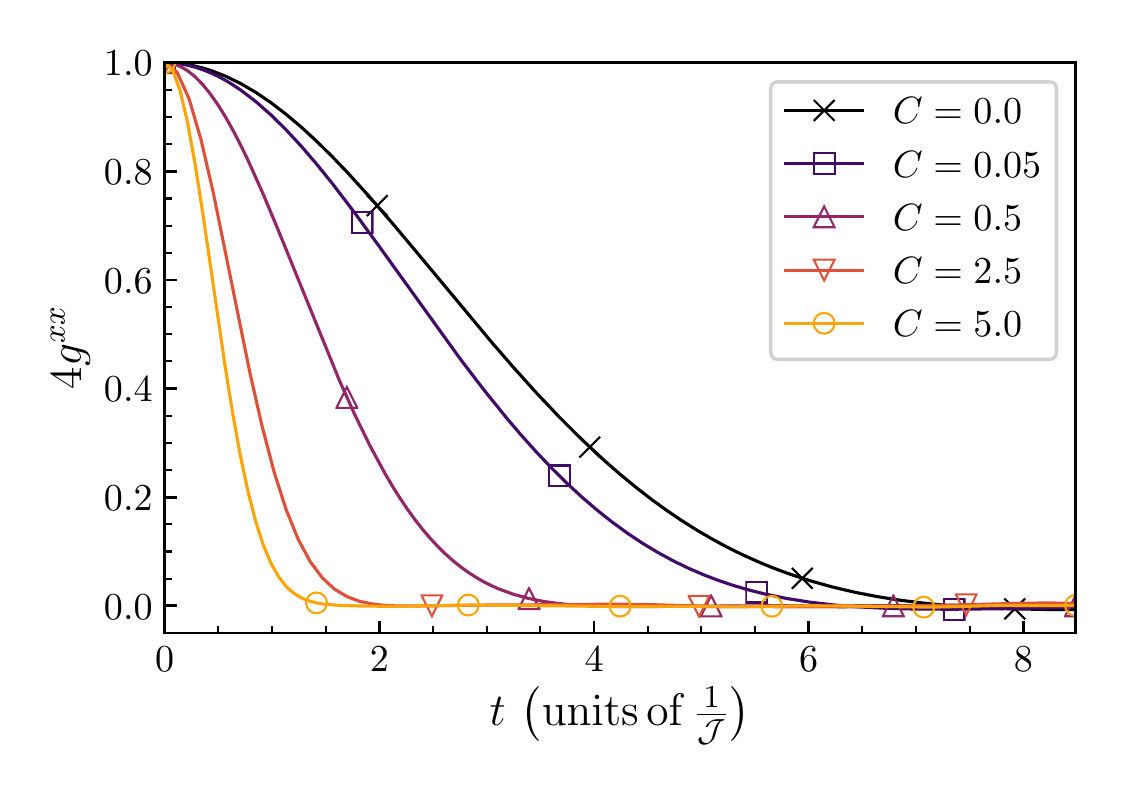}
	\caption{Transversal RWA results of the self-consistency problem \eqref{eqn:strongfield} for various noise strengths.}
	\label{fig:dipole_multiC_xx}
\end{figure}

\begin{figure}[ht]
	\centering
	\includegraphics[width=1.0\columnwidth]{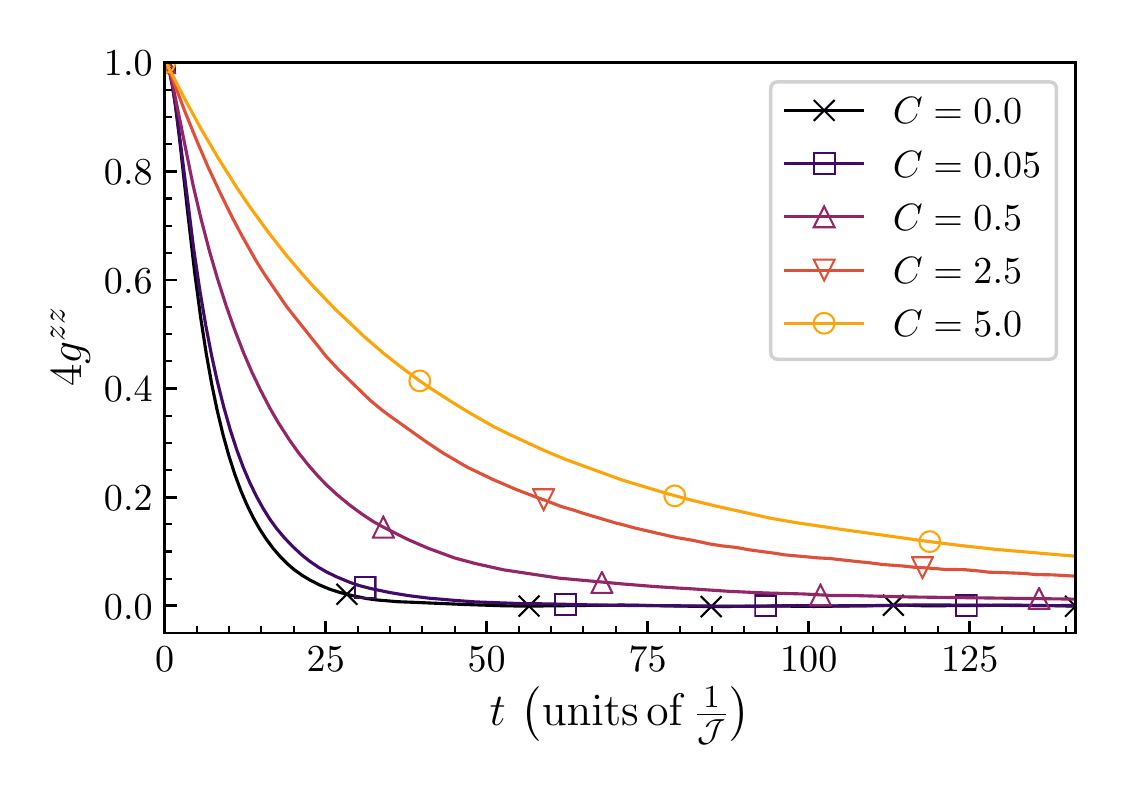}
	\caption{Longitudinal RWA results of the self-consistency problem \eqref{eqn:strongfield} for various noise strengths.}
	\label{fig:dipole_multiC_zz}
\end{figure}

Figs.~\ref{fig:dipole_multiC_xx} and \ref{fig:dipole_multiC_zz} show our numerical findings 
for the solutions of the self-consistent equations.
Analyzing them, we conclude two important facts:
\begin{itemize}
	\item[(i)] Because of the natural anisotropy \emph{and} the noise, the longitudinal signal decays considerably slower than the transversal signal. 
	Increasing the noise strength amplifies this difference. 
	\item[(ii)] The transversal signal decays very accurately following a Gaussian,
	while the longitudinal decay is weaker than exponential.
\end{itemize}

\begin{figure}
	\centering
	\includegraphics[width=0.46\textwidth]{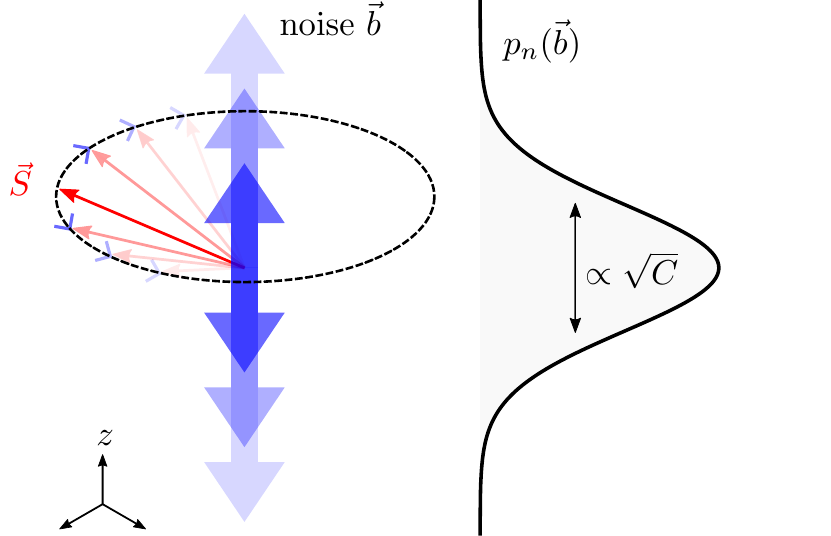}
	\caption{Illustration of a spin subjected to a Gaussian noise in $z$ direction. As the 
	speeds of rotation of the transversal components fluctuate more strongly 
	upon increasing noise strength $C$ the transversal components vanish more rapidly.
	In return, this weakens the processes which destroy the longitudinal component so that
	it is stabilized.}
	\label{fig:noise_effects}
\end{figure}

Considering the results, the first statement is rather obvious. For $C=0.0$, 
the longitudinal signal already survives significantly longer than the transversal one
due to the anisotropic factors in \eqref{eqn:strongfield}. Increasing the noise strength causes the spin to 
precess more and more quickly and randomly about the $z$ axis. 
As a result, the transversal spin components experience an even stronger
\emph{decoherence} than before and the corresponding autocorrelations decay faster. 
In contrast, the 
$z$ component of the spin is \emph{stabilized} by the additional rotations about the $z$ 
axis so that the longitudinal signal relaxes only very slowly.
Fig.~\ref{fig:noise_effects} schematically illustrates the behavior of the spins.

\begin{figure}
	\centering
	\includegraphics[width=1.0\columnwidth]{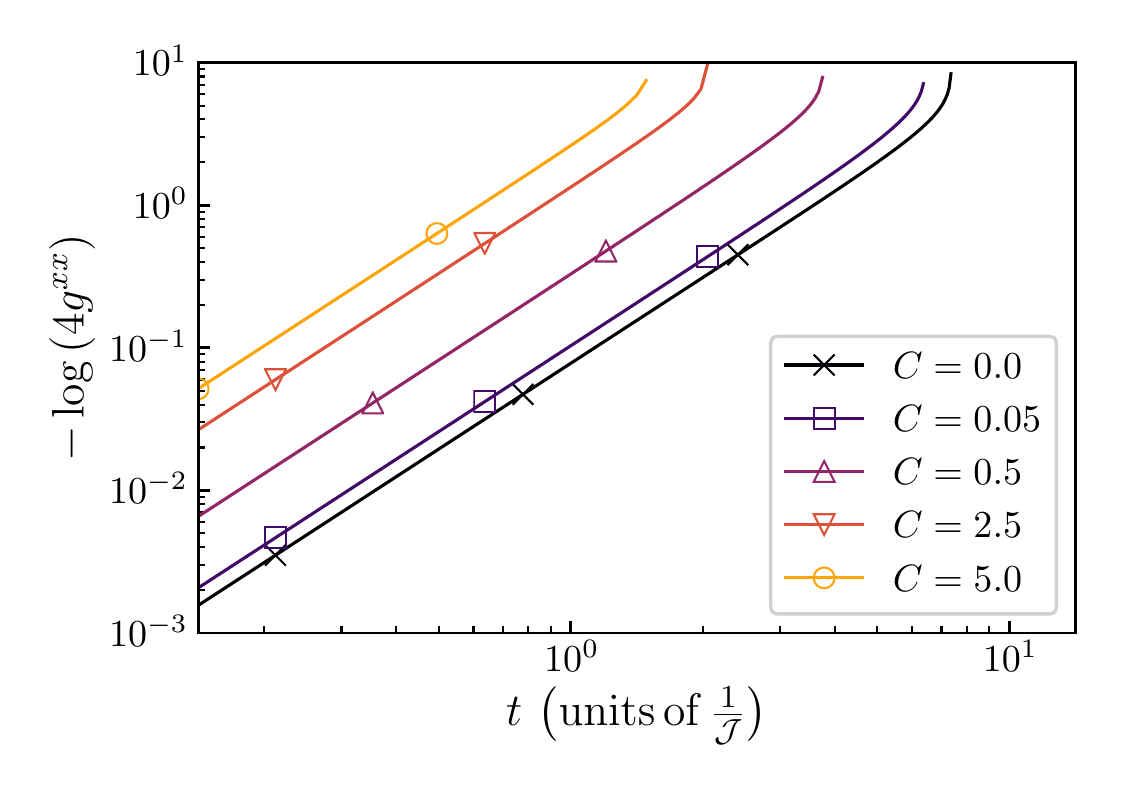}
	\caption{Logarithmic representation of the transversal RWA results for various noise strengths.}
	\label{fig:dipole_transversal_log}
\end{figure}

To corroborate the statement (ii), Fig.~\ref{fig:dipole_transversal_log} shows the 
transversal signal  in loglog vs.\ log representation for various noise strengths. 
All of the results show a linear behavior with a small upward curvature at the end. 
A closer observation of the curvature reveals that the autocorrelations fall slightly below zero 
just before they finally converge to zero. We emphasize that this negative dip is only very small, $\propto 10^{-3}$,
and hence not visible in the provided figures. Still, we ensured that it does not result from numerical 
inaccuracies. Interestingly, its height decreases upon increasing the noise width.

\begin{figure}
	\centering
	\includegraphics[width=1.0\columnwidth]{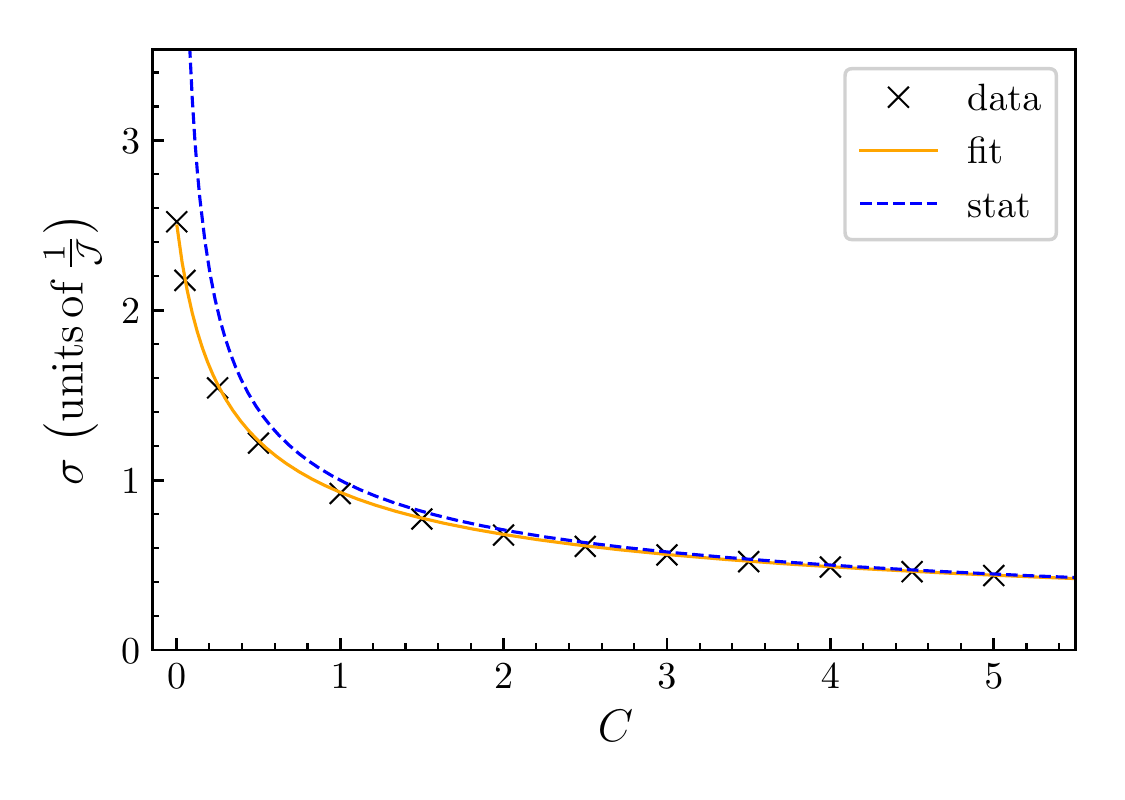}
	\caption{Standard deviations obtained from fitting the transversal signals versus the noise strengths. 
	The dashed blue line is the analytical prediction based on purely static noise, see Eq.\ \eqref{eqn:onlynoise}.
	The orange line results from a data fit according to \eqref{eqn:sigmaCfit}.}
	\label{fig:dipole_transversal_sigma}
\end{figure}

The clearly linear behavior in the plot has a slope $r=2$ so that it clearly indicates 
Gaussian behavior. Hence, we use the fit function
\begin{align}
	4g^{xx}_{\text{Gauss}}(t) &= \mathrm{e}^{-\frac{t^2}{2 \sigma^2}}
	\label{eqn:Gaussfit}
\end{align}
to extract the standard deviation $\sigma$ as function of the noise strength displayed in
Fig.~\ref{fig:dipole_transversal_sigma}.

This behavior can be understood by an analytical consideration \cite{merku02,stane13}. 
We consider purely static noise neglecting the mean-field contributions in the 
 RWA Hamiltonian \eqref{eqn:RWA-mf-Hamiltonian}:
\begin{align}
	\mb{H} &= \gamma_{\text{s}} b_i^z \mb{S}^z,
\end{align}
The analytical averaging yields the transversal signal 
\begin{subequations}
\begin{align}
	4g^{xx}_{\text{N}}(t) &= \mathrm{e}^{-\frac{t^2}{2 \sigma(C)^2}} 
	\\
	\sigma(C) &= \frac1{\mathcal{J}\sqrt{C}},
\end{align}
\label{eqn:onlynoise}%
\end{subequations}
which explains the Gaussian behavior of the transversal signal for large values of $C$. 
The analytical standard deviation is also depicted
in Fig.~\ref{fig:dipole_transversal_sigma} as blue dashed line. 
For increasing noise strength it describes the fitted data better and better.
This is expected because the larger the static noise is relative to the mean-fields, 
the better the static noise model captures the dynamics.
For small values of $C$ and in particular for $C=0.0$, 
the transversal signal still follows a Gaussian to good accuracy.
The mean-field contribution changes the standard deviations;
quite unexpectedly the mean-fields appear to counteract the static noise partly
 reducing the standard deviation.  It turns out that this effect is well
captured by the fit
\begin{align}
	\sigma_{\text{fit}}(C) &= \frac1{\mathcal{J}\sqrt{C+R}}
	\label{eqn:sigmaCfit}
\end{align}
as can be seen in Fig.~\ref{fig:dipole_transversal_sigma} with 
\begin{align}
	R &= \num{0.159\pm 0.001}
\end{align}
This quantifies the contribution of the mean-fields to the transversal spin dynamics.

\begin{figure}
	\centering
	\includegraphics[width=1.0\columnwidth]{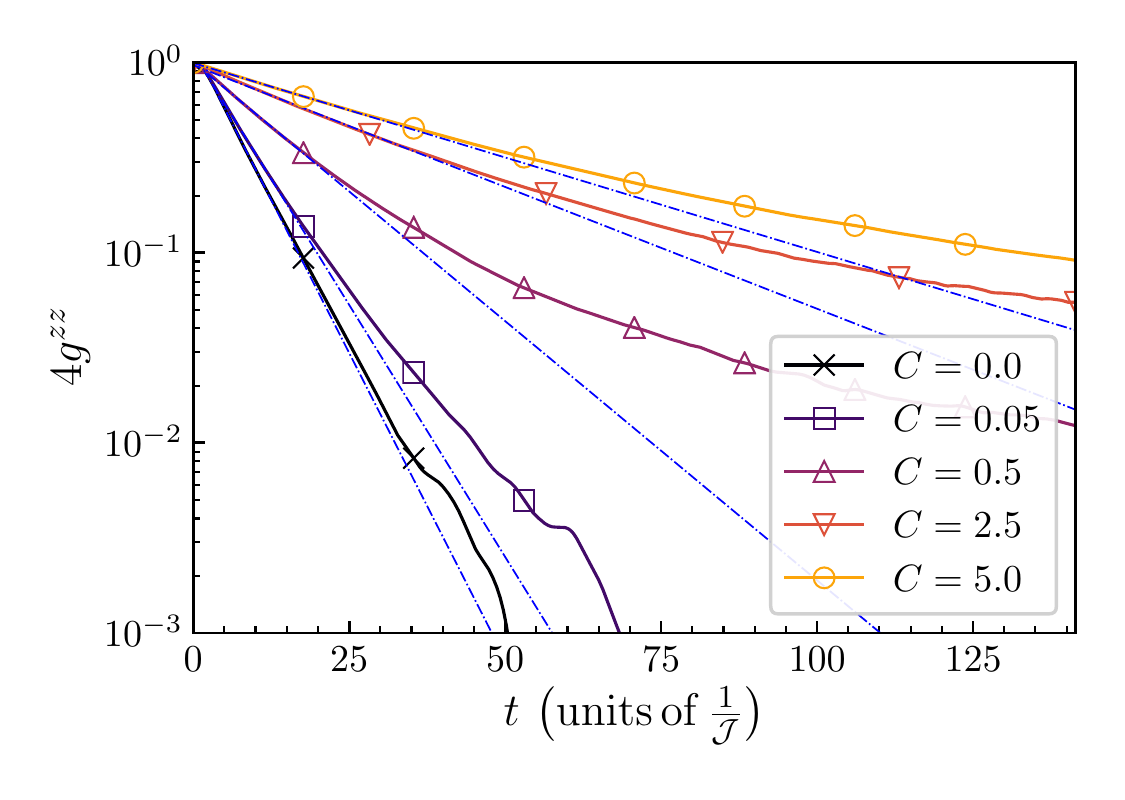}
	\caption{Logarithmic representation of the longitudinal correlation for various noise strengths. 
	The blue dashed-dotted lines correspond to 
	exponential functions with $f(t) = A\mathrm{e}^{-\alpha t}$. 
	They illustrate that the correlations show a positive curvature and thus decay weaker than exponentially.}
	\label{fig:dipole_longitudinal_log}
\end{figure}

The behavior of the longitudinal autocorrelations is more complex; Fig.~\ref{fig:dipole_longitudinal_log} shows
that the decay is weaker than exponentially as we stated before. We refer to Sect.\ \ref{ss:longtimebehavior} for 
a detailed examination of the behavior.

\subsection{Transition from weak to strong external magnetic field}

If we consider the case of a surface perpendicular to the
 external magnetic field, i.e., $\vartheta=0$, we can compare the results from Sect.\ \ref{ss:generalfield}
for arbitrarily strong magnetic fields to the results from the previous Sect.\ \ref{ss:rwa}
based on the {RWA}.
This allows us to study how well the RWA reproduces the exact result. In particular, we can 
determine above which magnetic fields the RWA is reliable and to which extent.
First, we compute the results of the self-consistency problem in the lab frame 
(exact spinDMFT) \eqref{eqn:weakfield} and in the Larmor rotating frame using 
RWA (RWA spinDMFT) \eqref{eqn:strongfield}. Then, we transform the exact spinDMFT results 
to the Larmor rotating frame via 
\begin{subequations}
\begin{align}
	g_{\text{rf}}^{xx}(t) &= g^{xx}(t) \cos(\omega_{\text{L}} t) - g^{xy}(t) \sin(\omega_{\text{L}} t), \\
	g_{\text{rf}}^{xy}(t) &= g^{xy}(t) \cos(\omega_{\text{L}} t) + g^{xx}(t) \sin(\omega_{\text{L}} t), \\
	g_{\text{rf}}^{zz}(t) &= g^{zz}(t),
\end{align}
\label{eqn:transform_rotframe}%
\end{subequations}
so that a quantitative comparison is possible. 

\begin{figure}[ht]
	\centering
	\includegraphics[width=\columnwidth]{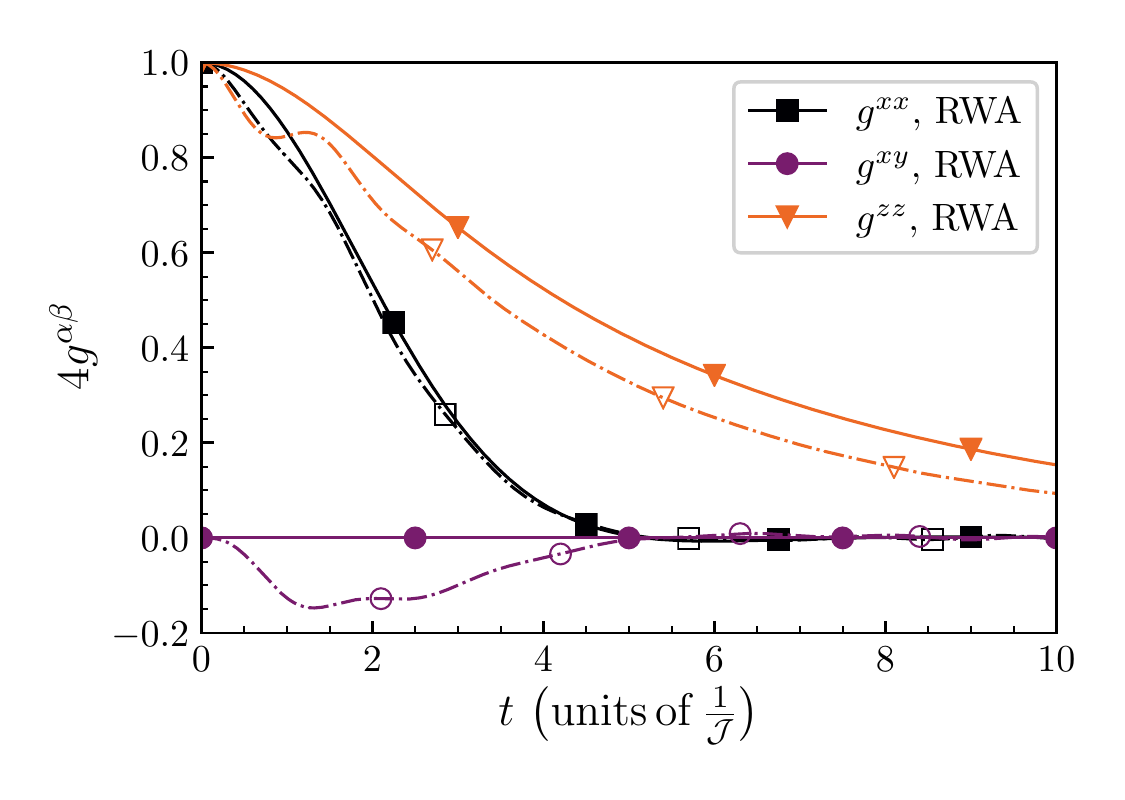}
	\caption{Comparison of the RWA spinDMFT results (solid line, filled markers) with the exact {spinDMFT} results 
	(dashed-dotted line, open markers) for a small dimensionless magnetic field $\widetilde{B} = 2.0$ in the rotating frame 
	at zero static noise $C=0.0$.}
	\label{fig:dipole_RWA_C0_B2}
\end{figure}

\begin{figure}[ht]
	\centering
	\includegraphics[width=\columnwidth]{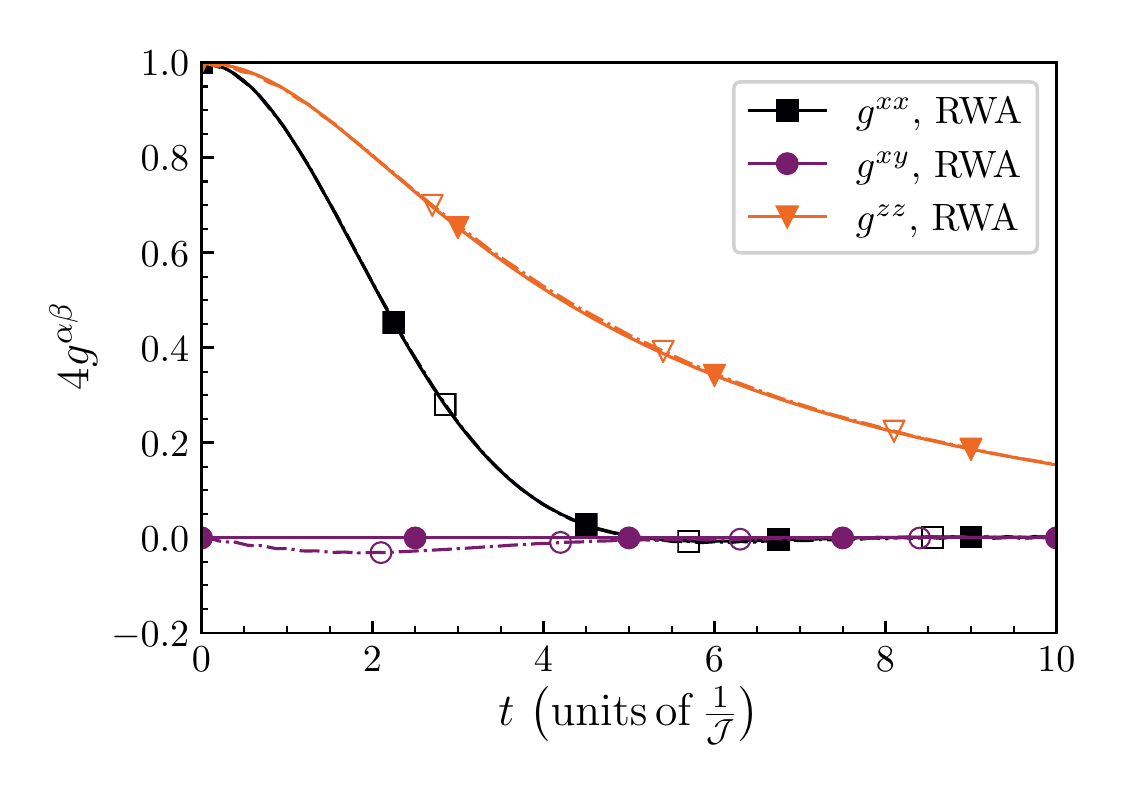}
	\caption{Same as Fig.\ \ref{fig:dipole_RWA_C0_B2} for moderate dimensionless
	magnetic field $\widetilde{B} = 10.0$.}
	\label{fig:dipole_RWA_C0_B10}
\end{figure}

\begin{figure}[ht]
	\centering
	\includegraphics[width=\columnwidth]{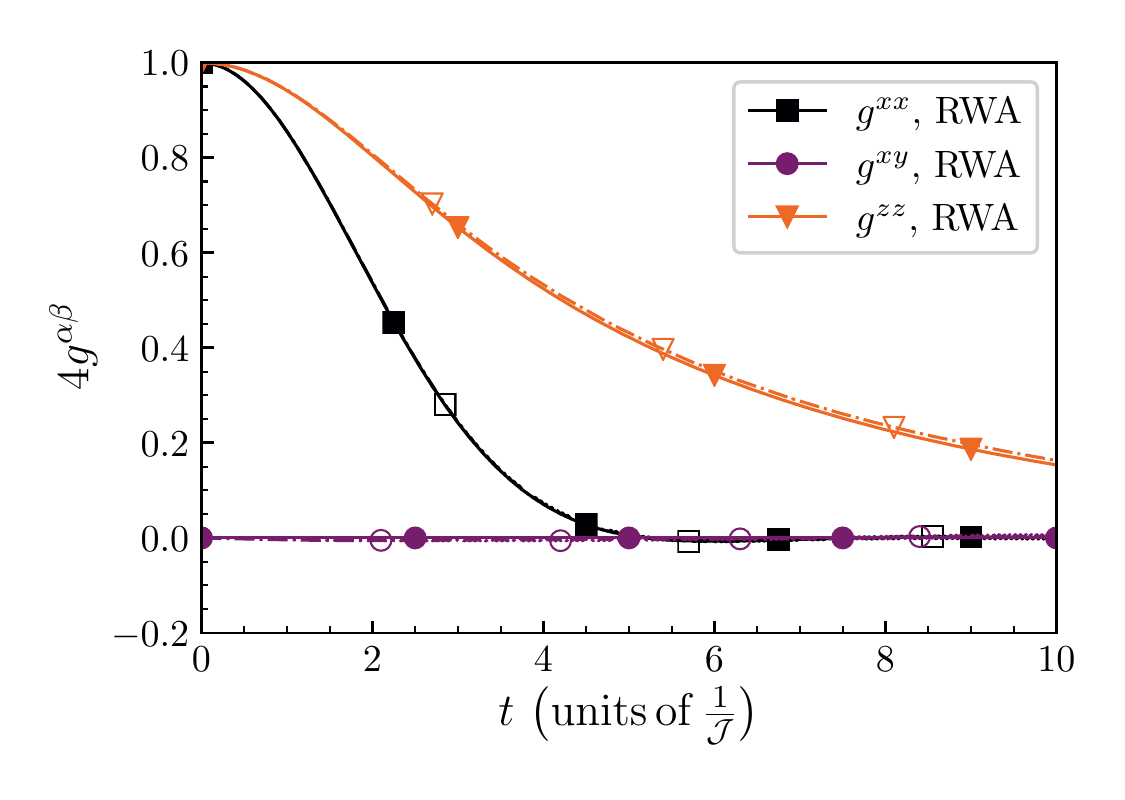}
	\caption{Same as Fig.\ \ref{fig:dipole_RWA_C0_B2} for large dimensionless
	magnetic field $\widetilde{B} = 50.0$.}
	\label{fig:dipole_RWA_C0_B50}
\end{figure}

In Fig.~\ref{fig:dipole_RWA_C0_B2} ($C=0.0$, $\widetilde{B}=2.0$), we observe considerable deviations between both approaches,
especially in $g^{xy}$ and $g^{zz}$: due to the moderately large magnetic field, the exact results
show deflections and shifts which are not present in RWA. 
Considering Fig.~\ref{fig:dipole_RWA_C0_B10} ($C=0.0$, $\widetilde{B}=10.0$), 
these deviations clearly shrink upon increasing magnetic field. The deviations {from the} RWA are difficult to discern. {They appear most strongly} in
$g^{xy}$. As we raise $\widetilde{B}$ further, see Fig.~\ref{fig:dipole_RWA_C0_B50} ($C=0.0$, $\widetilde{B}=50.0$), 
the deviations due to RWA are not visible anymore. The tiny shift between both results for the longitudinal autocorrelation
only stems from the discretization of time. It is a purely numerical effect
which grows with $\widetilde{B}$.

\begin{figure}[ht]
	\centering
	\includegraphics[width=\columnwidth]{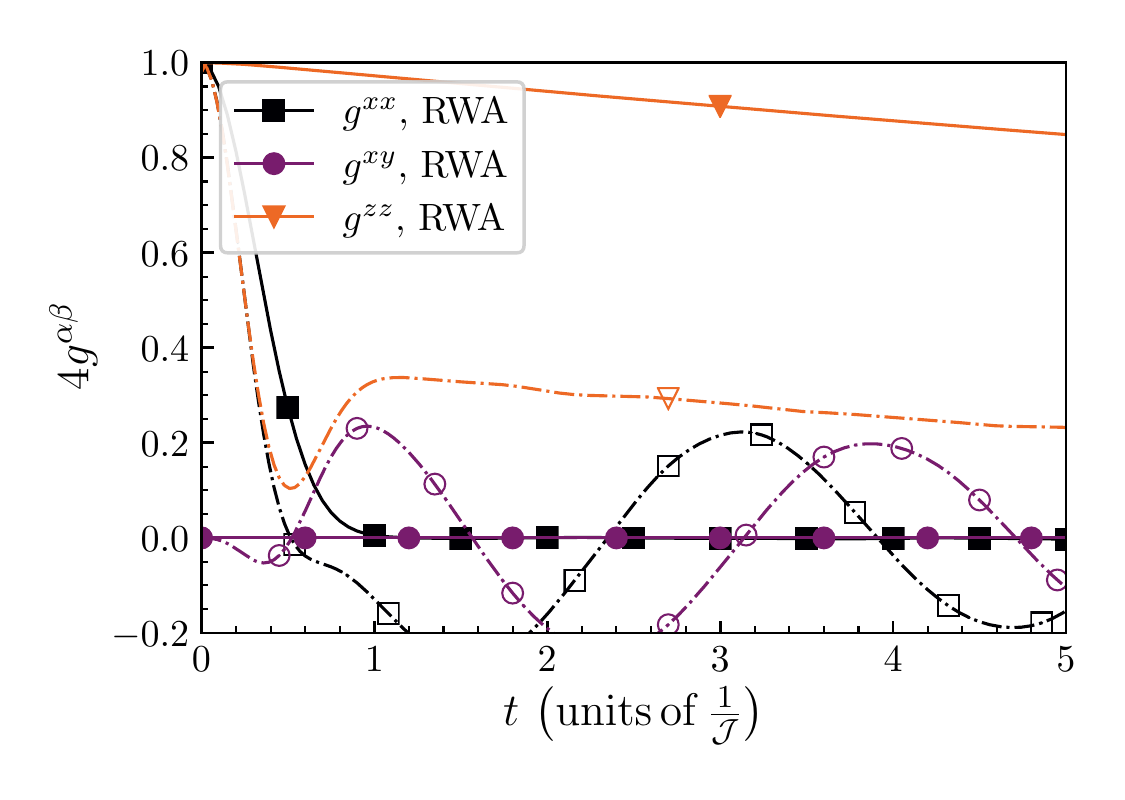}
	\caption{Same as Fig.\ \ref{fig:dipole_RWA_C0_B2} for finite static noise $C = 10.0$ and small 
	dimensionless magnetic field $\widetilde{B} = 2.0$.}
	\label{fig:dipole_RWA_C10_B2}
\end{figure}

\begin{figure}[ht]
	\centering
	\includegraphics[width=\columnwidth]{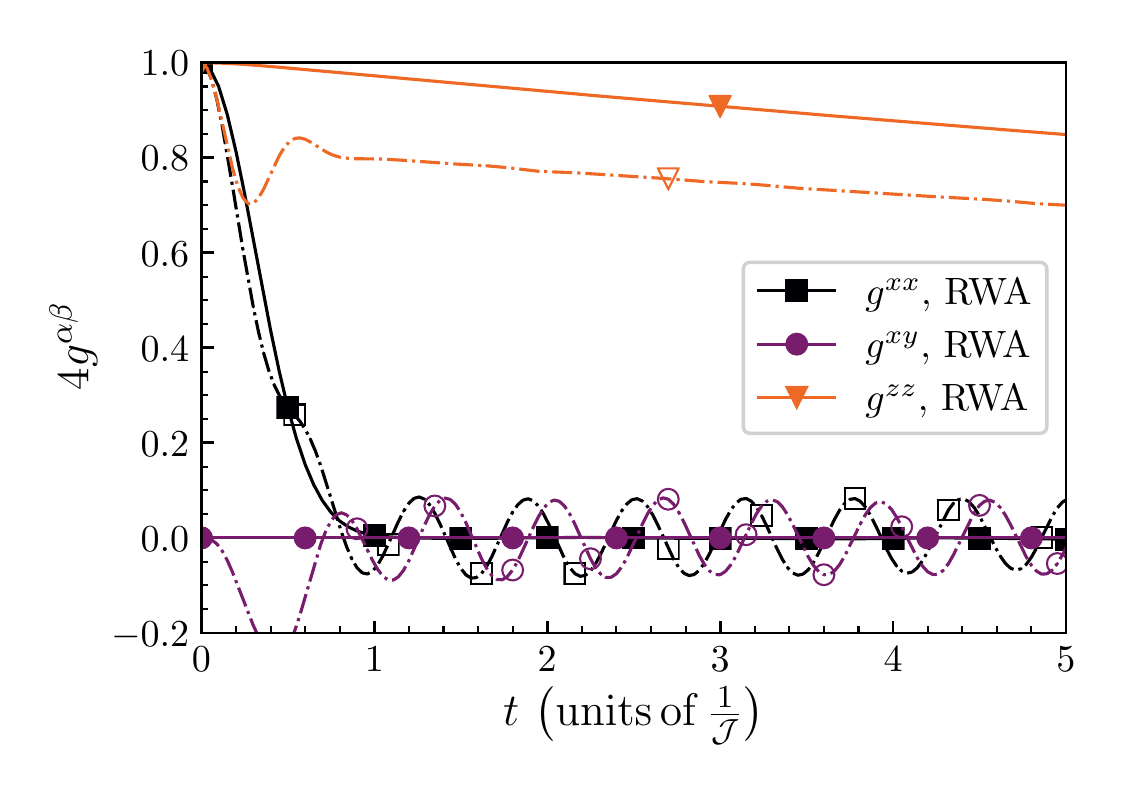}
	\caption{Same as Fig.\ \ref{fig:dipole_RWA_C0_B2} for finite static noise $C = 10.0$ and moderate 
	dimensionless {magnetic field} $\widetilde{B} = 10.0$.}
	\label{fig:dipole_RWA_C10_B10}
\end{figure}

\begin{figure}[ht]
	\centering
	\includegraphics[width=\columnwidth]{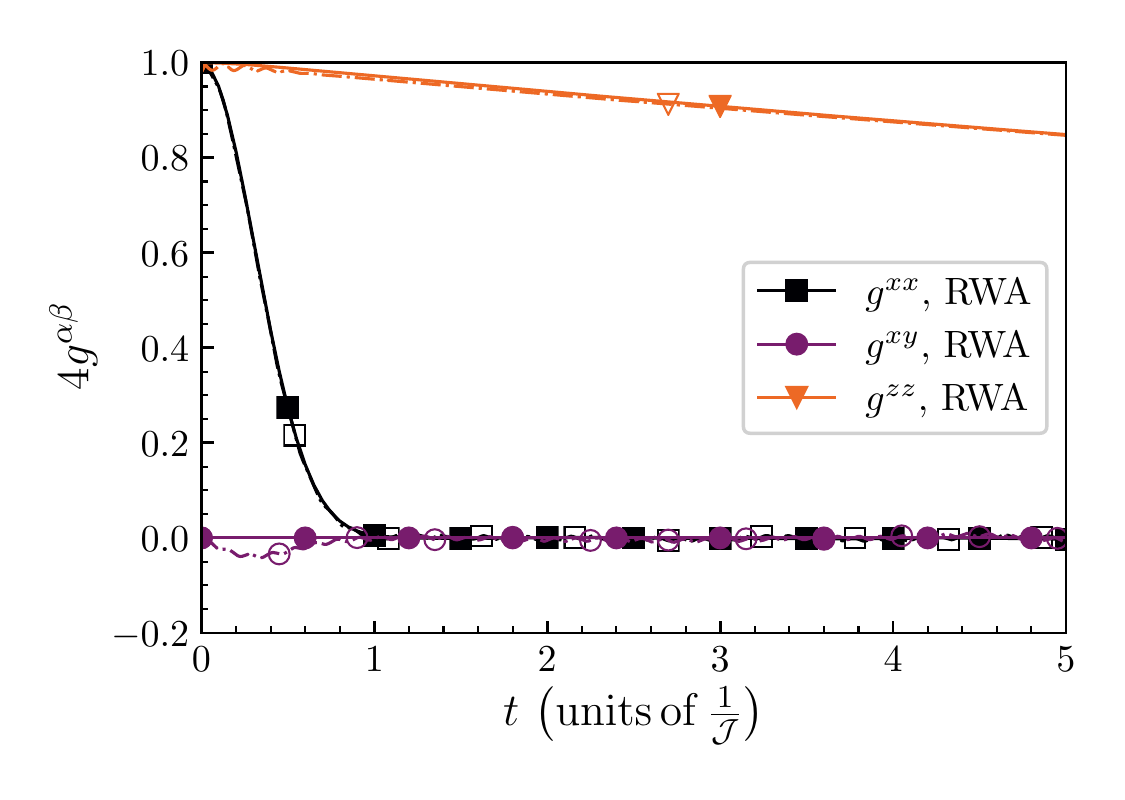}
	\caption{Same as Fig.\ \ref{fig:dipole_RWA_C0_B2} for finite static noise $C = 10.0$ and large 
	dimensionless magnetic field $\widetilde{B} = 50.0$.}
	\label{fig:dipole_RWA_C10_B50}
\end{figure}

If static noise is included, see Fig.~\ref{fig:dipole_RWA_C10_B2} for $C=10.0$, $\widetilde{B}=2.0$, 
we observe large deviations for all autocorrelations, even more than
what we showed in Fig.\ \ref{fig:dipole_RWA_C0_B2}. The exact transversal results strongly oscillate
in contrast to the RWA results and a huge shift between the two curves for the
longitudinal autocorrelations occurs. This implies that the presence of 
static noise requires larger magnetic fields for the RWA to be justified. 
Figures~\ref{fig:dipole_RWA_C10_B10} ($C=10.0$, $\widetilde{B}=10.0$) and \ref{fig:dipole_RWA_C10_B50} ($C=10.0$, $\widetilde{B}=50.0$) confirm this
conclusion displaying better and better {agreement} between the results of both approaches.
We argue that this behavior is physically highly plausible because the 
 RWA is justified if the energy scale of the magnetic field is 
larger than the energy scales of \emph{any} other interaction 
in the system, including the static noise.

\subsection{Long-time behavior}
\label{ss:longtimebehavior}

Now, we come back to the long-time behavior of the longitudinal autocorrelation. This is an interesting
issue because various ideas exist on the origin of the rather slow decay 
and its functional form \cite{kucsk18}.

\begin{figure}[ht]
	\centering
	\includegraphics[width=1.0\columnwidth]{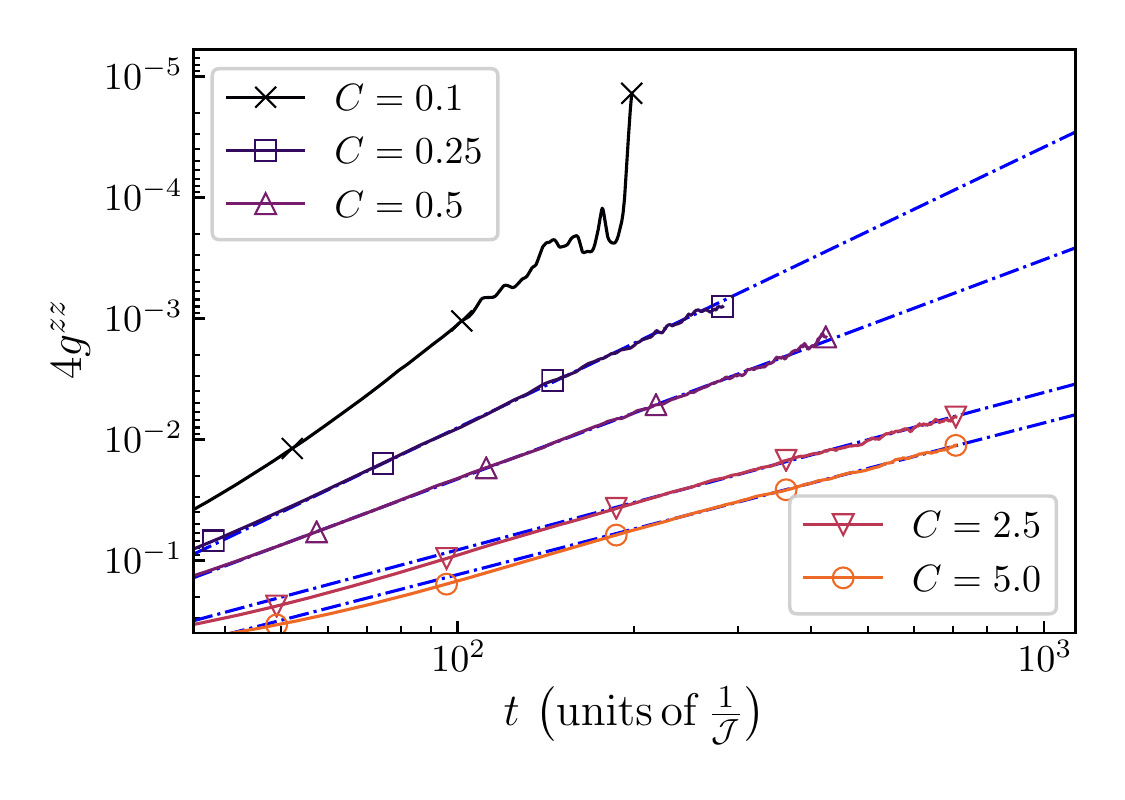}
	\caption{Numerical long-time results of the longitudinal autocorrelations in RWA for various noise strengths in Log vs. Log representation.
	The blue dashed-dotted lines correspond to power law fits as in \eqref{eqn:Polyfit}.}
	\label{fig:dip_rot_longtime}
\end{figure}

Fig.\ \ref{fig:dipole_longitudinal_log} shows that the autocorrelations do not decay in a Gaussian fashion at all.
Such decay would have led to a negative curvature downwards. Instead we discern a positive curvature
upwards which implies that the decay is even slower than exponential. The question arises which functional
dependencies describe this decay. Considering this, our most successful fitting attempt is a power law according to
\begin{align}
	4g^{zz}_{\text{fit:B}}(t) &= B t^{-m}
	\label{eqn:Polyfit}
\end{align}
with parameters $B$ and $m$. We checked also stretched exponentials since these were suggested in Ref.~\cite{kucsk18}.
But we did not achieve satisfactory fits for an ansatz according to 
\begin{align}
		4g^{zz}_{\text{fit:A}}(t) &= A\mathrm{e}^{-\alpha t^{\nu}}
		\label{eqn:StretchedExpfit}
\end{align}
with parameters $A$, $\alpha$, and the exponent $\nu$.
The longitudinal results including the power law fits can be seen in Fig.~\ref{fig:dip_rot_longtime}.
Note that much longer times are not easily accessible
for two reasons. First, the numerical effort increases as $t^2$ for the Monte-Carlo simulation and as $t^3$ for the diagonalization. 
Second, the smaller the autocorrelation is the more difficult it becomes to determine it
with good relative accuracy in view of the statistical way of computing it.
This is also the reason why we cannot go to longer times for $C\approx 0$.

\begin{figure}[ht]
	\centering
	\includegraphics[width=1.0\columnwidth]{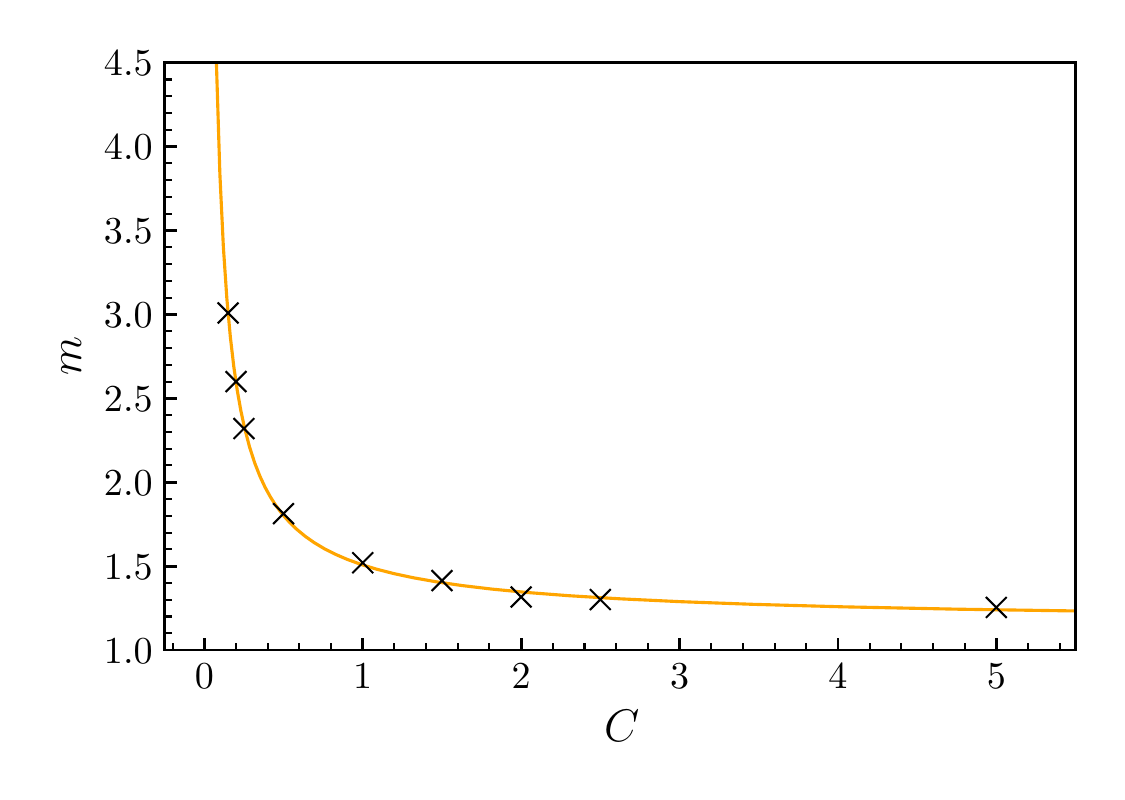}
	\caption{Fitted exponent $m$ from the fit of the longitudinal autocorrelations \eqref{eqn:Polyfit} vs.\ the noise strength and a fit (orange) 
	as in \eqref{eqn:PolyExponentFit}. The estimated parameters are $m_0 = \num{1.15(2)}$, $k = \num{0.35(2)}$ and $r = \num{0.87(3)}$.}
	\label{fig:dip_rot_longtime_polyfit}
\end{figure}

The exponent $m$ displays a pronounced dependence on the relative noise strength $C$
as depicted in Fig.\ \ref{fig:dip_rot_longtime_polyfit}. The dependence $m(C)$ can be 
described heuristically by 
\begin{align}
		m_{\text{fit}}(C) &= m_{0} + \frac{k}{C^{r}}.
		\label{eqn:PolyExponentFit}
\end{align}
This fit works surprisingly well in spite of the divergence for $C\to0$. Limited by the numerical accuracy, we can hardly say if we actually obtain 
this divergence or if it can be truncated, e.g., by replacing $C \to C + C_0$ in \eqref{eqn:PolyExponentFit}. This issue is associated to the question if 
the power law behavior solely results from the presence of the noise or if it is a valid feature of spinDMFT.

All in all, these results provide evidence that the longitudinal autocorrelations are
relatively long-lived. Clearly, their long-time behavior poses an interesting
question which calls for further research, both numerical and analytical.

\section{Conclusions}
\label{sec:conclusion}

In this paper, we introduced and justified a mean-field theory designed
to capture the spin dynamics in disordered dense spin systems. The key idea is that 
it is not sufficient to introduce a static mean-field but that the mean-field
is dynamic itself so that we call it ``spin dynamic mean-field theory'' (spinDMFT).
As usual, this approach becomes exact if each site has an infinite  number
of interaction partners, i.e., the coordination number becomes infinitely large.
{Historically}, the same limit led to the introduction of the fermionic
dynamic mean-field theory \cite{metzn89a,georg96}.

For spins, we established that the important correlations are the
autocorrelations and that these define the dynamic mean-fields to which each spin
is subjected. These mean-fields are normally distributed and the dynamic variances
of these normal distributions are given by the autocorrelations. This constitutes
the self-consistency {problem} which has to be solved for spinDMFT. 
We showed how this can be done stochastically.


{If the effective single-site problem is linear in the
spin operators it does not matter whether we consider
classical spins averaged over all directions or a quantum spin
given that the average length is scaled to be the same.
In this sense, the quantum spin system and the classical spin
system have the same spinDMFT.
For spin-$1/2$ this has to be the case since locally only
linear spin operators can appear. For larger spins, however,
higher powers may arise such as anisotropies of various kinds.
Then, the quantum spinDMFT and the classical spinDMFT are different.}


We gauged the advocated spinDMFT against numerical results for isotropic 
spin systems obtained by other 
methods, namely exact diagonalization, iterated equations of motion, and 
Chebyshev expansion. This can only be done for rather small spin clusters
in low dimensions so that the reproduction of the results by spinDMFT is particularly
challenging. {Nevertheless}, encouraging agreement could be established.

Subsequently, we applied spinDMFT to a two-dimensional ensemble of spins with dipolar 
interactions including local static noise, i.e., 
fluctuations of magnetic fields. This underlines that spinDMFT is capable {of dealing} with anisotropic
interactions of long range as well. We studied the case of an arbitrary external
magnetic field perpendicular to the  plane of spins and the {RWA}
for a large tilted magnetic field. For zero tilt, we compared both approaches
quantitatively. This allowed us to show quantitatively to which extent
the RWA is justified and above which magnetic field it yields reliable results.

We showed that the transversal autocorrelations behave essentially like Gaussians
in time. The longitudinal autocorrelations, however, display a more complex behavior 
with a rather slow decay towards long times. Evidence for power law behavior
is found. This certainly calls for further investigations.

We are confident that spinDMFT can be applied successfully
to many more physical systems aside from the ones that we mentioned
so far. Ample applications can be found for
nuclear magnetic resonance (NMR), electron spin resonance (ESR), quantum information
storage and processing based on spins in solid state systems, in particular in
nanostructures, and for all phenomena related to spin diffusion in such systems.

Conceptually, further issues to be addressed are the treatment
of explicitly time-dependent Hamiltonians and spatially inhomogeneous solutions
with distributions of mean-fields which vary along the samples.
Both extensions are of greatest interest, for example in the coherent control
of spin degrees of freedom.

A third fascinating issue consists in an extension of spinDMFT to finite
temperatures. So far, we derived the approach for disordered ensembles.
But in view of related developments for spin glasses \cite{gremp98,georg00a} and the
general analogy between real and imaginary timess a dynamic mean-field 
theory for spins at finite temperatures should also exist.
Its development would enhance the applicability of spinDMFT even further.

\acknowledgments

We are thankful to K. Rezai and A. Sushkov at Boston University 
for pointing out the experimental issues to us and for intensive discussions.
We also thank D. Manolopoulos and J. Stolze for useful discussions and J.S. also for the provision of data. 
Furthermore, we acknowledge the compute time provided on the 
Linux HPC cluster at TU Dortmund University (LiDO3),
partially funded by the German Research Foundation (DFG) in Project No. 271512359.
We gratefully acknowledge financial support by the DFG and the
Russian Foundation of Basic Research in the International Collaborative
Research Center TRR 160 and by the DFG in Grant No. UH 90/13-1 (T.G. and G.S.U.) 
as well as by the Konrad Adenauer Foundation (P.B.). Furthermore, we acknowledge financial 
support by the DFG and the TU Dortmund University within the funding program Open Access Publishing.


%



\begin{appendix}
\section{Dynamic spin correlations on a Bethe lattice at infinite temperature}
\label{app:bethe}

The aim of this appendix is to derive the scaling of the correlations
as function of the coordination number. This is a tremendous task
on arbitrary lattices, even at infinite temperature.
Therefore, we consider the Bethe lattice \cite{econo79} with nearest neighbor
coupling between spins $S=\tfrac12$, i.e., a Cayley tree
of infinite depth, so that each site has the same environment and the system
is homogeneous. The coordination number is $z$ and hence the so-called
branching ratio is $z-1$. First, we consider the spin-spin correlation.
Second, we deduce further correlations.
We take the results for the Bethe lattice as representative for more general lattices
including long-range interactions leading to large effective 
coordination numbers.

\subsection{Spin-spin correlations}

We want to show that the two-time pair correlation functions
$\langle \mb{S}_{i}^{\alpha}(t_1) \mb{S}_{j}^{\beta}(t_2) \rangle$ for
$ i \neq j$ are suppressed for $z\to\infty$. 
This property is required to justify the use of the central limit theorem in
\ref{subsec:distribution} and to treat the dynamics of the local fields
as independent from the dynamics of a single spin.
Mostly, we set $j=0$ and $t_2 = 0$ without loss of generality. 
The advantage of considering the Bethe lattice 
 is that the shortest distance between any pair of sites is unique
because the lattice does not have any loops except self-retracing paths.
We consider the general Hamiltonian
\begin{align}
  \mb{H}_{\text{BL}} &= \frac1{\sqrt{z}} \sum_{\langle i,j\rangle} \sum_{\alpha\beta} J^{\alpha\beta} \mb{S}^{\alpha}_{i} \mb{S}^{\beta}_{j}
  \label{eqn:Hbethe}
\end{align}
with arbitrary couplings $J^{\alpha\beta}=J^{\beta\alpha}$ allowing for spin anisotropy.
The factor $z^{-\frac12}$ is denoted separately to explicitly keep track of the 
scaling with $z$. It must be chosen in this way to keep the energy scale
of the dynamics, $\mc J_2$ in Eq.\ \eqref{eq:momenta_def}, constant for $z\to\infty$.

The argument runs as follows. First, 
we use the Heisenberg equations of motion to set up a system of differential
equations for the temporal evolution of the correlation functions. 
Second, we postulate the scaling of the correlations. 
Third, we show that the postulated scaling is consistent
with the initial conditions and with the differential equations, i.e., 
the scaling is fulfilled by the initial conditions at $t=0$  
and by the differential equations.
As starting point, we calculate the time derivative of 
the general pair correlation function,
\bs
\begin{align}
  &\frac{\mathrm{d}}{\mathrm{d}t} \langle \mb{S}^{\rho}_{k}(t) \mb{S}^{\gamma}_{0}(0) \rangle 
	= \texttt{i} \langle \mc{L} \mb{S}^{\rho}_{k}(t) \mb{S}^{\gamma}_{0}(0) \rangle
	\\	
& \quad  = \frac{-1}{\sqrt{z}} \sum_{j,\langle j,k\rangle} \sum_{\alpha\beta\varphi} 
J^{\alpha\beta} \epsilon_{\beta \rho \varphi}
  \langle \mb{S}^{\alpha}_{j}(t) \mb{S}^{\varphi}_{k}(t) \mb{S}^{\gamma}_{0}(0) 
	\rangle,
  \label{eqn:gderiv}
\end{align}
\es
where the Levi-Civita tensor occurs due to the spin algebra.
The complexity of the expectation value is increased by the Liouville operator $\mc{L}$
which consists in the commutation with the Hamiltonian.
This leads to an an additional time-dependent spin operator in the expectation value. 

The same behavior is observed for higher derivatives: 
the application of $\mc{L}$ results in expectation values with incremented or decremented number of
spin operators by one. The decrement occurs if the additionally generated
spin operator hits an already present spin operator at the same site
and of the same component due to $(\mb S^\alpha)^2=1/4$.
The quickly growing number of products of spin operators makes the book
keeping tedious. A solution consists in considering the correlation function
of a general cluster at time $t$ with a single spin at time $0$ and site $0$
\be
	g^{\gamma}(\mb{C},t) := \langle \mb{C}(t) \mb{S}_{0}^{\gamma}(0) \rangle .
	\label{eqn:clustercorrel}
\ee
Here, $\mb{C}$ denotes an arbitrary product of spin operators at different sites 
\be
	\mb{C} := \mb{C}(c,\alpha) = \prod_{r \in c} \mb{S}^{\alpha_{r}}_{r}
\ee
where $c$ is a set of lattice sites $r \in \{0,...,N\}$ 
and $\alpha$ is a set of components $\alpha_r \in \{x,y,z\}$.
The time derivative of such cluster correlations \eqref{eqn:clustercorrel} 
reads
\be
	\frac{\mathrm{d}}{\mathrm{d}t} g^{\gamma}(\mb{C},t) = 
	\frac1{\sqrt{z}} \sum_{\mb{C}'} J(\mb{C},\mb{C}') g^{\gamma}(\mb{C}',t),
	\label{eqn:clusterheisen}
\ee
where the sum runs over all clusters $\mb{C}'$ that can be reached from $\mb{C}$ by one application of the Liouville operator. All factors that are independent of the coordination number, e.g., the couplings $J^{\alpha\beta}$ and factors $1/4$ from 
products of the same spin operators, are collected 
in $J(\mb{C},\mb{C}')$. Hence, this
generalized coupling $J(\mb{C},\mb{C}')$ does not contribute to the scaling
which we want to determine.

By \eqref{eqn:clusterheisen} we have formally defined the system of 
differential equations describing the dynamics of all  cluster correlation functions. 
The starting conditions read as
\bs
\begin{align}
	g^{\gamma}(\mb{C},0) &= 0, & \forall \ \mb{C} &\neq \mb{S}^{\gamma}_{0}, 
	\\
	g^{\gamma}(\mb{S}^{\gamma}_{0},0) &= \frac14. & &
\end{align}
\label{eqn:clusterstartcond}%
\es

\begin{figure}
	\centering
	\includegraphics[width=4.5cm]{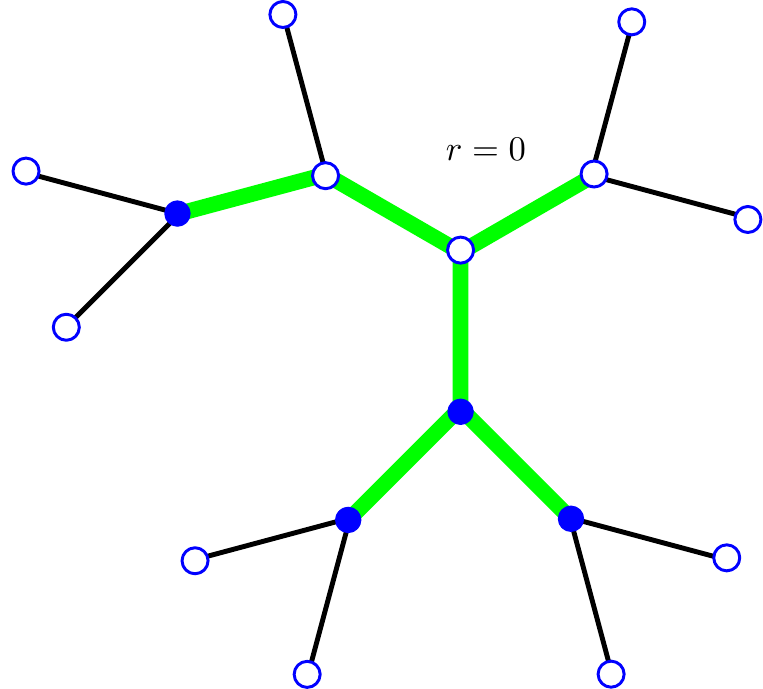}
	\caption{Cutout of a Bethe lattice with $z=3$. As an example we consider a 
	cluster of four spin operators situated at the filled dots representing occupied sites.
	The open dots represent empty sites; their number defines $\kappa_2(c)=3$.
	The covering of the cluster is defined by the green links; their number
	defines $\kappa_1=6$ so that $\kappa=9$.}
	\label{fig:BetheExample}
\end{figure}

It is challenging to keep track of the more and more complex clusters. To this
end, we define a measure of the cluster size or complexity and link it to
the scaling. Since $S=\tfrac12$, at each site $r$ of a cluster the active operator
can be the identity or one of the three spin components. All products with 
more factors can be reduced to this case. We call a site ``occupied''
if one of the three spin components is present at this site. Otherwise,
we call it ``unoccupied''. We introduce the measure $\kappa(c)$ which 
consists of two components
\be
\label{eq:kappa_def1}
\kappa(c) := \kappa_1(c) + \kappa_2(c).
\ee
The first term $\kappa_1(c)$ measures the overall size of the cluster $c$.
Consider a covering of the minimum number of bonds needed to link
all sites in $c$ and the origin $0$, see the green bold links in
Fig.\ \ref{fig:BetheExample}. We stress that this covering
is unique, i.e., there is only one such covering due to the properties of
the Bethe lattice where two sites are linked by one specific path. 
There are no loops except
self-retracing paths. The second term $\kappa_2(c)$
counts the number of empty, unoccupied sites which are touched by this covering.

We show below that $\kappa(c)$ is a lower bound for the minimum number $n(\mb C)$ 
of {applications} of the Liouville operator $\mc L$ needed to generate $\mb C(c,\alpha)$ 
from an initial cluster $\mb C_0(\{0\},\alpha_0)$ with a single spin operator 
$\mb S_0^{\alpha_0} $ at the origin. For the sake of completeness, 
we mention that in the exceptional cases
\begin{align}
	\mb{C} &= \mb{S}^{\rho}_{0}, & \rho&\neq \gamma,
\end{align}
$n = \kappa(c) + 2 = 2$ holds for reaching $\mb{S}^{\rho}_{0}$
from $\mb{S}^{\gamma}_{0}$.
From the assertion $n(\mb C)\gtrapprox \kappa(c)$ 
and the starting conditions \eqref{eqn:clusterstartcond} we deduce
\begin{align}
	\frac{\mathrm{d}^n}{\mathrm{d}t^n} g^{\gamma}(\mb{C},t) 
	\Bigr\rvert_{t=0} &= 0, & \forall\; n &< \kappa(c).
\end{align}
This motivates our central assertion 
that the cluster correlation functions are scaling
 with $z$ according to
\be
	g^{\gamma}(\mb{C},t) \propto z^{-\frac{\kappa(c)}{2}}.
	\label{eqn:gscaling}
\ee
Obviously, this agrees with the starting conditions
\be
\label{eq:start_scale}
	g^{\gamma}(\mb{S}^{\gamma}_{0},0) = \frac14 \propto z^{0},
\ee
while all other cluster correlation functions  vanish at $t=0$.

To validate the claim \eqref{eqn:gscaling} for arbitrary times, we show that it is
consistent with the equations of motion \eqref{eqn:clusterheisen}. 
A single application of the Liouville operator $\mc L$ to a cluster
$\mb{C}$ generates a sum of multiple clusters $\mb{C}'$. Since we consider 
nearest-neighbor interaction, each of these clusters $\mb{C}'$  differs
from $\mb{C}$ only at one  link $(i,j)$ where $i$ and $j$ are adjacent. Therefore, 
it is sufficient to study the possible processes on this link
\be
	\mb{C}_{(i,j)} \stackrel{\mc{L}}{\longrightarrow} \mb{C}'_{(i,j)}
\ee
where we denote the subcluster of $\mb C$ or $\mb{C}'$ on this link 
by $\mb{C}_{(i,j)}$ and $\mb{C}'_{(i,j)}$, respectively.

\begin{figure}[ht]
	\centering
	\includegraphics[width=7.5cm]{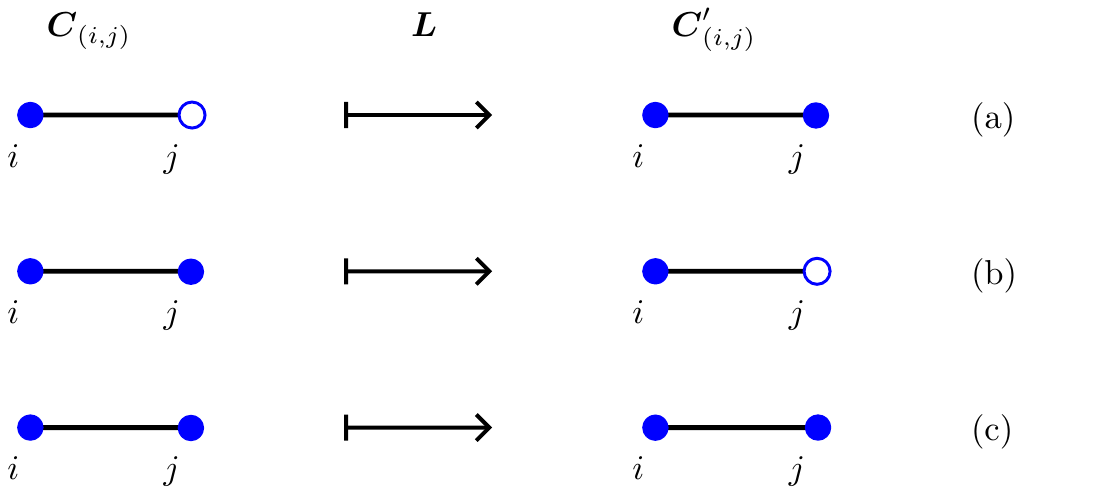}
	\caption{Possible link processes for a single application of $\mc{L}$
	and their effect on the occupation of the involved sites.}
	\label{fig:link_processes}
\end{figure}

\begin{figure}[ht]
	\centering
	\includegraphics[width=7.5cm]{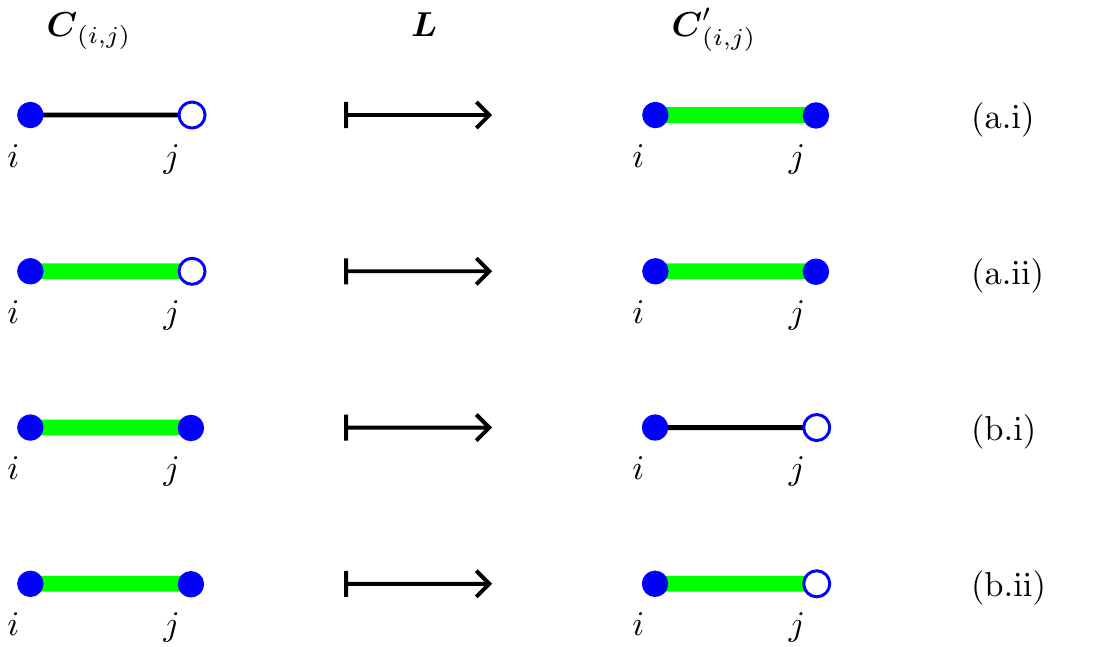}
	\caption{Relevant link processes for a single application of $\mc{L}$ 
	including the effect on the covering which is shown in green.}
	\label{fig:link_processes_covering}
\end{figure}

Fig.~\ref{fig:link_processes} shows how the effect of $\mc L$ on the link
\be
[ \mb{H}_{\text{BL},(i,j)}, \mb{C}_{(i,j)} ] = \sum \mb{C}'_{(i,j)}
\ee
 can be categorized in three different types.
The relevant commutators read as
\begin{subequations}
\begin{align}
	[ \mb{S}_{i}^{\alpha}\mb{S}_{j}^{\beta}\,,\, \mb{S}_{i}^{\rho} ]
	&= \texttt{i} \epsilon_{\alpha\rho\omega} \mb{S}_{i}^{\omega} \mb{S}_{j}^{\beta} 
	\\
	[ \mb{S}_{i}^{\alpha}\mb{S}_{j}^{\beta}\,,\, \mb{S}_{i}^{\rho} \mb{S}_{j}^{\beta} ]
		&= \frac{\texttt{i}}{2} \epsilon_{\alpha\rho\omega} \mb{S}_{i}^{\omega} 
		\\
	[ \mb{S}_{i}^{\alpha}\mb{S}_{j}^{\beta}\,,\, \mb{S}_{i}^{\rho} \mb{S}_{j}^{\delta} ]
	&= 0.
\end{align}
\label{eqn:commutationprocesses}%
\end{subequations}
Interestingly, the last commutator yields zero for $S=\tfrac12$, 
so that this process does not contribute. 
The remaining two processes (a) and (b) are analyzed further.
Since $\kappa(c)$ depends on the covering of $c$, we distinguish whether
the considered link $(i,j)$ is part of this covering or not. 
This leads to two subcases for processes (a) and (b) as shown in 
 Fig.~\ref{fig:link_processes_covering}.

As an example, we derive that the first process (a.i) 
preserves the asserted scaling.
The extension of the covering by one link leads to
\begin{align}
	\kappa_1(c') &= \kappa_1(c) + 1
\end{align}
while the number of unoccupied sites remains the same
\begin{align}
	\kappa_2(c') &= \kappa_2(c).
\end{align}
In order to assess the complete scaling one has to count how
often a process can occur. We call this its multiplicity $m(c,z)$.
The multiplicity of process (a.i) can be bounded from above
by counting at maximum $z$ neighbors of each occupied
site in $c$. There are $\kappa_1(c) - \kappa_2(c) + 1$ of such sites
so that 
\be
	m(c,z) \le z (\kappa_1(c) - \kappa_2(c) + 1)
\ee
holds. The indicators $\kappa_i$ depend on the size of the cluster, but
not on the coordination number so that the scaling resulting from the
differential equation \eqref{eqn:clusterheisen} is
\bs
\ba
\label{eq:first_z}
	g^{\gamma}(\mb{C}^{\text{(a.i)}},t) &\propto z^{-\frac12} m(c,z) 
	z^{-\frac{\kappa(c')}{2}}
	\\
	&\propto z^{-\frac{\kappa(c)}{2}}
	\\
	&\propto g^{\gamma}(\mb{C},t),
\end{align}
\es
where $g^{\gamma}(\mb{C}^{\text{(a.i)}},t)$
denotes the sum of the contributions of the clusters $\mb C'$ to the correlation
of cluster $\mb C$ via the link process (a.i). Clearly, the asserted scaling of 
$g^{\gamma}(\mb{C},t)$ is confirmed. Note that the first factor $z^{-\frac12}$
in \eqref{eq:first_z} stems from the overall scaling on the right hand side of
Eq.\ \eqref{eqn:clusterheisen}.
In conclusion, we showed that the first process is consistent with the claimed scaling
\eqref{eqn:gscaling}. 

\begin{table}[ht]
	\begin{tabular}{|p{2cm} | c | c | c | c |}
		\toprule
		{process $(x.y)$} & {$\kappa_1$} & {$\kappa_2$} & {$m(c,z) \propto$} &  
		$g^\gamma(\mb{C}^{\text{(x.y)}},t)\propto$ \\
		\colrule
		(a.i) 	&	$+1$	&	$0$	  &	$z$	& $z^{-\kappa(c)/2}$	\\
		(a.ii)	&	$0$   &	$-1$	&	$1$	& $z^{-\kappa(c)/2}$	\\
		(b.i) 	&	$-1$	&	$0$ 	&	$1$	& $z^{-\kappa(c)/2}$	\\
		(b.ii)	& $0$ 	&	$+1$	&	$1$	& $z^{-(\kappa(c)+1)/2}$\\
		\botrule
	\end{tabular}
	\caption{Induced scaling of the contributions of the link process
	to the differential equation for $\mb C$. The number $-1,0,+1$ indicate the increments
	in $\kappa_1$ and $\kappa_2$. The entries of the column of the multiplicity 
	 provides the scaling factors in $m(c,z)$. The last column
	shows the resulting scaling of the contribution of $\mb{C}'$
	to $\mb{C}$.}
	\label{tab:link_process_scaling}
\end{table}

We do not repeat the line of argument for the processes (a.ii), (b.i), and (b.ii).
Their effects on the scaling are summarized in Tab.~\ref{tab:link_process_scaling}. 
While the processes (a.ii) and (b.i) yield the same scaling as (a.i), 
the last process (b.ii) is even suppressed by an additional factor 
$z^{-\frac12}$. Hence, we have derived that
none of the processes violates the asserted scaling. 
Since this scaling holds initially, see Eq.~\eqref{eq:start_scale}, we deduce 
by continuous induction via the differential equations \eqref{eqn:clusterheisen}
that the scaling \eqref{eqn:gscaling} holds at all times on the Bethe lattice at 
infinite temperature. As mentioned in the beginning,
we assume that this behavior is generic, i.e., that it applies to general
lattices and clusters. This justifies the application of spinDMFT 
for dense systems by which we mean systems with large effective
coordination numbers.

A direct corollary applies to the two-time pair autocorrelations 
$\langle \mb{S}_{i}^{\alpha}(t_1) \mb{S}_{j}^{\beta}(t_2) \rangle$.
In this particular case, the cluster reads as
\be
	\mb{C} = \mb{S}_{i}^{\alpha}
\ee
so that only the site $i$ belongs to $c$ and
\be
	\kappa_1(c) = \kappa_2(c) = \norm{i-j},
\ee
where one must keep in mind that the site $j$ needs
to be {connected to} a link as well without being occupied.
This implies 
\be
	\kappa(c) = \kappa_1(c) + \kappa_2(c) = 2\norm{i-j}
\ee
where $\norm{i-j}$ is the taxicab distance between $i$ and $j$, i.e., 
the number of links required to reach $i$ from $j$. 
Then the general scaling \eqref{eqn:gscaling} reads as
\be
\label{eq:general_scaling}
	\langle \mb{S}_{i}^{\alpha}(t_1) \mb{S}_{j}^{\beta}(t_2) \rangle 
	\propto z^{-\norm{i-j}},
\ee
which {entails} that the two-time pair correlations ($i\ne j$) are suppressed at least by
$z^{-1}$ relative to the autocorrelations.
Hence, any pair of spins is uncorrelated in the limit $z\to \infty$. 
We used this insight to justify the application of the central limit theorem 
in Sect.~\ref{subsec:distribution}.

\subsection{Correlations of the local-environment fields}

Here we draw further conclusions from the  scaling derived in the preceding
section. We recall definition \eqref{eq:Vdef} of the local-environment fields which 
take the form 
\begin{align}
	\mb{V}^{\alpha}_{i} &= \frac1{\sqrt{z}}\sum_{k,\langle k,i\rangle} 
	\sum_{\gamma} J^{\alpha \gamma} \mb{S}^{\gamma}_{k}
\end{align}
on the Bethe lattice. Their autocorrelation reads
\begin{widetext}
\begin{subequations}
\begin{align}
	\langle \mb{V}^{\alpha}_{i}(t_1) \mb{V}^{\beta}_{i}(t_2) \rangle &= 
	\frac1{z} \sum_{\gamma\rho} \left( J^{\alpha\gamma} J^{\beta\rho}\right)^2 
	\sum_{\substack{k,\langle k,i\rangle \\ l,\langle l,i\rangle}}
	\langle \mb{S}^{\gamma}_{k}(t_1) \mb{S}^{\rho}_{l}(t_2) \rangle 
	\\
	&=  \frac1{z} \sum_{\gamma\rho} \left( J^{\alpha\gamma} J^{\beta\rho}\right)^2 
	\sum_{k,\langle k,i\rangle} 
	\Bigl( \langle \mb{S}^{\gamma}_{k}(t_1) \mb{S}^{\rho}_{k}(t_2) \rangle
	+ \sum_{\substack{l,\langle l,i\rangle 	\\ l\neq k}} 
	\langle \mb{S}^{\gamma}_{k}(t_1) \mb{S}^{\rho}_{l}(t_2) \rangle \Bigr) 
	\\
	&=  \frac1{z} \sum_{\gamma\rho} \left( J^{\alpha\gamma} J^{\beta\rho}\right)^2 
	\sum_{k,\langle k,i\rangle} \Bigl( \langle \mb{S}^{\gamma}_{k}(t_1) 
	\mb{S}^{\rho}_{k}(t_2) \rangle 	+ \mc{O}\left(z^{-1}\right) \Bigr),
	\label{eqn:Vautocorrsimpl}
\end{align}
\end{subequations}
\end{widetext}
where in the last line we neglected the second term in the brackets because
it is suppressed by the factor $1/z^\norm{k-l} = z^{-1}$ relative to the first term. 
We emphasize that the total autocorrelation
is not suppressed since the summation and the factor $z^{-1}$ compensate each other.
Equation \eqref{eqn:Vautocorrsimpl} is vital 
to simplify the self-consistency problem in \ref{sss:selfconscond}.

\begin{figure}[ht]
	\centering
	\includegraphics[width=0.47\textwidth]{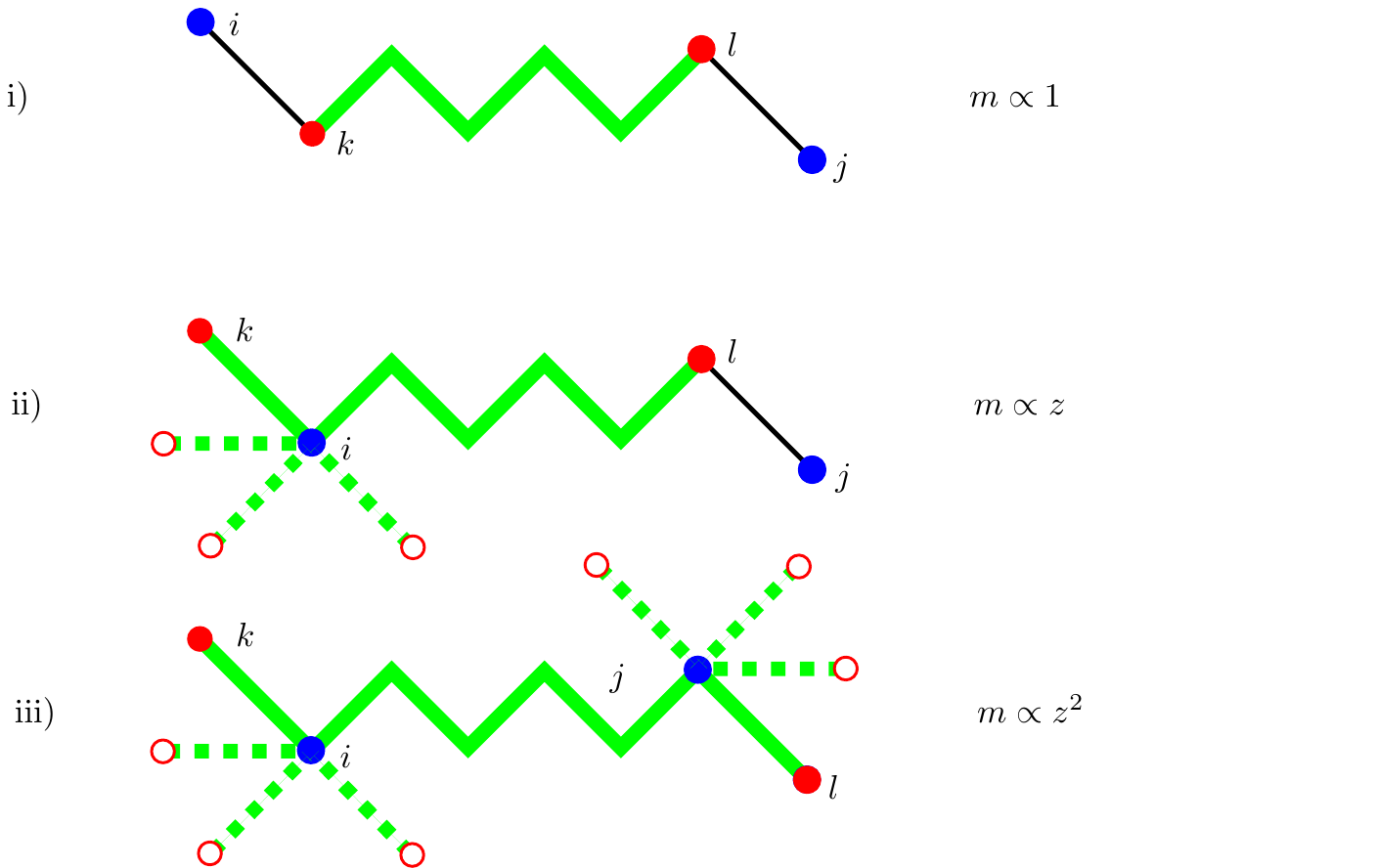}
	\caption{Sketch to visualize the three cases occurring in 
	the double sum in equation \eqref{eqn:VVcorrelation}. The distance between the red dots $k$ 
	and $l$ defines the scaling of the expectation value in the double sum.
	The green zigzag lines indicates that the corresponding adjacent sites are connected 
	via arbitrarily many bonds. By the green dashed lines with open red dots we indicate that there 
	are multiple options for the sites $k$ or $l$.}
	\label{fig:VVcorrelation_multiplicity}
\end{figure}

Similarly, we analyze possible correlations of the local fields at different 
sites. Inserting the definitions into the pair correlation functions one obtains
\begin{align}
		\langle \mb{V}^{\alpha}_{i}(t_1) \mb{V}^{\beta}_{j}(t_2) \rangle = \frac1{z}
		\sum_{\gamma\rho}\! \left( J^{\alpha\gamma} J^{\beta\rho}\right)^2 \!\!\!
		\sum_{\substack{k,\langle k,i \rangle \\ l,\langle l,j\rangle}} \!
		\langle \mb{S}^{\gamma}_{k}(t_1) \mb{S}^{\rho}_{l}(t_2) \rangle
\label{eqn:VVcorrelation}
\end{align}
for $i \neq j$. Inspecting the Bethe lattice, one sees that
the above double sum consists of three types of terms
with different multiplicity $m$:
\begin{enumerate}[label=\roman*)]
	\item $\norm{k-l} = \norm{i-j} - 2$, \quad $m\propto 1$,
	\item $\norm{k-l} = \norm{i-j}$, \quad $m \propto z$,
	\item $\norm{k-l} = \norm{i-j} + 2$,\quad  $m\propto z^2$.
\end{enumerate}
Fig.~\ref{fig:VVcorrelation_multiplicity} visualizes these cases.
The first case occurs if $k$ and $l$ are neighbors of $i$ and $j$ {and both are part of the 
direct path connecting $i$ with $j$.} The second case occurs if one
index $k$ or $l$ stands for a neighbor {that is not part of this path while the other one 
lies on the path linking $i$ and $j$}. The third case, finally, occurs if both indices $k$ or $l$ stand for neighbors
{that are not part of the path connecting $i$ with $j$}. 
The scalings stem from the fact that there are always $z-1 \propto z$ ways
to choose a neighbor {that is not on the path} while there is only one unique
 neighbor {on the path}. In the special case $\norm{i-j} = 1$ the first case
 modifies to $\norm{k-l} = \norm{i-j} = 1$ with $m=1$.

Using the above multiplicities in combination
with the scaling of the spin-spin correlations \eqref{eq:general_scaling}
we obtain that the pair correlations scale like
\begin{align}
	\langle \mb{V}^{\alpha}_{i}(t_1) \mb{V}^{\beta}_{j}(t_2) \rangle &\propto
	\begin{cases}
		z^{-1} & \text{if $\norm{i-j}=1$,} \\
		z^{-\norm{i-j}+1} & \text{if $\norm{i-j}>1$,} \\
	\end{cases}
\end{align}
so that they are suppressed by at least the factor $z^{-1}$. Hence,
no correlations between the local-environment fields need
to be accounted for. By self-consistency, this extends to the 
second moments of the local mean-fields.

\begin{figure}[ht]
	\centering
	\includegraphics[width=0.43\textwidth]{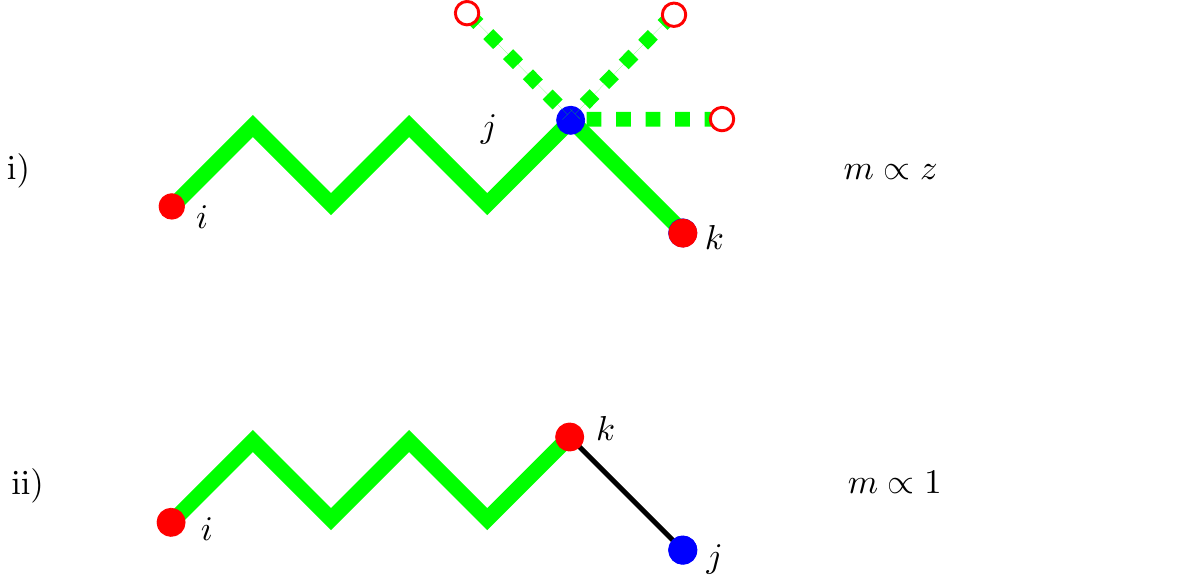}
	\caption{Sketch to visualize the two cases occurring in 
	the site sum in equation \eqref{eqn:SVcorrelation}. The distance between the red dots $k$ 
	and $i$ defines the scaling of the expectation value on the right hand side. The green zigzag lines 
	indicate that the corresponding adjacent sites are connected 
	via arbitrarily many bonds. By the green dashed lines we indicate that there 
	are multiple options for the site $k$.}
	\label{fig:SVcorrelation_multiplicity}
\end{figure}

In a similar fashion, we show that the spin dynamics of $\vec{\mb{S}}_i$
is uncorrelated to its corresponding local-environment field $\vec{\mb{V}}_i$:
\bs
\begin{align}
	\langle \mb{S}^{\alpha}_{i}(t_1) \mb{V}^{\beta}_{i}(t_2) \rangle &= 
	\frac1{\sqrt{z}} \sum_{\gamma} J^{\beta\gamma} 
	\overbrace{\sum_{k, \langle k,i \rangle}}^{\propto z} 
	\underbrace{\langle \mb{S}^{\alpha}_{i}(t_1) \mb{S}^{\gamma}_{k}(t_2) \rangle}_{\propto z^{-1}} \\
	&\propto z^{-\frac12},
\end{align}
\es
as well as to any other local-environment field $\vec{\mb{V}}_j$ ($j\neq i$):
\bs
\begin{align}
	\langle \mb{S}^{\alpha}_{i}(t_1) \mb{V}^{\beta}_{j}(t_2) \rangle &= 
	\frac1{\sqrt{z}} \sum_{\gamma} J^{\beta\gamma} 
	\sum_{k, \langle k,j \rangle}
	\langle \mb{S}^{\alpha}_{i}(t_1) \mb{S}^{\gamma}_{k}(t_2) \rangle \\
	&\propto z^{-\norm{i-j}+\frac12},
\end{align}
\label{eqn:SVcorrelation}%
\es
which both tend to 0 for $z\to \infty$. For the last conclusion, one has to distinguish two cases again, see Fig.~\ref{fig:SVcorrelation_multiplicity}. 
First, site $i$ can be nearer to $j$ than to $k$. Second, site $i$ can be nearer to $k$ than to $j$. The latter case 
is dominant with a multiplicity of $m\propto 1$ and a scaling of $\propto z^{-\norm{j-1}+1}$. Together 
with the prefactor $1/\sqrt{z}$ this yields the provided scaling.

\section{Error analyses}
\label{app:errors}

In this appendix, we discuss numerical errors and issues which arise in 
evaluating the mean-field moments.
An assessment of the errors resulting from statistics, finite discretization, and convergence by iteration is provided. Subsequently,
we explain how the numerical effort can be reduced by
exploiting time translation invariance and we discuss 
the definiteness of the covariance matrices.

\subsection{Statistical Error}
\label{app:staterr}

By the self-consistent equations derived in Sect.~\ref{sss:selfconscond} 
 the  moments of the mean-fields are linked to spin
expectation values which are calculated by path integrals 
averaged over the distribution of the mean-field time series $\vec{\mc{V}}$.
Numerically, we estimate the path integrals using a Monte-Carlo method. The autocorrelations
\be
	g^{\alpha \beta}_{\vec{\mc{V}}}(t_1,t_2) := 
	\langle \mb{S}^{\alpha}(t_1) \mb{S}^{\beta}(t_2) \rangle\esc_{\vec{\mc{V}}}
\ee
are computed for $M$ time series $\vec{\mc{V}}$ and averaged 
\be
	g_M^{\alpha \beta}(t_1,t_2) := \frac1M \sum_{\vec{\mc{V}}} 
	g^{\alpha \beta}_{\vec{\mc{V}}}(t_1,t_2)
	\label{eqn:sampleav}
\ee
which converges to $g^{\alpha \beta}(t_1,t_2)$ for $M\to\infty$.
Since the time series are drawn independent of each other 
 the variance of $g_M^{\alpha \beta}(t_1,t_2)$  is
given by the variance of a single time series divided by $M$
\be
	\sigma^2 \left( g^{\alpha \beta}_M(t_1,t_2) \right) = 
	\frac1{M} \sigma^2 \left( g_{\vec{\mc{V}}}^{\alpha \beta}(t_1,t_2) \right),
\ee
where $\sigma^2(\dots)$ denotes the variance of the quantity in the bracket. 

\begin{figure}
	\centering
	\includegraphics[width=0.5\textwidth]{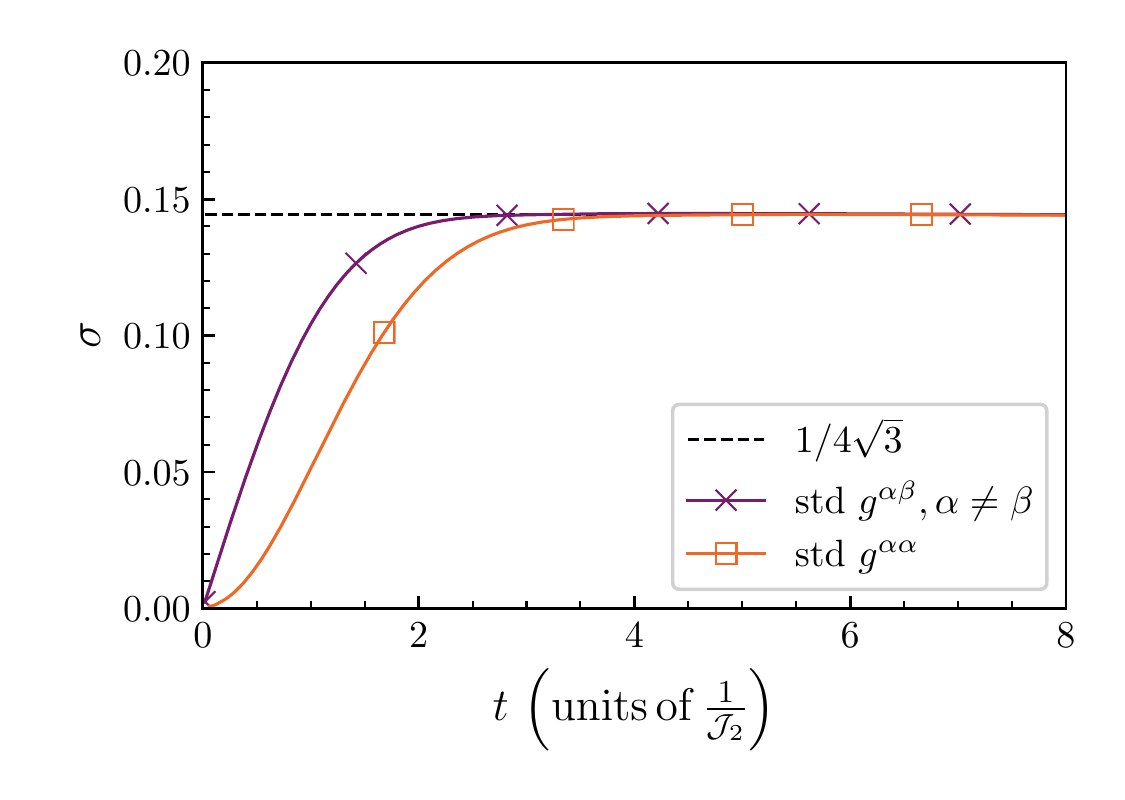}
	\caption{Numerical result for the standard deviations
	$\sigma\left(g_{\vec{\mc{V}}}^{\alpha\beta}(t,0)\right)$ as function of $t$ for the isotropic 
	Heisenberg model without any magnetic field. 
	For $t=0$, the autocorrelations are fixed  and 
	hence their standard deviation vanishes.
	For large $t$ it converges apparently to $1/4\sqrt{3}$, indicated by the
	dashed line.}
	\label{fig:std_isotropic}
\end{figure}

\begin{figure}
	\centering
	\includegraphics[width=0.5\textwidth]{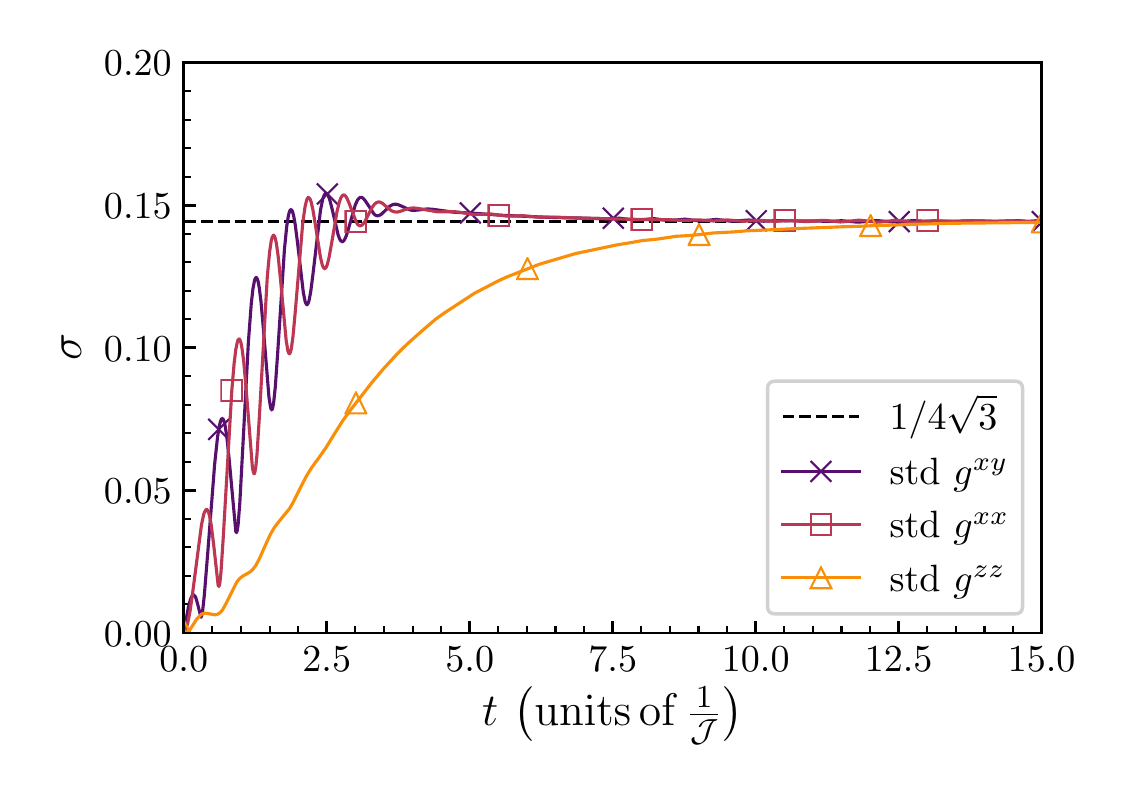}
	\caption{Same as Fig.\ \ref{fig:std_isotropic} for the dipole 
	model in the laboratory frame with $\vartheta=0$ at $\widetilde{B} = 5.0$.}
	\label{fig:std_dipolelab}
\end{figure}

\begin{figure}
	\centering
	\includegraphics[width=0.5\textwidth]{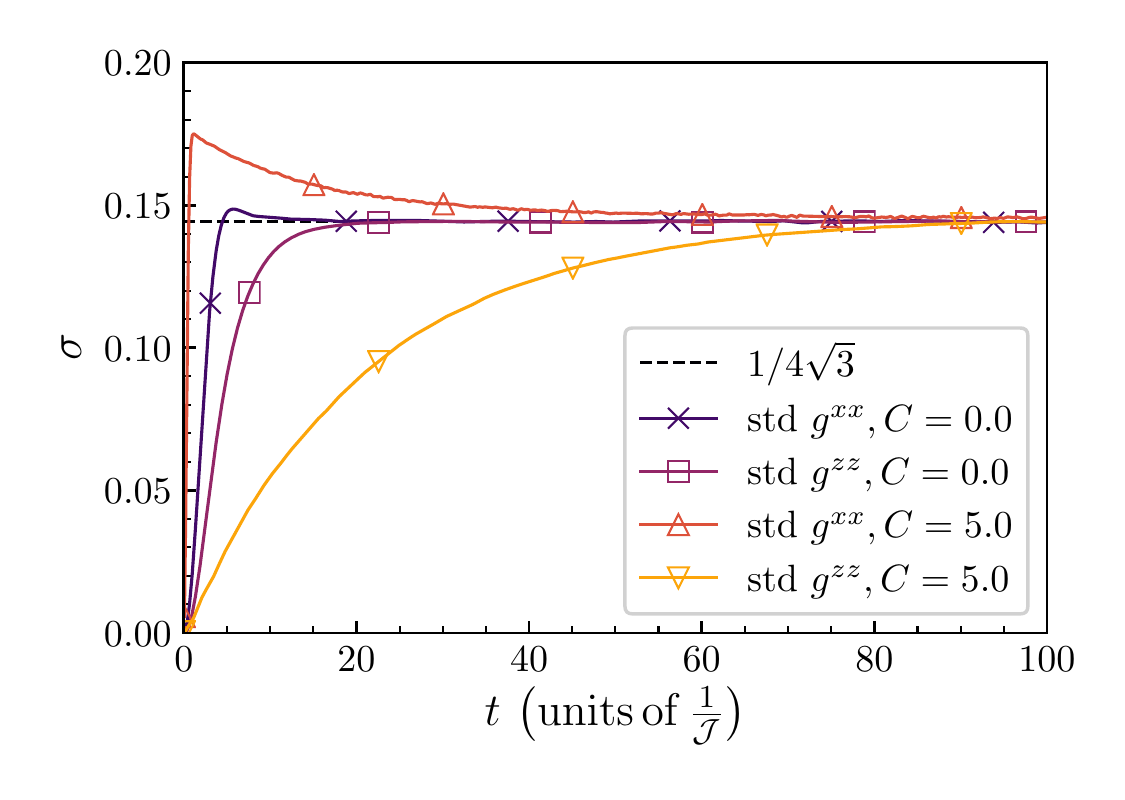}
	\caption{Same as Fig.\ \ref{fig:std_isotropic}
	for the dipole model in the rotating frame in RWA and various noise strengths.}
	\label{fig:std_dipolerot}
\end{figure}

The standard deviation $\sigma$ of a single time series 
depends on many parameters and cannot be 
calculated analytically in a simple way. But it is {clear that}
its value is bounded by the maximum value of the autocorrelations, i.e., 
by $\tfrac14$. Figs.~\ref{fig:std_isotropic}, \ref{fig:std_dipolelab}, and \ref{fig:std_dipolerot} 
show generic time dependencies
of $\sigma$ computed for different physical situations.
Interestingly, we find that all of them converge to the value $1/(4\sqrt{3})$ for 
$t\to\infty$. The rapidity of this convergence depends on the actual decay of the 
corresponding autocorrelation. This behavior can be understood by the 
following argument. The autocorrelations vanish for large $t$. Hence,
the variance equals the quadratic mean
\be
	\lim_{t \to \infty} \sigma^2 \left( g^{\alpha \beta}(t,0) \right) 
	= \lim_{t \to \infty} \overline{  \bigl((g^{\alpha \beta}(t,0)\bigr)^2 }.
\ee
In a next step, we consider the vector-valued signal
\bs
\begin{align}
	\vec{g}^{\,\beta}_{\vec{\mc{V}}}(t,0) &:= 
	\langle \mb{U}^{\dagg}(t,0) \vec{\mb{S}}(0) \mb{U}(t,0) \mb{S}^{\beta}(0) \rangle\esc_{\vec{\mc{V}}} 
	\\ 
	&= \left\langle \left(\dul{R}_{\vec{\mc{V}}}(t,0) \vec{\mb{S}}(0)\right) 
	\mb{S}^{\beta}(0) \right\rangle\pesc 
	\\
	&= \frac14 \dul{R}_{\vec{\mc{V}}}(t,0) \vec{\mathrm{e}}_{\beta},
\end{align}
\es
where $\dul{R}_{\vec{\mc{V}}}(t,0)$ denotes the orthogonal rotation matrix 
which describes the rotation of the initial spin vector due
to its temporal evolution subjected to the fluctuating time series $\vec V(t)$.
Obviously, the square of $\vec{g}^{\,\beta}_{\vec{\mc{V}}}(t,0)$ is constant
\be
	\left(\vec{g}^{\beta}(t,0) \right)^2 = \frac1{16}
	\label{eqn:vectorsignalsqav}
\ee
before and thus also after averaging. 

Next, it is plausible and in accord with all previous results
that the decoherence is sufficiently strong to have the 
spin vector $\dul{R}_{\vec{\mc{V}}}(t,0) \vec{\mb{S}}(0)$  lose all
 information about its initial direction. If it points in $z$ direction at $t=0$
it will point into any direction of the unit sphere after sufficiently long time.
Therefore, each component $\alpha$ of the rotated spin vector
 has the same variance for $t \to \infty$ and contributes equally 
 to the squared vector in \eqref{eqn:vectorsignalsqav}. This
allows us to conclude
\begin{align}
	\lim_{t \to \infty} \sigma \left( g^{\alpha \beta}(t,0) \right) &= 
	\lim_{t \to \infty} \sqrt{ \overline{\left(\vec{g}^{\beta}(t,0) \right)^2}/3} 
	&= \frac1{4\sqrt{3}}
\end{align}
in perfect agreement with the numerical findings. 
This enables us to derive the reliable estimate 
\be
	\sigma \left( g_M^{\alpha \beta}(t_1,t_2) \right) 
	\approx \frac1{4\sqrt{3M}}.
\ee
for the statistical error from averaging over $M$ times series.
It becomes even exact for $t\to\infty$.


\subsection{Discretization Error}
\label{app:discerr}

The goal of this appendix is to 
estimate the error resulting from the discretization of
time, i.e., the error resulting from working with a finite time step $\delta t>0$
instead of taking $\delta t\to 0$. Certainly, this error will
depend on details of the model. We discuss it for both the 
isotropic Heisenberg model and the dipole model without and with {RWA}. 

We consider the discretization error of the diagonal autocorrelations 
$\alpha=\beta$ up to some maximum time
$t_\text{max}$ which we determine such that 
\begin{align}
	\lvert g^{\alpha\alpha}(t_{\text{max}})\rvert \approx \frac1{100}
\end{align}
holds, i.e., the autocorrelation has decreased to $1/25$ of its initial value.
For simplicity, we focus on the autocorrelation that decays slowest in each physical scenario, since it is
most susceptible to the discretization error which
accumulates in the course of its temporal evolution. Mostly this is the longitudinal autocorrelation. 

To be specific, we compute $g^{\alpha\alpha}(t)$ on the time interval 
$t \in [0,t_{\text{max}}]$ for various step widths
\begin{align}
	\delta t (\nu) &= \delta t(0) 2^{-\nu}, & \nu &\in [0, 1, 2, ..., \nu_{\text{max}}].
\end{align}
The results for $g^{\alpha\alpha}(t)$ for $\nu < \nu_{\text{max}}$ are compared to the 
`best' solution for $\nu_{\text{max}}$, i.e., the solution from 
the finest discretization which is used as reference.
We define the discretization error 
\begin{align}
	\overline{\Delta q}(\nu) &= \sum_{l=0}^{L(\nu)} \frac{\lvert g^{\alpha\alpha}_{\nu_{\text{max}}}(l\delta t(\nu)) - g^{\alpha\alpha}_{\nu}(l\delta t(\nu)) \rvert}{L(\nu)+1}
\end{align}
where $L(\nu) = t_{\text{max}}/\delta t(\nu)$ is the number of time steps. 
We stress that the input data for $\overline{\Delta q}(\nu)$ 
are not exact, but subjected also to the statistical error discussed
in the previous section.

\begin{figure}
	\centering
	\includegraphics[width=0.5\textwidth]{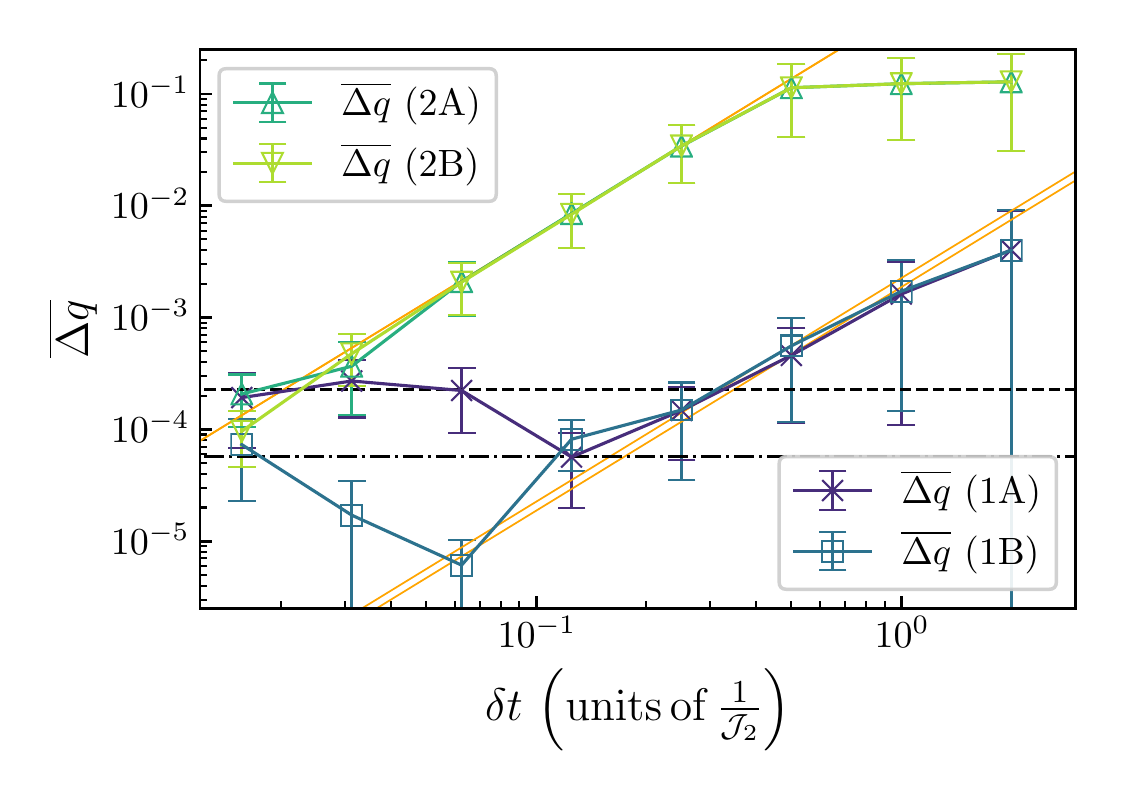}
	\caption{Numerical analysis of the time discretization error for the isotropic model. 
	Data are shown for zero magnetic field (1) and for finite magnetic field $\gamma_{\text{s}}B=5.0\mathcal{J}_2$ (2)
	and, furthermore, for $M=\num{4e5}$ (A) and for $M=\num{6.4e6}$ (B) time series, respectively. The dashed line (A) and the 
	dashed-dotted line (B) represent the statistical errors $\sigma$ for the corresponding $M$.
	The thin orange lines follow $\delta t^2$ and are given for comparison.}
	\label{fig:disc_error_iso}
\end{figure}

\begin{figure}
	\centering
	\includegraphics[width=0.5\textwidth]{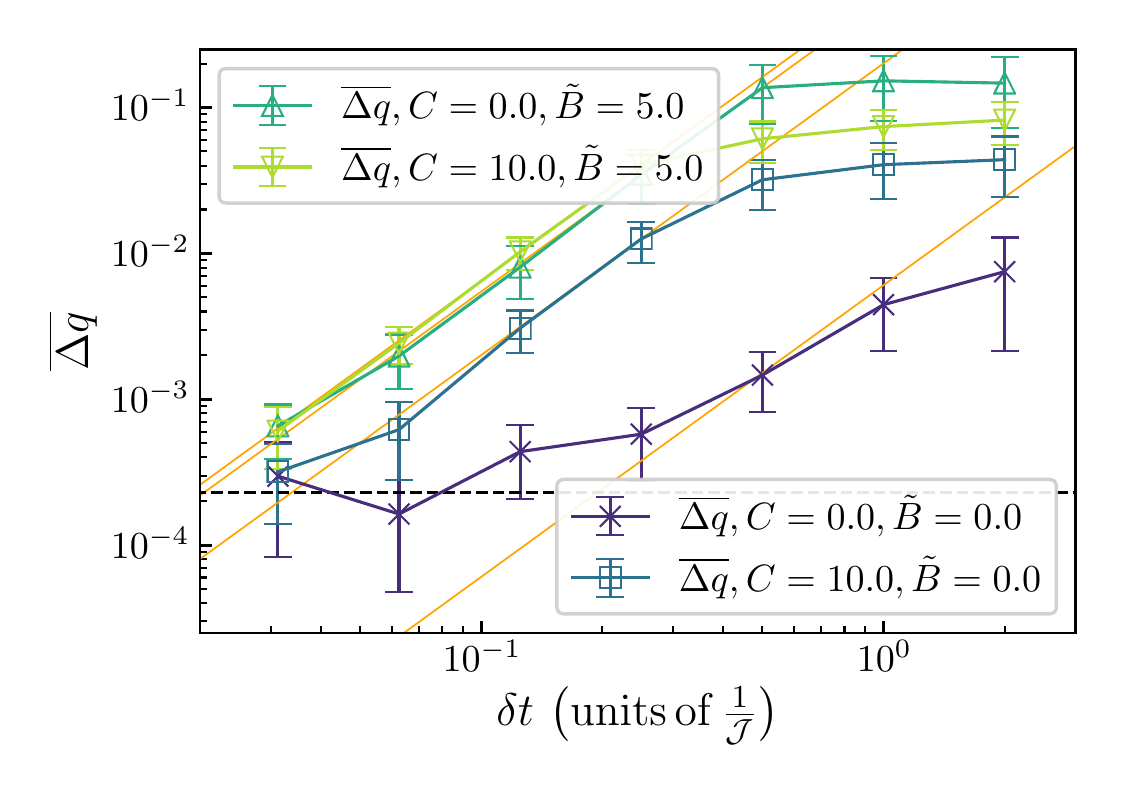}
	\caption{Numerical analysis of the time discretization error for the lab-frame dipole model with $\vartheta=0$ for various magnetic fields $\tilde{B}$ 
	and noise widths $C$. The dashed line corresponds to the statistical error $\sigma$ for the sample size $M=\num{4e5}$.
	The thin orange lines follow $\delta t^2$ and are given for comparison.}
	\label{fig:disc_error_diplab}
\end{figure}

\begin{figure}
	\centering
	\includegraphics[width=0.5\textwidth]{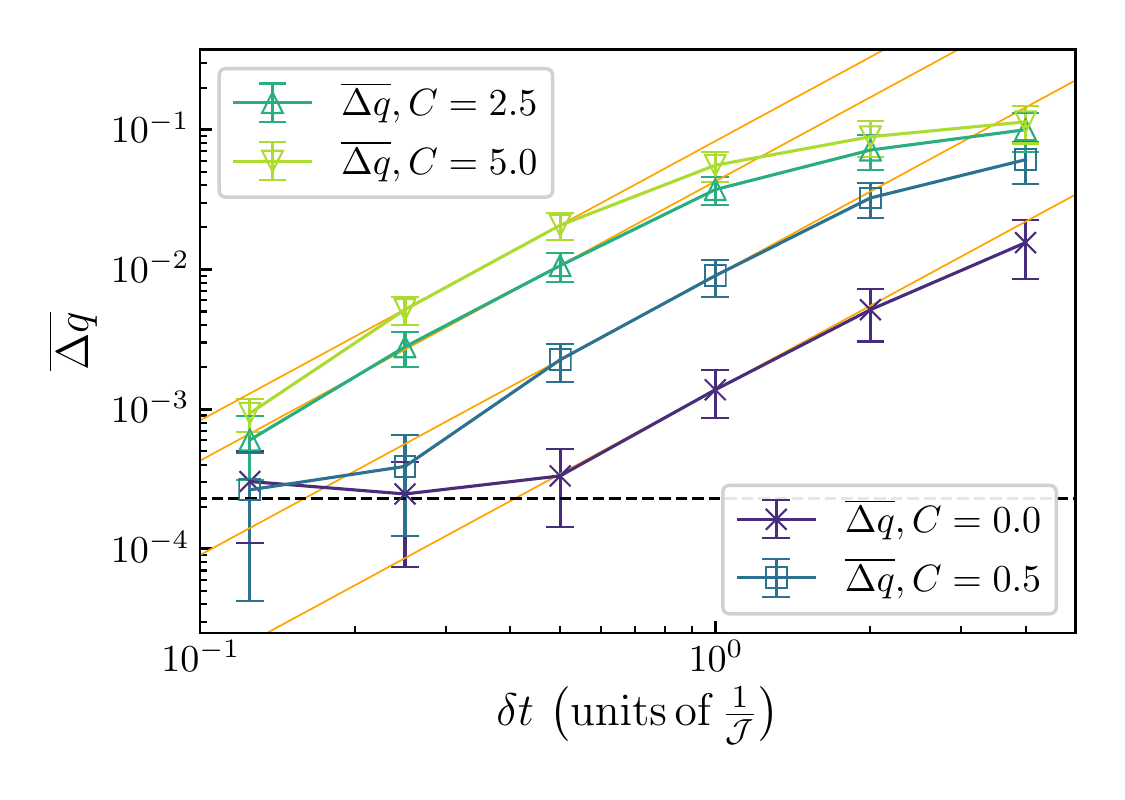}
	\caption{Numerical analysis of the time discretization error for the RWA dipole model for various noise widths $C$. 
	The dashed line corresponds to the statistical error $\sigma$ for the sample size $M=\num{4e5}$.
	The thin orange lines follow $\delta t^2$ and are given for comparison.}
	\label{fig:disc_error_diprot}
\end{figure}

Figs.~\ref{fig:disc_error_iso}, \ref{fig:disc_error_diplab}, and \ref{fig:disc_error_diprot} show the results for the three different
models investigated in the main text. The qualitative behavior is 
very similar. For small values of $\delta t$  the deviation $\overline{\Delta q}$
is clearly dominated by the statistical error. Hence the curves level off
displaying roughly plateaus with some fluctuations.
For large values of $\delta t$ the curves also level off displaying
a plateau. This stems from errors so large that the deviations 
are of order $2/4$ because the fluctuating diagonal autocorrelations $g^{\alpha\alpha}$ are
bounded by $1/4$.

Thus, the relevant regime is at intermediate time steps between these two plateaus. 
Here the effect of the discretization 
can be discerned. The double logarithmic plots are consistent
with the conclusion $\overline{\Delta q} \propto \delta t^2$
as can be read off by comparing to the straight lines resulting
from the quadratic power law.  This can be easily understood
by the approximation we have to use for the unitary time 
evolution operators which propagate the system from $t$ to $t+\delta t$.
The employed second-order CFET and the trapezoidal rule entail
a discretization error scaling like $\delta t^3$. But since this
error accumulates over time one has to multiply this scaling
by the number of step from $t=0$ to $t=t_\text{max}$ so that 
we obtain
\be
	\overline{\Delta q} \propto L \delta t^3 \propto \delta t^2.
\ee
This explains the observed quadratic scaling of the discretization 
error with the time step size $\delta t$.

Aside from the above discussed scaling further conclusions on the
influence of the discretization can be drawn. Inspecting Figs. \ref{fig:disc_error_iso} and \ref{fig:disc_error_diplab}
shows that a finite magnetic field
increases $\overline{\Delta q}$. The reason is that the
Larmor precessions need to be resolved. If the step size $\delta t$
is too long, approaching the Larmor period, sizable discretization
errors occur. As a rule of thumb, $\delta t$ 
should be at maximum a tenth of the Larmor precession period
\be
	T = \frac{2 \pi}{\gamma_{\text{s}} B}.
\ee

Since the time discretization error accumulates over time longer lasting correlations
imply larger discretization errors. This can be seen for instance
 in Fig. \ref{fig:disc_error_diprot} where the dipole model is studied in RWA.
The error $\overline{\Delta q}$ increases with the noise strength 
because the decay of the longitudinal autocorrelation 
is slowed down upon increasing $C$.
Finally, we point out that a small discretization error is
desirable. But in view of the efficiency of the total algorithm
it does not pay to reduce the discretization error below
the statistical error. Hence, the parameters should be set
such that $\overline{\Delta q} \approx \sigma(g)$ holds.

\subsection{Termination condition for the iteration}
\label{app:abortcon}

We determine the solution of the self-consistency conditions
iteratively. If the algorithm is stable, the autocorrelation functions 
$g^{\alpha\beta}_{(i)}(t)$ of the iteration $i$
converge to the exact results for $i\to\infty$. 
In practice, it is necessary to define a
termination condition to  decide when the iterations
can be stopped. For this we use
\be
	\Delta I^{\alpha\beta} \left(i\right) = \frac1{L+1} \sum_{l=0}^{L} 
	\lvert g^{\alpha\beta}_{(i)}(t_l) - g^{\alpha\beta}_{(i-1)}(t_l) \rvert.
\ee
This quantity measures the difference between the results of iteration $j+1$ and $j$, where $j \geq 1$ . The execution of the code is stopped
 if $\Delta I^{\alpha\beta} \left(i\right) $ falls below a certain threshold.
Since the iteration error itself is limited by the statistical accuracy
this threshold cannot be chosen smaller than the statistical error. 
It needs to be set above the statistical standard deviation to
achieve reliable termination. In our numerics, it turned out that
\be
	\Delta I_{\text{threshold}} = 2 \sigma \left( g_M^{\alpha \beta}(t_1,t_2) \right) 
	= \frac1{ 2\sqrt{3M}}
	\label{eqn:threshold}
\ee
is a reasonable choice. It avoids unnecessary iterations while it
is strict enough to yield sufficient convergence.

\begin{figure}
	\centering
	\includegraphics[width=\columnwidth]{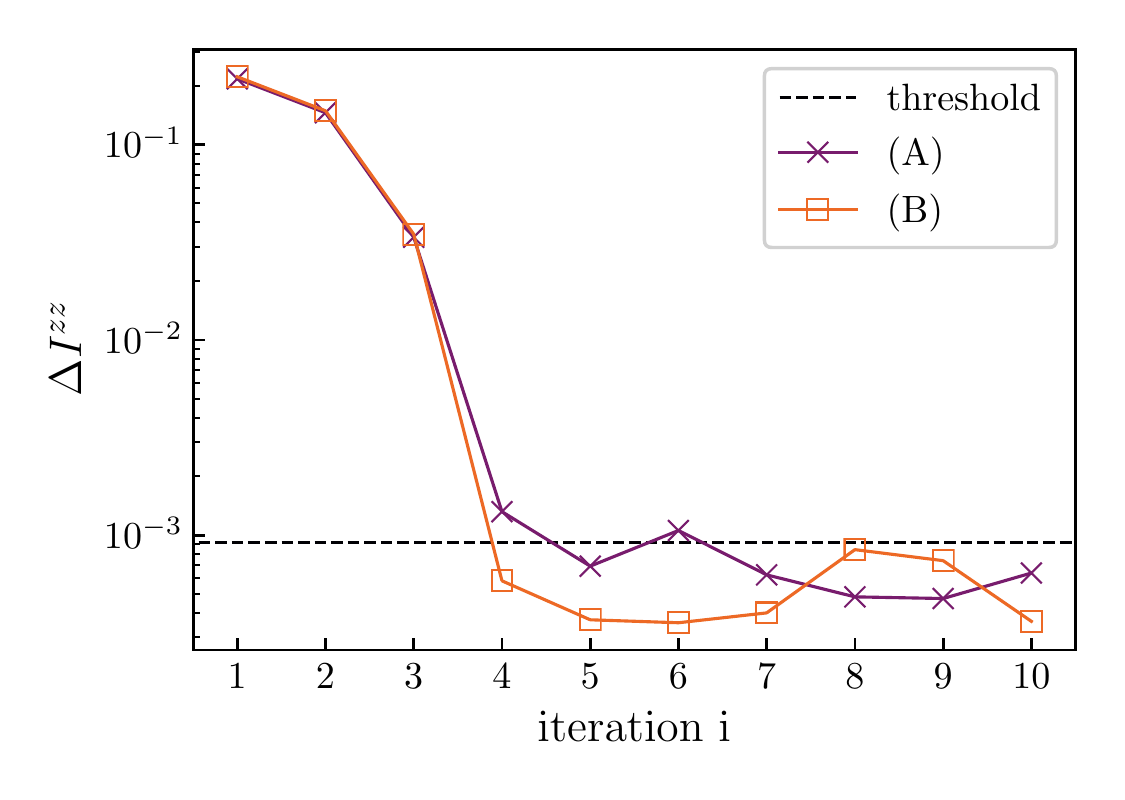}
	\caption{Iteration error $4\Delta I^{zz}$ as function of the iteration number $i$. 
	Case (A) starts from an exponential initial diagonal autocorrelation and case (B) 
	from a 	Gaussian initial diagonal autocorrelation.
	The error threshold 
	\eqref{eqn:threshold} is displayed as horizontal dashed line for $M=\num{1e5}$ time series.
	The termination condition is fulfilled at $i=5$ for case (A) and 
	at $i=4$ for case (B).}
	\label{fig:iter_error}
\end{figure}

Figure~\ref{fig:iter_error} depicts the iteration error of $g^{zz}$ 
for the isotropic {Heisenberg model} without magnetic field.
The iterations started from two different initial diagonal autocorrelations
 $g^{\alpha\alpha}$ while the cross autocorrelations  $g^{\alpha\beta}$
with $\alpha\ne \beta$ are set to zero.
The iteration error decreases very fast in the beginning. At about the fourth 
iteration it reaches the magnitude of the threshold. 
Beyond the fourth iteration, statistical fluctuations stemming from
the averaging over $M$ time series occur and dominate the iteration 
error $\Delta I^{xx}$. For optimum computational efficiency, the code
should  terminate  before the statistical fluctuations take over.

We emphasize that the ``converged'' results obtained in this way
are independent of the initially chosen autocorrelations. The deviation
between the iterated results from the exponential and the Gaussian
initial autocorrelation is of the same magnitude as the error threshold
\eqref{eqn:threshold}. For this reason, we consider the chosen iterative
algorithm  robust enough to determine {physically} meaningful
solutions.

\subsection{Time-Translation Invariance}
\label{app:timetrans}

For both the isotropic and  the dipolar spin systems we considered a 
time-independent Hamiltonian so that the systems are 
invariant under time translation 
\be
	\mb{U}(t_2,t_1) = \mathrm{e}^{-\texttt{i}\mb{H}\left(t_2 - t_1\right)} 
	= \mb{U}(t_2-t_1,0)
\ee
and so are all two-time spin autocorrelations 
\be
	\langle \mb{S}^{\alpha}(t_1) \mb{S}^{\beta}(t_2) \rangle 
	= \langle \mb{S}^{\alpha}(t_1-t_2) \mb{S}^{\beta}(0) \rangle.
\ee

Applying spinDMFT, the situation is slightly more complicated. 
The autocorrelations with respect to a single time series $\vec{\mc{V}}_i$
are not time-translation invariant because the mean-field is dynamic, i.e.,
it depends on time.
But the {physically} meaningful expectation values are obtained
 by averaging over the distribution of possible mean-fields. 
In the path integral the mean-field $\vec{\mc{V}}_{i}$ can be shifted
in time so that the time evolution operators
and the distribution functional $p[\vec{\mc{V}}_i]$ are shifted. 
Time-translation invariance is ensured if and only if 
the distribution remains unchanged by this
shift. Specifically, this is equivalent to the condition 
\be
	\overline{V^{\alpha}(t_1)V^{\beta}(t_2)} =
	\overline{V^{\alpha}(t_1-t_2)V^{\beta}(0)},
\ee
for the second moments of the mean-fields, i.e., 
they are time-translation invariant themselves.
As stated in the main text, the {time invariance} of the
moments and hence of the distribution implies time 
translation invariance of the spin-spin autocorrelations.
So time-translational self-consistent solutions are possible,
but it cannot be excluded that solutions exist which are not
time-translationally invariant. Here, we do not consider them
because they cannot occur in a quantum system at infinite temperature
 although time-crystalline behavior cannot be
excluded generally for specific situations. 

In practice, the averaging based on Monte-Carlo sampling
allows for small unphysical violations of time translational invariance
due to the finite statistical error: the
self-consistently computed moments and autocorrelations may not be exactly 
time-translation invariant, even though the initially inserted moments 
have this property.
To avoid violations of time translational invariance, we 
 enforce it in each iteration step. A simple implementation consists
in setting
\begin{subequations}
\begin{align}
	t_1 &= \Delta t, &t_2 &= 0, &\text{for} \quad t_1&>t_2, \\
	t_2 &= \Delta t, &t_1 &= 0, &\text{for} \quad t_2&>t_1,
\end{align}
\end{subequations}
where $\Delta t = |t_1 - t_2|$, instead of considering all pairs of times $t_1,t_2$.

The reduced set of data together with the assumption of time translation invariance 
provides all required information. 
Due to the time-translational invariance the covariance matrix 
has constant matrix elements on all diagonals 
\begin{align}
	\dul{M}^{\alpha\beta} &=
	\begin{pmatrix}
	a & b & c &  \\
\tilde{b} & a & b & \ddots\\
	\tilde{c} & \tilde{b} & a & \ddots\\
	 & \ddots & \ddots & \ddots \\
	\end{pmatrix},
	\label{eqn:covmatToeplitz}
\end{align}
where
\begin{subequations}
\begin{align}
	a &= \overline{V^{\alpha}(0)V^{\beta}(0)}, & & \\
	b &= \overline{V^{\alpha}(0)V^{\beta}(\delta t)}, &
	\tilde{b} &= \overline{V^{\alpha}(\delta t)V^{\beta}(0)}, \\
	c &= \overline{V^{\alpha}(0)V^{\beta}(2\delta t)}, &
	\tilde{c} &= \overline{V^{\alpha}(2\delta t)V^{\beta}(0)}.
\end{align}
\end{subequations}
This procedure has two advantages: (i)  time-translation invariance 
is built-in by construction; (ii) the autocorrelations need only be computed at 
the time differences $\Delta t$. Hence,  the effort of 
steps \ref{item:EstimateSpin} and \ref{item:SampleAverage} 
of the numerical procedure is reduced from $\mc{O}(L^2)$ to $\mc{O}(L)$
where $L$ is the number of time steps.

\subsection{Definiteness of the Covariance Matrix}
\label{app:definitecov}

In the self-consistency conditions the two-time correlations
are taken as matrix elements of a covariance matrix. The 
physical justification is given in Sect.~\ref{sec:approach}.
In addition, we discuss whether this identification is
mathematically possible.  To this end, the quantum correlations must provide
(i) real symmetric matrix elements and the resulting matrix must be (ii)
positive semi-definite.

In Sect.~\ref{subsec:numproc} we already showed property (i)
 at infinite temperature. Property (ii) is equivalent {to} 
\be
	\sum_{\alpha\beta} \sum_{t,t' \in I} \langle \mb{V}_{i}^{\alpha}(t) 
	\mb{V}_{i}^{\beta}(t') \rangle \, \lambda^{{\alpha}}_t \lambda^{{\beta}}_{t'} \geq 0,
\ee
for arbitrary real coefficients  $\lambda_t^{{\alpha}}$ and $t \in I$ where $I$ is an arbitrary set of time
instants. By shifting the sums into the expectation value we find
\be
	\Bigl\langle \Bigl( \underbrace{\sum_{\alpha} \sum_{t \in I} \mb{V}_{i}^{\alpha}(t) \lambda^{{\alpha}}_t}_{:= \mb{B}} \Bigr)^2 \Bigr\rangle \geq 0.
	\label{eqn:possemi}
\ee
Since $\mb{B}$ is a Hermitian operator, its square is a non-negative operator and
thus any of its expectation values {are} non-negative.
This proves the required positive semi-definiteness
and applies to the expectation values of the  path integral \eqref{eq:correlation} 
and extends to the averages \eqref{eqn:sampleav}.

There is, however, a subtlety in the implementation described in the previous section.
We guaranteed time translation invariance by using
\be
	\frac1{M}\sum_{\vec{\mc{V}}} g^{\alpha\beta}_{\vec{\mc{V}}}(t_1,t_2) 
	\approx \frac1{M}\sum_{\vec{\mc{V}}} g^{\alpha\beta}_{\vec{\mc{V}}}(\Delta t,0).
\ee
For $M=\infty$ this is relation holds exactly true. But the statistical fluctuations
at finite $M$ imply that the covariance matrix computed by Eq.\ \eqref{eqn:covmatToeplitz}
need not be non-negative. If we refrained from setting all matrix elements
on a diagonal constant, but used the two-time autocorrelations, $\dul{L}$ would be non-negative.
Enforcing time translation invariance breaks the positive semi-definiteness
to the extent of the statistical standard deviation. Since we keep this deviation
low the computed covariance matrices are very close to being non-negative.
Due to the Monte Carlo approach to the averaging all numerical results are
subjected to the  statistical error so that the statistical inaccuracies
of the approximate covariance matrix are no source of additional deviations.

Practically, we diagonalize the approximate covariance matrix and set the
{eigenvalues} which are negative to zero for the next iteration step.
We stress that the modulus of these negative {eigenvalues} is of
the order of the statistical error, i.e., small in a systematically controlled
way. Therefore, the employed procedure provides consistent physical solutions
for the dynamic autocorrelations within the discussed errors.

\end{appendix}

\end{document}